\documentclass[11pt, a4paper]{article}
\pdfoutput=1
\usepackage{jcappub,amsmath,amssymb,amsfonts,hyperref}

\usepackage{bm}
\usepackage[table]{xcolor}
\usepackage{amsmath}
\usepackage{amssymb}
\usepackage{bbold}
\usepackage{pgfplots}
\usepackage{wasysym}

\usepackage{graphicx}
\usepackage{feynmp}
\usepackage{color}
\usepackage{epsfig}

\usepackage[utf8]{inputenc}

\numberwithin{equation}{section}

\definecolor{MyBlue}{rgb}{0.15,0.15,0.70}

\hypersetup{
colorlinks=true,
citecolor=MyBlue,
linkcolor=MyBlue,
urlcolor=MyBlue
}

\usepackage{amssymb}
\usepackage{amsmath}
\usepackage{amsfonts}
\usepackage{upgreek}
\usepackage{latexsym}

 \newcommand{\SSS}{Schwarzschild\,\,}

\newcommand{\nn}{\nonumber}

\def\lsim{\raise 0.4ex\hbox{$<$}\kern -0.8em\lower 0.62
ex\hbox{$\sim$}}

\def\gsim{\raise 0.4ex\hbox{$>$}\kern -0.7em\lower 0.62
ex\hbox{$\sim$}}

\def\lbar{{\hbox{$\lambda$}\kern -0.7em\raise 0.6ex
\hbox{$-$}}}

\newcommand\p{\partial}

\newcommand\ee{\end{equation}}
\newcommand\be{\begin{equation}}
\def\bea{\begin{array}}
\def\eea{\end{array}}\def\ea{\end{array}}
\newcommand\ees{\end{eqnarray}}
\newcommand\bees{\begin{eqnarray}}
\def\nn{\nonumber}

\def\dslash{\hspace{-1mm}\not{\hbox{\kern-2pt $\partial$}}}
\def\Dslash{\not{\hbox{\kern-4pt $D$}}}
\def\pslash{\not{\hbox{\kern-2.1pt $p$}}}
\def\kslash{\not{\hbox{\kern-2.3pt $k$}}}
\def\qslash{\not{\hbox{\kern-2.3pt $q$}}}

\def\p1{{\bf p}_1}
\def\p2{{\bf p}_2}
\def\k1{{\bf k}_1}
\def\k2{{\bf k}_2}

\newcommand{\dddM}{\kern 0.2em \raise 1.9ex\hbox{$...$}\kern -1.0em \hbox{$M$}}
\newcommand{\dddQ}{\kern 0.2em \raise 1.9ex\hbox{$...$}\kern -1.0em \hbox{$Q$}}
\newcommand{\dddI}{\kern 0.2em \raise 1.9ex\hbox{$...$}\kern -1.0em\hbox{$I$}}
\newcommand{\dddJ}{\kern 0.2em \raise 1.9ex\hbox{$...$}\kern-1.0em
\hbox{$J$}}
\newcommand{\dddcalJ}{\kern 0.2em \raise 1.9ex\hbox{$...$}\kern-1.0em
\hbox{${\cal J}$}}

\newcommand{\dddO}{\kern 0.2em \raise 1.9ex\hbox{$...$}\kern -1.0em
\hbox{${\cal O}$}}
\def\dddz{\raise 1.5ex\hbox{$...$}\kern -0.8em \hbox{$z$}}
\def\dddd{\raise 1.8ex\hbox{$...$}\kern -0.8em \hbox{$d$}}
\def\dddbd{\raise 1.8ex\hbox{$...$}\kern -0.8em \hbox{${\bf d}$}}
\def\ddbd{\raise 1.8ex\hbox{$..$}\kern -0.8em \hbox{${\bf d}$}}
\def\dddx{\raise 1.6ex\hbox{$...$}\kern -0.8em \hbox{$x$}}
\newcommand\spart{\;\raise1.0pt\hbox{/}\hskip-6pt\partial}
\newcommand\spartb{\;\overline{\raise1.0pt\hbox{/}\hskip-6pt\partial}}

\newcommand{\PotE}{\Xi}

\newcommand{\gr}[1]{\mathbf{#1}}
\newcommand{\beqa}{\begin{eqnarray}}
\newcommand{\eeqa}{\end{eqnarray}}
\newcommand{\dd}{\mathrm{d}}

\newcommand{\tinbar}{\bar t_{\rm in}}
\newcommand{\rin}{r_{\rm in}}
\newcommand{\rinbar}{\bar r_{\rm in}}

\newcommand{\hattin}{\hat t_{\rm in}}
\newcommand{\hattinbar}{\hat{\bar t}_{\rm in}}

\newcommand{\hatrinbar}{\hat{\bar r}_{\rm in}}

\newcommand{\troisj}[6]{\left(\begin{array}{ccc}
      #1 & #2 & #3 \\
      #4 & #5 & #6\end{array}\right)}

\begin{document}

\title{ Are we living near the center of a local void?}
%\title{A scale-free model of bigravity}

\author[a]{Giulia Cusin,}
\author[b]{Cyril Pitrou,}
\author[b]{Jean-Philippe Uzan}

\affiliation[a]{D\'epartement de Physique Th\'eorique and Center for Astroparticle Physics,  
Universit\'e de Gen\`eve, 24 quai Ansermet, CH--1211 Gen\`eve 4, Switzerland}
\affiliation[b]{Institut d'Astrophysique de Paris, Universit\'e  Pierre~\&~Marie Curie - Paris V\\
CNRS-UMR 7095, 98 bis, Bd Arago, 75014 Paris, France\\
Sorbonne Universit\'es, Institut Lagrange de Paris, 98 bis Bd Arago, 75014 Paris, France}

\emailAdd{giulia.cusin@unige.ch}
\emailAdd{pitrou@iap.fr}
\emailAdd{uzan@iap.fr}

\date{today}

\abstract{
The properties of the cosmic microwave background (CMB) temperature and polarisation anisotropies measured by a static, off-centered observer located in a local spherically symmetric void, are described. In particular in this paper we compute, together with the standard 2-point angular correlation functions,  the off-diagonal correlators, which are no more vanishing by symmetry. While the energy shift induced by the off-centered position of the observer can be suppressed by a proper choice of the observer velocity, a lensing-like effect on the CMB emission point remains. This latter effect is genuinely geometrical (e.g. non-degenerate with a boost) and reflects in the structure of the off-diagonal correlators. At lowest order in this effect, the temperature and polarisation correlation matrices have non-vanishing diagonal elements, as usual, and all the off-diagonal terms are excited. This particular signature of a local void model allows one, in principle, to disentangle geometrical effects from local kinematical ones in CMB observations.}


%\notoc
\setcounter{tocdepth}{2}

\maketitle

%
%\vspace*{2cm}
%
%\center{\Large\bf Are we living near the center of a local void ?}
%
%\vskip 0.4cm
%\vskip 0.7cm
%\centerline{\large Giulia Cusin$^a$, Cyril Pitrou$^b$, Jean-Philippe Uzan$^b$}
%\vskip 0.3cm
%\centerline{\em $^a$D\'epartement de Physique Th\'eorique and Center for Astroparticle Physics,}  
%\centerline{\em Universit\'e de Gen\`eve, 24 quai Ansermet, CH--1211 Gen\`eve 4, Switzerland}
%
%\vspace{3mm}
%\centerline{\em $^b$ Institut d'Astrophysique de Paris, Universit\'e  Pierre~\&~Marie Curie - Paris VI}
%\centerline{\em CNRS-UMR 7095, 98 bis, Bd Arago, 75014 Paris, France}
%\centerline{\em Sorbonne Universit\'es, Institut Lagrange de Paris, 98 bis Bd Arago, 75014 Paris, France}
%
%\vskip 1.9cm

\newpage

\section{Introduction}

In the standard lore of the construction of a cosmological model~\cite{pu-book,Ruth_book} the universe on large scale is assumed to be spatially homogeneous and isotropic. In this framework a class of privileged fundamental observers is naturally identified. This class of reference observers is a theoretical construct and any (real) observer should be able to (1) identify this privileged cosmological reference frame and (2) determine his peculiar velocity with respect to this frame, using his observations.

This has probably been best achieved with the analysis of the cosmic microwave background (CMB). In the standard interpretation, the observed large amplitude of the CMB temperature dipole is interpreted as the Doppler effect associated to our motion with respect to the CMB rest frame, assumed to coincide with the one of the fundamental observers. Assuming that the whole CMB dipole is of Doppler origin (i.e. it arises from the boost of the CMB monopole), one concludes~\cite{kogut93,fixsen96,hinshaw09} that our velocity is $v=(369\pm0.9)~\rm{km}\cdot{\rm s}^{-1}$ in the direction $(l,b)=(263^{\rm o}.99\pm0^{\rm o}.14,48^{\rm o}.26\pm0^{\rm o}.03)$. Besides this dominant effect, a boost induces other observable effects on the CMB: (1) a {\em modulation}, which gives rise to an amplification of the apparent temperature in the direction of the motion (similar to the dipole as a boosting of the monopole);
(2) an {\em aberration} effect, which shifts the apparent position of fluctuations toward the velocity direction and changes the angular scale, hence shrinking the anisotropy on one half of the sky and stretching it on the other half; (3) a quadrupole induced by the dipole~\cite{KK03}. Finally (4) a boost affects polarization since it generates $B$-modes from $E$-modes. 

Both modulation and aberration also induce couplings among neighboring multipoles. Indeed, the observed temperature $\tilde \Theta$ can be related to the one in the CMB frame $\Theta$ by~\cite{CvL02} 
\be\label{primiere}
\tilde \Theta({\tilde{\bm n}}) = \frac{\Theta({{\bm n}})}{\gamma\left(1-\tilde{\bm n}\cdot{\bm v}/c \right)}\,,
\ee
with $\gamma=(1-\beta^2)^{-1/2}$ and $\beta=v/c$. The multiplicative factor in eq. (\ref{primiere}) has the effect of inducing couplings on all scales between neighboring multipoles of the correlation function. The detectability of these effects, was discussed in Refs.~\cite{Kosowsky:2010jm,BR06,Amendola:2010ty} and the effects were shown to be observable by the {\em Planck} satellite.  Such a measurement was later performed by {\em Planck} \cite{Aghanim:2013suk} and the result confirmed this standard kinematic interpretation.

Despite this strong case for a Doppler interpretation, the possibility that the \emph{anomalous} amplitude of the dipolar modulation might have a non-kinematical origin has been considered, raising the more fundamental question that a Doppler-like modulation can in fact have a geometrical origin, i.e. that it would originate from our universe not being spatially homogeneous and/or isotropic, e.g. because of the existence of a local void. Indeed, in full generality one expects that both effects (i.e. the kinematical and the geometrical ones) have to be considered and ideally one should be able to disentangle them from CMB observations.

In particular, the idea that the CMB dipole can arise from a large scale  isocurvature perturbation was considered in Ref.~\cite{Langlois:1995ca}. Such a perturbation was modeled by considering a spherically symmetric spacetime of the Lema\^{\i}tre-Tolman-Bondi (LTB) family as a perturbation of a Friedmann-Lema\^{\i}tre-Robertson-Walker (FLRW) spacetime. More recently, Ref.~\cite{Roldan:2016ayx} argued that a large scale dipolar gravitational potential could mimic a Lorentz boost. In particular, because of lensing such a gravitational potential can induce mode couplings similar to aberration and modulation. Anyway this requires both a fine-tuning of the radial profile of the potential and a primordial dipolar potential.\\

The goal of this article is to fully characterize the effect of kinematics (local boost) and geometry (local void) on temperature and $E$-, $B$-modes of polarization. To that purpose, we consider two models:
\begin{itemize}
\item a standard model in which the universe is described by a FLRW spacetime, allowing for a boost of the observer with respect to the cosmological frame. As emphasized earlier, this has been extensively studied but it will serve as a reference for comparison. In this analysis the small parameter in which analytical results are expanded is the boost velocity $\beta$;
\item a model of universe consisting of a spherical void with an overdense central region described by a Kottler spacetime and embedded in a FLRW universe. This construction is known as a Swiss-cheese model~\cite{ES45a,ES45b,Fleury:2014gha,Fleury:2013sna}. The Kottler spacetime is the generalization of the Schwarzschild (\emph{Sch}) spacetime to the case of a non-vanishing cosmological constant. For simplicity, when deriving analytic expressions, we shall assume that at late time the universe is fully matter dominated and we will thus describe the void by a \emph{Sch} spacetime. We shall not assume the observer to seat at the center of symmetry so that he will observe an axially symmetric spacetime. The last scattering surface is described as a constant time hypersurface lying in the FLRW region. In this analysis, two small parameters come into play:  (1) the ratio between the radial displacement of the observer from the center of the void $D$ and the radius of the void $\chi_h$ (noted $\hat D$) and (2) the ratio between the radii of the void and of the last scattering surface ($\chi_h/\chi_{\text{LSS}}$) which we show to be proportional to $\sqrt{\hat r_S} \equiv \sqrt{r_S/\chi_h}$ where $r_S$ is the Schwarzschild radius ($r_S\equiv 2GM$) of the \emph{Sch} region. 
\end{itemize}
In both cases our goal is to compute the 2-point angular correlation functions and the off-diagonal correlators. In particular, we shall compute analytically  all the effects induced on the CMB (temperature and polarisation) related to the off-center position of an observer in the void. For an observer who does not seat at the center of symmetry, light deflection  generates $B$-modes from $E$-modes at first order in lensing.\footnote{On the other hand, for a spherically symmetric situation, $B$-modes are generated from $E$-modes as a coupling between the lensing potential and CMB polarisation. This is considered as a second order effect, but also linear in the lensing.} This was already investigated in Ref.~\cite{Goto:2011ru}, an analysis that will be refined in our study. \\

Technically, while the first model has been studied in various works, the second requires to go through a series of technical steps. 
\begin{enumerate}
\item First, we have to describe the geometry of the void and how it is matched to the outside FLRW region. Among the matching conditions, we find that  the radius of the boundary $r_h(t)$ seen from the  void is expanding with the cosmological scale factor $a(t)$ as 
\be
r_h(t) = a(t) \chi_h\,,
\ee
where $\chi_h$ is the constant comoving radius of the void as seen from the FLRW region.

\item Second, the geodesic equation for an off-center observer needs to be solved and the resulting trajectory expressed in terms of the angle $\theta_{\text{obs}}$ between the off-center direction and the direction of observation. Two main effects have to be considered: (1) the bending of the geodesic and more generally its deformation due to the propagation in the \emph{Sch} region. This has the effect of deforming the last scattering surface located in the FLRW region in two ways: an orthoradial displacement similar to lensing, and a radial displacement similar to time-delay or Shapiro potential effect; (2) the modification of the energy of the emitted photons which adds to the Sachs-Wolfe effect, located on the last scattering surface. Both the deformation of the photon trajectory and energy modulation are non-local effects. If the local velocity of the observer is carefully chosen, we find that there is no effect at order $\sqrt{\hat r_S}$. Furthermore at order $\hat r_S$ there is no additional energy modulation, and only the effects of lensing-like deflection and radial displacement come into play. The lensing-like effect is the dominant contribution and its leading order term when expanded in powers of $\hat D$ is
\be\label{MainResult}
\theta_{\text{LSS}} - \theta_{\text{obs}} \simeq \frac{\hat r_S}{\hat D} \tan \frac{\theta_{\text{obs}}}{2}\,.
\ee

\item Finally, by taking into account this dominant effect, we are able to find analytic expressions for  angular power spectra of temperature and $E$- and $B$-modes of polarization, as well as for the off-diagonal correlators. For instance, we find that the off-diagonal correlator of the observed temperature anisotropy field $\tilde{\Theta}$  has non-vanishing matrix elements of the form 
\be
\langle \tilde{\Theta}_{\ell m}\tilde{\Theta}_{\ell+L\,m+M}^* \rangle\,,
\ee 
with no restriction on the value of $L$, due to the off-center position of the observer inside the local void. As a consequence of eq. (\ref{MainResult}), the ratio between these off-diagonal correlators and the isotropic diagonal correlators $C_\ell$ (which are the usual correlators generated in a perturbed FLRW geometry) are typically proportional to the geometrical factor $\hat r_S/\hat D$. This has to be compared with the kinematical effect of a local boost which generates correlators of  these types only for $L=1$ at lowest order in the boost parameter $\beta$.
\end{enumerate}

The paper is organized as follows.  Light propagation in the void model is detailed in section \ref{geometry}. In particular, in section \ref{geometrys} we describe the geometry of our void model, in section \ref{geodesic2} we detail  the general method to solve for the geodesic, and in \ref{geodesic3} we present an analytic method to determine the emission point on the last scattering surface (LSS) given the reception time and direction. In section \ref{numerics} the results of the analytic analysis are compared with the numerical resolution. In section \ref{boost}  we calculate the contributions of the lensing-like deflection and radial modulation, together with the energy modulation, at order ${\hat r_S}^{1/2}$ and $\hat{r}_S$. Separating the contribution which is non-degenerate with the effect of a boost, so as to isolate geometrical contributions from kinematical ones, we show that no geometrical contributions are present at order ${\hat r_S}^{1/2}$. The explanation of this result is discussed in section \ref{discussion physics}. In section \ref{CMBoff} we analyze the CMB sky seen by an off-center observer in the void and we calculate the temperature and polarization correlation functions, still focussing on contributions non-degenerate with the effects of a boost of the observer. Conversely, in section \ref{Boost} we turn to a FLRW model and we calculate correlation functions of CMB observables for an observer whose reference frame is in motion with respect to the CMB rest frame. To facilitate the reading, several technical details and intermediate calculations are relegated in the appendices. 

\section{Light propagation from emission until reception}\label{geometry}

\subsection{Spacetime description}\label{geometrys}

Outside the local void, the geometry is described by the standard spatially Euclidean FLRW metric
\begin{equation}
\dd s^2=-\dd T^2+a^2(T)\left[\dd\chi^2+\chi^2\dd\Omega^2\right] = a^2(\eta)\left[-\dd\eta^2+\dd\chi^2+\chi^2\dd\Omega^2\right]\,,\label{FL}
\end{equation}
where $T$ and $\eta$ denote cosmic time and conformal time respectively, and where $\dd \Omega^2=\dd\theta^2+\sin^2\theta \,\dd\phi^2$. From Einstein equations, it follows that the scale factor $a(T)$ satisfies the Friedmann equation
\be\label{Friedmann}
H^2=\frac{8\pi G}{3}\rho+\frac{\Lambda}{3}\,,\hspace{1 em}\text{with}\hspace{1 em}H\equiv \frac{1}{a}\frac{\dd a}{\dd T}\,.
\ee

Inside the hole the geometry is described by the extension of the \SSS (\emph{Sch}) metric to the case of a nonzero cosmological constant, namely the Kottler solution. In spherical coordinates $(t, r, \theta, \phi)$ it can be written as
\be
\dd s^2=-A(r)\dd t^2+A^{-1}(r)\dd r^2+r^2\dd\Omega^2\,,\qquad
A(r)\equiv 1-\frac{r_S}{r}-\frac{\Lambda r^2}{3}\,,
\ee
and $r_S\equiv 2 G M$ is the Schwarzschild radius associated with the mass $M$ at the center of the hole. This solution describes the vicinity of a gravitationally bound object such as a galaxy or a cluster of galaxies and therefore it should only be valid for $r>r_{\text{phys}}$  where $r_{\text{phys}}$ is the physical size of the object.

Two spacetimes can be glued together on an hypersurface $\Sigma$, if the Israel junction conditions \cite{Israel:1966rt} are satisfied. Explicitly, both geometries must induce: (a) the same 3-metric and (b) the same extrinsic curvature on $\Sigma$. In our model, the symmetry of the problem imposes that the junction hypersurface is a world sheet comoving 2-sphere defined by $\chi=\chi_h=cnst$  in FLRW coordinates and by $r=r_h(t)$ in Kottler coordinates. The first junction condition implies
\be\label{1junction}
r_h(t)=a(T) \chi_h\,,\hspace{2 em}\frac{\dd T}{\dd t}=\kappa (t)\,,\qquad\kappa(t)\equiv\sqrt{\frac{A^2[r_h(t)]-\dot{r}_h^2(t)}{A\left[r_h(t)\right]}}\,,
\ee
while the second junction condition is satisfied only if
\be\label{2junction}
\kappa(t)=A[r_h(t)]\,.
\ee
Note that these conditions can be extended to the  case of a FLRW geometry with curved spatial sections. Putting eqs. (\ref{1junction}) and (\ref{2junction})  together, we get the equation governing the dynamics of the hole boundary, i.e.
\be\label{dynamicsbounday}
\dot{r}_h(t) =A[r_h(t)]\sqrt{1-A[r_h(t)]}\,\quad \Longrightarrow \quad \frac{\dd u_h}{\dd t}=-u_h(t)^2A(u_h(t))\sqrt{1-A(u_h(t))}\,,
\ee
where for future convenience we introduced $u_h \equiv 1/r_h$. Furthermore, the junction conditions (\ref{1junction}, \ref{2junction})  together with the Friedmann equation (\ref{Friedmann}) imply that the Kottler and the FLRW regions have the same cosmological constant and that
\be\label{Mofrho}
M=\frac{4 \pi}{3}\rho\,a^3 \chi_h^3\,\quad\Longrightarrow \quad H(T) = \sqrt{\frac{r_S}{r_h^3(t(T))}} = \sqrt{\frac{r_S}{a^3(T) \chi_h^3}}\,,
\ee
thus requiring that the matter filling the FLRW outside region is pressureless and scales as $\rho = \rho_0 (a_0/a)^3$. For details on the derivation of the matching conditions above, see Ref. \cite{Fleury:2013sna}. Finally, in order to simplify the analysis of light propagation we shall assume in the remainder of this article that the cosmological constant vanishes ($\Lambda = 0$). Hence our description of the local void with this model corresponds only to a matter dominated era and the hole is a \emph{Sch} region.

\subsection{Propagation of CMB light rays}\label{geodesic2}

We consider an observer lying inside the \emph{Sch} region who receives a photon emitted by a point source on a constant cosmic time hypersurphace  $\Sigma_{\text{LSS}}$ lying in the FLRW region. We identify this hypersurface with the LSS, described  in FLRW coordinates as the hypersurface $\eta=\eta_{\text{LSS}}={\rm cnst}$.

The photon is emitted with wave vector $k^{\mu}_{\text{LSS}}$, enters into the hole with wave vector $k^{\mu}_{\text{in}}$ and reaches the observer with wave vector $k^{\mu}_{\text{o}}$. We respectively denote with $\mathcal{E}_{\text{LSS}}$, $\mathcal{E}_{\text{in}}$ and $\mathcal{E}_{\text{o}}$ the corresponding events. The coordinates of the first event are expressed with respect to the FLRW frame, e.g. in spherical coordinates $(\eta_{\text{LSS}}, \chi_{\text{LSS}}, \theta_{\text{LSS}}, \phi_{\text{LSS}})$,  while the coordinates of the last event $\mathcal{E}_{\text{o}}$ are expressed in the \emph{Sch} spherical coordinate system, e.g. $(t_{\text{o}}, r_{\text{o}}, \theta_{\text{o}}, \phi_{\text{o}})$. The coordinates of $\mathcal{E}_{\text{in}}$ can be either expressed with respect to the FLRW spherical coordinates, e.g.  $(\eta_{\text{in}}, \chi_{\text{in}}, \theta_{\text{in}}, \phi_{\text{in}})$ or with respect to the \emph{Sch} spherical coordinate system, e.g. $(t_{\text{in}}, r_{\text{in}}, \theta_{\text{in}}, \phi_{\text{in}})$. 

\begin{figure}[ht!]
    \centering
    % \subfigure[\label{chiLSS}]
     {\includegraphics[scale=0.42,angle=0]{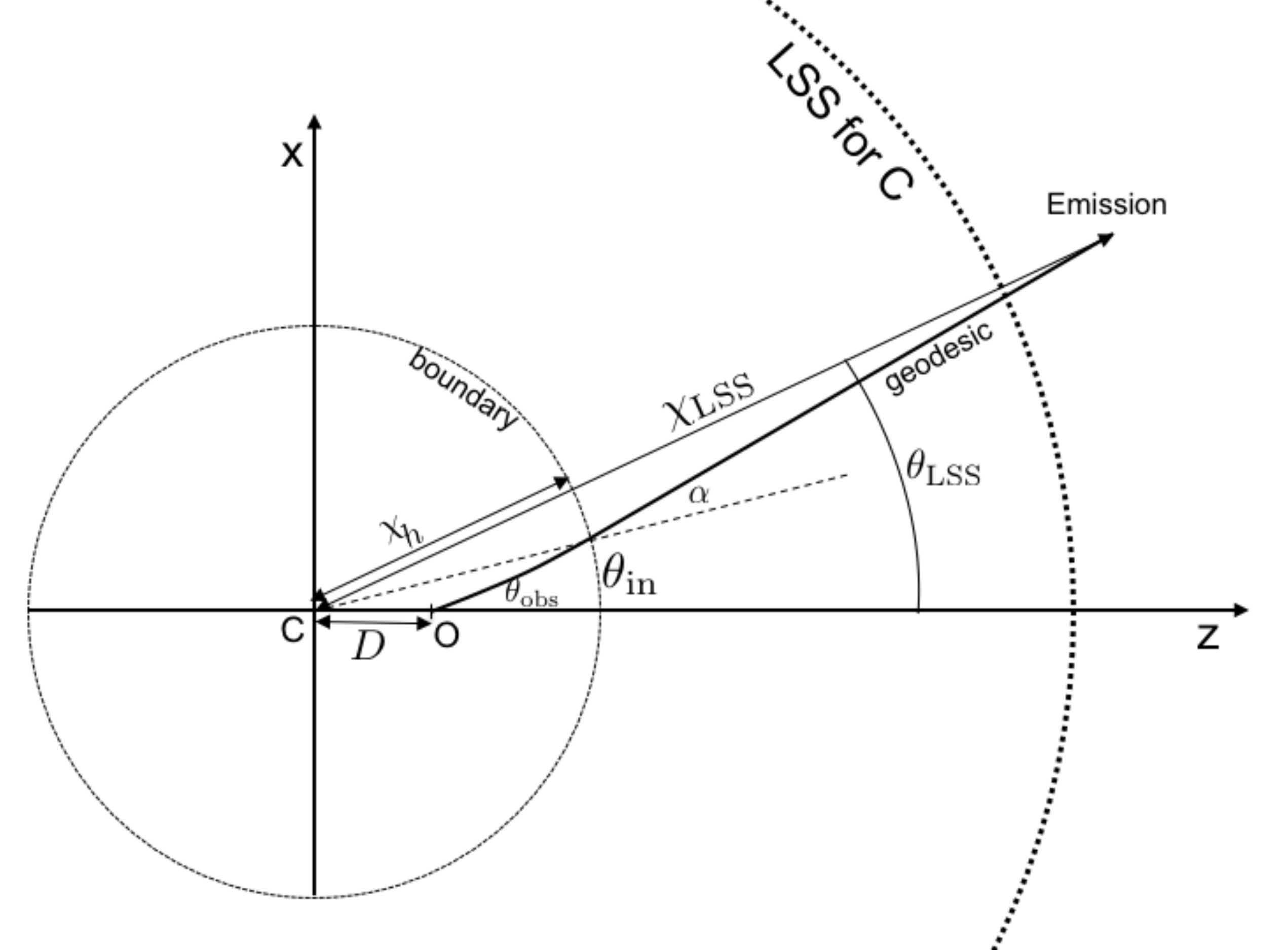}}
     \caption{\label{fig1} \small Schematic representation of our void model in the $x-z$ plane. $C$ is the center of the void and the observer $O$ is located at a distance $r=D$ (in Schwarzschild coordinates) from the center.}
  \end{figure}

For convenience, our calculations trace a photon backward in time. Starting from $\mathcal{E}_o$ we first determine $\mathcal{E}_{\text{in}}$  and second $\mathcal{E}_{\text{LSS}}$. In this way we associate to a given $\mathcal{E}_o$  (i.e. to a given angle at which the photon is received) the point on the LSS from which the photon is emitted, identified by the FLRW coordinates of $\mathcal{E}_{\text{LSS}}$. Furthermore, since the trajectory of the light ray is necessarily contained in a plane, we can work in the $x-z$ plane, that is ($\phi_{\text{o}} = \phi_{\text{in}}=\phi_{\text{LSS}} =0\,\,{\rm or}\,\,\pi$, and $k^\phi_{\text{o}}=0$). Moreover, we choose the position of the observer at time $t_o$ at which the photon is received on the $\hat{z}$ axis, i.e. $\mathcal{E}_o=(t_o, r_{\text{o}}, \theta_{\text{o}}, \phi_{\text{o}})\equiv(t_o, D, 0, 0)$. A schematic view of our void model is presented in Fig. \ref{fig1}.

\subsubsection{Geodesics inside the local void}\label{preliminary}

In the \emph{Sch} region, the existence of two Killing vectors associated with staticity ($\partial/\partial t$) and spherical symmetry ($\partial/\partial \theta$) implies the existence of two conserved quantities $\tilde{E} \equiv -k_t$ and $L\equiv -k_\theta$ and to determine the trajectory of the photon we do not need to solve the geodesic equation ($k^\nu \nabla_\nu k^\mu = 0$). Using these conserved quantities together with the condition $k^\mu k_\mu=0$, the wave vector components can be written as
\be\label{geodesicKottler}
k^{\phi}=0\,,\quad A(r)k^t=\tilde{E}\,,\hspace{2 em} r^2k^\theta =- L\,,\quad \left(k^r\right)^2+\left(\frac{L}{r}\right)^2 A(r)=\tilde{E}^2\,.
\ee \normalsize
In particular, at the position of reception by the observer ($r=D$), the wave vector is given by
\be\label{ko}  \small
\left(k^{\phi}\right)_o=0\,,\quad \left(k^t\right)_o=\frac{\tilde{E}}{A(D)}\,,\quad\left(k^{\theta}\right)_o=-\frac{L}{D^2}\,,\quad\left(k^r\right)_o=\pm \sqrt{\tilde{E}^2-A(D)\left(\frac{L}{D}\right)^2}\,.
\ee \normalsize
For an observer with four-velocity $u^{\mu}$ ($u_{\mu}u^{\mu}=-1$) the spatial direction of light propagation is defined as the opposite of the direction in which the signal is measured $n^\mu$ ($n^\mu n_\mu = 1$ and $n^\mu u_\mu=0$). Hence the wavevector is decomposed as\footnote{In other terms, introducing a space projector $\mathcal{S}_{\mu\nu}=g_{\mu\nu}+u_{\mu}u_{\nu}$, we have $n^{\mu}\equiv -1/E\,\mathcal{S}^{\mu}_{\nu}k^{\nu}$.}
\be\label{directionobs}
k^{\mu}=E \left(u^{\mu}-n^{\mu}\right)\,.
\ee
In particular, for a \emph{Sch} static observer at position $r$, whose velocity is $u^\mu = 1/\sqrt{A(r)}(\partial_t)^\mu$, the energy $E$ and direction $n^\mu$ of the photon are
\be\label{EnergyAndDirection}
E= \frac{\tilde E}{\sqrt{A(r)}}\,,\quad n_\mu e_\theta^\mu \equiv \sin \theta_{\text{rad}} = \frac{b}{r} \sqrt{A(r)} \,,\quad  n_\mu e_r^\mu \equiv \cos \theta_{\text{rad}}= \sqrt{1-\frac{b^2}{r^2} A(r)}\,,
\ee
where $e_r^\mu\equiv \sqrt{A(r)}(\partial_r)^\mu$ and $e_\theta^\mu\equiv 1/r(\partial_\theta)^\mu$ are respectively the unit radial and unit orthoradial vectors, and  we denote with  $b=L/\tilde E$ the impact parameter. The angle $\theta_{\text{rad}} $ is the angle between the direction of propagation $n^\mu$ and the radial unit vector $e_r^\mu$. In particular, for the observer this angle corresponds to the angle of observation with respect to the $\hat{z}$ axis, that is with the direction connecting the center of the \emph{Sch} region to its position, and we note $\theta_{\text{obs}} \equiv (\theta_{\text{rad}})_{\text{o}}$. Recalling $k^\mu = \dd x^\mu/\dd v$, and using the notation $u=1/r$, eqs. (\ref{geodesicKottler}) lead to differential equations for the geodesic
\be\label{radialphoton}
\left(\frac{\dd u}{\dd t}\right)^2=\frac{u^4}{\epsilon_1^2}P(u)A^2(u)\,,\hspace{1 em}r_S^2\left(\frac{\dd u}{\dd\theta}\right)^2=P(u)\,,\hspace{1 em}\frac{1}{E^2}\left(\frac{\dd u}{\dd v}\right)^2=\frac{u^4}{\epsilon_1^2}P(u)\,,
\ee
with
\be
A(u)\equiv1-r_S u\,,\hspace{1 em}P(u)\equiv \epsilon_1^2-r_S^2u^2 A(u)=\epsilon_1^2 \cos^2 \theta_{\text{rad}}\,,\qquad \epsilon_1\equiv r_S/b\,.
\ee

The radius $r_{\text{in}}$ (or $u_{\text{in}}$) and the time $t_{\text{in}}$ at entrance are determined by comparing the radial dynamics of the first relation in eqs. (\ref{radialphoton}) with the one of the boundary, eq. (\ref{dynamicsbounday}). Once $u_{\text{in}}$ is known, then from integrating the second relation in eqs. (\ref{radialphoton})\,, $\theta_{\text{in}}$ can be determined. Details are gathered in appendix  \ref{AppGeodesicInside}. Finally, once $\mathcal{E}_{\text{in}}$ is determined, the components of the wave vector at the crossing of the void boundary, $(k^{\mu})_{\text{in}}$  are found from eq. (\ref{geodesicKottler}) to be
\be\label{kcompin} \small
\left(k^{\phi}\right)_{\text{in}}=0\,,\quad \left(k^t\right)_{\text{in}}=\frac{\tilde{E}}{A(r_{\text{in}})}\,,\quad\left(k^{\theta}\right)_{\text{in}}=-\frac{L}{r_{\text{in}}^2}\,,\quad\left(k^r\right)_{\text{in}}=-\sqrt{\tilde{E}^2-A(r_{\text{in}})\left(\frac{L}{r_{\text{in}}}\right)^2}\,.
\ee
\normalsize
\subsubsection{Matching of geodesics on the boundary}\label{matching coordinates}

In the previous section, we have determined $\mathcal{E}_{\text{in}}$ and $k^{\mu}_{\text{in}}$ in terms of the \emph{Sch} coordinate system. However, in order to proceed solving the geodesic equation outside the hole, we need to express these quantities in terms of the FLRW coordinate system $(\eta, \chi, \theta, \phi)$.  We choose the FLRW axes parallel to the \emph{Sch} ones in such a way that the angular coordinates inside and outside the hole can be identified, thus justifying our use of the same notation for angles\footnote{This is always possible since the two spacetimes are locally rotationally invariant.}. We remind that the photon trajectory lies in the plane $\phi = 0$ (or $y=0$). Furthermore, from the matching conditions all points on the boundary have $\chi = \chi_h$ so we need only to determine the cosmic (or conformal) time at entrance.

From the first matching condition, eq. (\ref{1junction}), we can immediately extract the value of the scale factor at crossing time $a(T_{\text{in}})=a_{\text{in}}=r_{\text{in}}/\chi_h$. Then from integrating the Friedmann  equation, we get
\be\label{Tinetain}
T_{\text{in}}=\frac{2}{3}\sqrt{\frac{r^{3}_{\text{in}}}{r_S}} \,,\qquad \eta_{\text{in}} = 2 \chi_h \sqrt{\frac{r_{\text{in}}}{r_S}}\,,
\ee
where we choose both cosmic and conformal time to vanish at the singularity. From these simple relations we can calculate the FLRW coordinates of the photon at crossing, once the \emph{Sch} coordinates of the entrance point $\mathcal{E}_{\text{in}}$ are known. 

The first junction condition ensures that the affine connection does not diverge on $\Sigma$. Integrating the geodesic equation ${\rm d}k^{\mu}=-\Gamma_{\alpha\beta}^{\mu}k^{\alpha}k^{\beta} {\rm d}v$ from $v_{\text{in}}^-$ to $v_{\text{in}}^+$ we get that $k^{\mu}$ is continuous at $\mathcal{E}_{\text{in}}$ . Therefore we just need to convert its components from the \emph{Sch} coordinate system to the FLRW one. To this purpose, we first express the normal and tangential vector to the surface $\Sigma$ in both coordinate systems to obtain the relations between coordinates valid on the boundary
\be
a \dd \eta = \dd t -\frac{\sqrt{1-A(r)}}{A(r)} \dd r\,, \,\qquad a\dd \chi = \frac{1}{A(r)}\dd r - \sqrt{1-A(r)}\dd t\,.
\ee
From this we deduce the continuity relations
\begin{align}\label{kofk}
&k^{\chi}_{\text{in}}=\frac{1}{a_{\text{in}}}\left[-\sqrt{1-A(r_{\text{in}}})k^t_{\text{in}}+\frac{1}{A(r_{\text{in}})}k^r_{\text{in}}\right]\,,\\
&k^{\eta}_{\text{in}}=\frac{1}{a_{\text{in}}}\left[ k^t_{\text{in}}-\frac{\sqrt{1-A(r_{\text{in}})}}{A(r_{\text{in}})}k^r_{\text{in}}\right]\,.
\end{align}
The components $k^{\theta}$ and $k^\phi$ of the wave-vector are the same in both coordinate systems.

\subsubsection{Friedmann-Lema\^{\i}tre-Robertson-Walker region}

To integrate the geodesic equation in the FLRW region, it is convenient to work in Cartesian coordinates $x^i\equiv (x, y, z)$. Since we are in the  plane $y=0$, we have $z=\chi\cos\phi$, $x=\chi\sin\phi$. 
The vectors $\partial_i$ are Killing vectors associated to homogeneity, and it follows that $g(\partial_i, k)=k_i$ are constants of motion. Therefore, we can relate the components of the wave vector on the LSS to its components at entrance in the \emph{Sch} region through
\be\label{scaling}
(k^i)_{\text{LSS}}=\left(\frac{a_{\text{in}}}{a_{\text{LSS}}}\right)^2(k^i)_{\text{in}}\,.
\ee
Only energies of photons are affected by expansion,  while the direction of propagation is constant in the FLRW region.
Recalling that we are dealing with a null geodesic, $\left(k^{\eta}\right)^2= \left(k^{x}\right)^2+ \left(k^{z}\right)^2$ the trajectory is solved as
\begin{equation}\label{LSS}
z_{\text{LSS}}=z_{\text{in}}+\frac{1}{\sqrt{1+\mathcal{Q}^2}}\left(\eta_{\text{in}}-\eta_{\text{LSS}}\right)\,,\qquad x_{\text{LSS}}=x_{\text{in}}+\frac{\mathcal{Q}}{\sqrt{1+\mathcal{Q}^2}}\left(\eta_{\text{in}}-\eta_{\text{LSS}}\right)\,, 
\end{equation}
where $\mathcal{Q}\equiv \left(k^x/k^z\right)_{\text{in}}$ from which we immediately get
\be\label{phiLSS}
\chi_{\text{LSS}}=\sqrt{x_{\text{LSS}}^2+z_{\text{LSS}}^2}\,,\hspace{2 em}\theta_{\text{LSS}}=\arctan\left(\frac{x_{\text{LSS}}}{z_{\text{LSS}}}\right)\,.
\ee
We have therefore completely determined the coordinates of the event $\mathcal{E}_{\text{LSS}}$ starting from the position $r=D$ of the observer and the direction $\theta_{\text{obs}}$ under which the photon is observed. The components of the wave vector $k^{\mu}$ on the LSS are completely determined by eq. (\ref{scaling}) together with the null geodesic condition.

\subsection{Analytic results}\label{geodesic3}

\subsubsection{General method}

We find analytic expressions for the photon trajectory, by performing a perturbative expansion in the dimensionless parameter $\hat r_S \equiv r_S/\chi_h$. For simplicity, we build dimensionless quantities for all lengths and times using notation of the type $\hat t \equiv t/\chi_h$, $\hat r = r/\chi_h$, $\hat H \equiv H \chi_h$. 

We take the radial geodesic which follows the $z$ axis, i.e. characterized by $\theta_{\text{obs}}=0$ and thus $b=0$, as a reference. All related quantities for this geodesic are denoted with an overbar, for instance the \emph{Sch} coordinates at boundary crossing  are $(\bar t_{\text{in}},\bar r_{\text{in}},\bar \theta_{\text{in}},\bar \phi_{\text{in}}=0)$. A photon following this geodesic, takes the minimum time to reach the observer once it has crossed the boundary, ${\Delta \bar t}\equiv t_o-\bar{t}_{\text{in}}$ (no deflection is present). We choose the normalization of the scale factor such that $a(\bar{T}_{\text{in}})=1$ and we set $t_o=0$. From the junction condition we therefore get $\bar r_{\text{in}} \equiv r(\bar{t}_{\text{in}})=\chi_h$ or $\hat{\bar{r}}_{\text{in}}=1$ and a simple expression for the Hubble factor at entrance time
\be\label{Hofrs}
\hat H(\bar{T}_{\text{in}})=\sqrt{\hat r_S}\,.
\ee

For a general geodesic corresponding to a direction of observation $\theta_{\text{obs}} \neq 0$, we introduce the following quantities 
$\hat{\delta r}_{\rm in} \equiv \hat r_{\rm in} - \hatrinbar$ and $\hat{\delta t}_{\rm in} \equiv \hat t_{\rm in} - \hattinbar$ which correspond to the differences of radius and time at crossing with respect to the reference geodesic. By construction $\hat{\delta r}_{\rm in} \leq 0$ and $\hat{\delta t}_{\rm in}\leq 0$. In appendix \ref{AppExpand} we show how these quantities can be determined introducing a perturbative expansion in powers of $\sqrt{\hat r}_S$ and  comparing the radial motion of the photon and of the boundary.   Up to first order in $\sqrt{\hat r_S}$ we get \small
\begin{eqnarray}
\delta \hat{r}_{\text{in}}&\simeq& \sqrt{\hat{r}_S}(\delta t_{\text{in}})^{(0)}\,,\qquad \left(\hat{\delta t}_{\rm in}\right)^0\equiv\hat D
(\cos\theta_{\rm obs}-1)- \left[\sqrt{1-\hat D^2 \sin^2 \theta_{\rm
      obs}}-1\right]\,,\label{rin0light}\\
\hat{\delta t}_{\rm in} &\simeq& \left(\hat{\delta t}_{\rm in}\right)^0 - \sqrt{\hat r_S} \left(\hat{\delta t}_{\rm in}\right)^0 \frac{1}{\sqrt{1-\hat D^2 \sin^2 \theta_{\rm obs}}}\,. \label{hatdtlight}
\end{eqnarray}\normalsize
Similarly, we can expand the angle $\theta_{\text{in}}$ at entrance and the conformal time at entrance $\eta_{\text{in}}$ and the results are gathered in appendix \ref{AppExpand}. 

\subsubsection{Deformation of the last scattering surface}

We define $\alpha$ the angle  between the direction of propagation of the photon in the FLRW region and the radial direction at the crossing of the hole boundary. This definition corresponds to $\alpha \equiv (\theta_{\text{rad}})_{\text{in}}$, see Fig.~\ref{fig1}. The angle $\alpha$ can be obtained from
\be\label{aalpha}
\sin\alpha=\chi_h\left(\frac{k^{\theta}}{k^{\eta}}\right)_{\text{in}}=\frac{L}{r_{\text{in}}a_{\text{in}}}\frac{1}{k^{\eta}_{\text{in}}}\,,
\ee
where $L$ is defined below eq. (\ref{EnergyAndDirection}). Appendix \ref{AppExpand} details how this angle can be expanded in powers of $\hat r_S$. Once the angle $\alpha$ is expressed in terms of $\theta_{\text{obs}}$, the position of the emission point on the last scattering surface can be determined. Reminding that the geodesic motion is considered in the $y=0$ plane, the Cartesian coordinates of  $\mathcal{E}_{\text{LSS}}$ are 
\be
\hat{z}_{\text{LSS}}=\cos\theta_{\text{in}}+\cos(\theta_{\text{in}}+\alpha) \left(\hat{\eta}_{\text{in}}-\hat{\eta}_{\text{LSS}}\right)\,,
\ee
\be
\hat{x}_{\text{LSS}}=\sin\theta_{\text{in}}+\sin(\theta_{\text{in}}+\alpha) \left(\hat{\eta}_{\text{in}}-\hat{\eta}_{\text{LSS}}\right)\,,
\ee
from which we easily get the associated spherical coordinates thanks to eq.~(\ref{phiLSS}).

It is possible to obtain expressions exact in the parameter $\hat{D}$ for the coordinates of the event $\mathcal{E}_{\text{LSS}}$. However, since these expressions are not particularly compact and intelligible, we perform an expansion in powers of $\hat{D}$, up to quadratic order. Explicit results are given in appendix \ref{AppExpand}.

\subsubsection{Centering the coordinates on the observer}\label{offset}

The radial distance to the LSS calculated in the previous section  and explicitly given by eq. (\ref{chiLSSS}) has an angular dependence, $\hat{\chi}_{\text{LSS}}(\theta_{\text{obs}})$. If we take the  $\hat{z}$ axis as azimuthal direction and we perform a dipolar decomposition of $\hat{\chi}_{\text{LSS}}$, we find that at order $\hat{r}_S^{(0)}$ only a dipolar modulation remains.  At this lowest order, this modulation is just a consequence of the fact that our spherical coordinate system is not centered on the observer.  At order $\hat{r}_S^{(1/2)}$ we have a dipolar and a quadrupolar modulation while at order $\hat{r}_S$ all the multipoles are excited.  We introduce  the following offset along the $\hat{z}$ axis. 
\be\label{Doff}
{D_{\rm off}}=D-\sqrt{\hat{r}_S} D(1-\hat{D})+r_S\left(1-\frac{5}{2} \hat{D}^2\right)\,.
\ee
This $D_{{\rm off}}$ is defined as the quantity needed to eliminate the dipole in radial modulation and by definition it corresponds to the radial position of the observer measured in the FLRW system of coordinates\footnote{For the \emph{Sch} coordinates, the size of the hole at reception is slightly larger than $\chi_h$, which is its size when the photon following the reference geodesic crosses the boundary. For the reference geodesic the dimensionless difference between crossing time and reception time is at lowest order $1-\hat D$  and thus at reception the boundary dimensionless radius has increased approximately by $\hat H(1-\hat D) = \sqrt{\hat r_S}(1-\hat D) $. The observer is thus located at a fraction $D/[\chi_h (1 + \sqrt{\hat r_S}(1-\hat D) )] \simeq \hat D -\hat D \sqrt{\hat r_S}(1-\hat D)$ of the hole radius. Since the hole has a constant radius $\chi_h$ in the outside FLRW coordinates, the offset of the observer in these coordinates is approximately $D-\sqrt{\hat{r}_S} D(1-\hat{D})$ thus explaining the form of the offset at order $\sqrt{\hat r_S}$.}. We consider a shifted system of coordinates (from now on denoted as ``tilde coordinates") defined by $\tilde{x}=x$, $\tilde{z}=z-D_{\text{off}}$. Following a definition analogous to eq. (\ref{phiLSS}), the new spherical coordinates of the last scattering surface are then found to be
\small 
\begin{align}
\tilde{\theta}_{\text{LSS}}&=\theta_{\text{obs}}-\hat{r}_S^{1/2} \hat{D}\sin\theta_{\text{obs}}-\hat{r}_S \left(\frac{3}{2} \hat{D}^2 \sin^2\theta_{\text{obs}}-\frac{1}{\hat{D}}\right)\tan \frac{\theta_{\text{obs}}}{2}+\mathcal{O}(\hat{r}_S \hat{D}^3)\,,\label{angle}\\
\hat{\tilde{\chi}}_{\text{LSS}}&=1+\frac{2}{\sqrt{\hat{r}_S}}-\hat\eta_{\text{LSS}}-\hat{D}+\hat{D} \,\hat{r}_S^{1/2}+\mathcal{O}(\hat{r}^{1/2}_S \hat{D}^3)+ \nn\label{radialmod}\\
&+\hat r_S\left[-\hat{D}^2+\frac{1}{2}-\frac{3}{2} \cos\theta_{\text{obs}}+\frac{7}{4} \hat{D}^2 +\frac{3}{4} \hat{D}^2 \cos^2\theta_{\text{obs}}+2\log\left(\cos\frac{\theta_{\text{obs}}}{2}\right)\right]+\mathcal{O}(\hat{r}_S \hat{D}^3)\,.
\end{align}
\normalsize
From eq. (\ref{radialmod}) we see that up to order $\hat{r}_S^{1/2}\hat{D}^3$ no angular modulation is left. Actually, this is even true when computing the full expression of the corrections at order $\hat{r}_S^{1/2}$, since corrections affect only the average radius at that order. Beyond this leading order, it is also possible to check that when decomposing  $\tilde{\chi}_{\text{LSS}}$ in spherical harmonics, no dipolar modulation is present up to order $\hat{r}_S$ either, thus justifying our choice (\ref{Doff}) for the offset.

\subsubsection{Lensing-like displacement and radial modulation}

We decompose the radial modulation in the standard way separating the average over angles and the radial modulation around this average
\be \label{Shapirodef}
\tilde{\chi}_{\text{LSS}}(\theta_{\text{obs}})=\langle \tilde{\chi}_{\text{LSS}}\rangle \left(1+d(\theta_{\text{obs}})\right)\,,
\ee
while $d$ represents the radial modulation. Explicitly we find\small
\be\label{Shapiro}
\frac{\langle \tilde{\chi}_{\text{LSS}}\rangle}{\chi_h}=\frac{2}{\sqrt{\hat{r}_S}} +{\cal O}(1)\,,\qquad d=\hat{r}_S^{3/2}\left[-\frac{3}{4} \cos\theta_{\text{obs}}+\frac{3}{16} \hat{D}^2 \cos 2\theta_{\text{obs}}+\log\left(2\cos\frac{\theta_{\text{obs}}}{2}\right)\right]\,.
\ee\normalsize
The angular dependence of the radial modulation appears only at order $\hat r_S$. The lowest order of the distance to the last scattering surface in units of the hole radius is $2/\sqrt{\hat r_S}$ meaning that the parameter $\sqrt{\hat r_S}/2$ is in fact the ratio between the radius of the hole ($\chi_h$) and the distance to the LSS ($\tilde \chi_{\text{LSS}}$). The expansion in $\sqrt{\hat r_S}$ is thus an expansion in the size of the hole. 

We define the lensing-like displacement seen by a comoving observer at the center of the tilde system of coordinates as the difference between the angular position of the source calculated with respect to the center of the tilde system of coordinates,  $\tilde{\theta}_{\text{LSS}}$, and the angle at which the comoving observer detects the photon, $\theta_{\text{obs}}$. Explicitly
\be
\tilde{\Gamma}\equiv \tilde{\theta}_{\text{LSS}}-\theta_{\text{obs}}\,.
\ee
We immediately find a lensing-like deflection angle
\be\label{lensing}
\tilde{\Gamma}=-\hat{r}_S^{1/2} \hat{D}\sin\theta_{\text{obs}}-\hat{r}_S \left(\frac{3}{2} \hat{D}^2 \sin^2\theta_{\text{obs}}-\frac{1}{\hat{D}}\right)\tan \frac{\theta_{\text{obs}}}{2}+\mathcal{O}(\hat{r}_S \hat{D}^3)\,.
\ee

\subsubsection{Energy shift}

In order to compute how the energy of photons gets modulated from the last scattering surface to the observer, we assume that the CMB temperature, i.e. the distribution of energies on the LSS, is formed entirely inside the last scattering surface and then propagates in the unperturbed FLRW spacetime. In the standard lore where there is no local void, this point of view can be assumed if the effect of the cosmological constant is neglected so that there is no integrated Sachs-Wolfe effect and all gravitational effects can be effectively described by the temperature at emission.\footnote{\textcolor{black}{Besides the integrated Sachs-Wolfe effect, also the effect of CMB lensing is neglected in our treatment: what we want to study is the effect of \emph{geometrical} lensing and time-delay on the CMB map.}} In order to take into account the energetic effect of the local void, we consider this scenario and we compute the energy effects introduced by the propagation through the hole.

Denoting with $E_{\text{CMB}}(\theta_{\text{obs}})$ the signal that would be measured without the local void,  with the usual angular dependence of the standard cosmology, the energy measured by the observer in the local void is of the form
\be\label{vac}
E_{\text{o}}(\theta_{\text{obs}}) = \PotE (\theta_{\text{obs}}) E_{\text{CMB}}(\theta_{\text{obs}})\,.
\ee 
where the lowest orders of the modulating factor $\PotE (\theta_{\text{obs}})$  are reported in eq. (\ref{Eobs}) of appendix \ref{Eobsobs}.
In particular we note that at order $\sqrt{\hat r_S}$ only a dipolar modulation is present.

\subsection{Comparison with numerical results}\label{numerics}

\begin{figure}[ht!]
    \centering
    % \subfigure[\label{chiLSS}]
     {\includegraphics[scale=0.34]{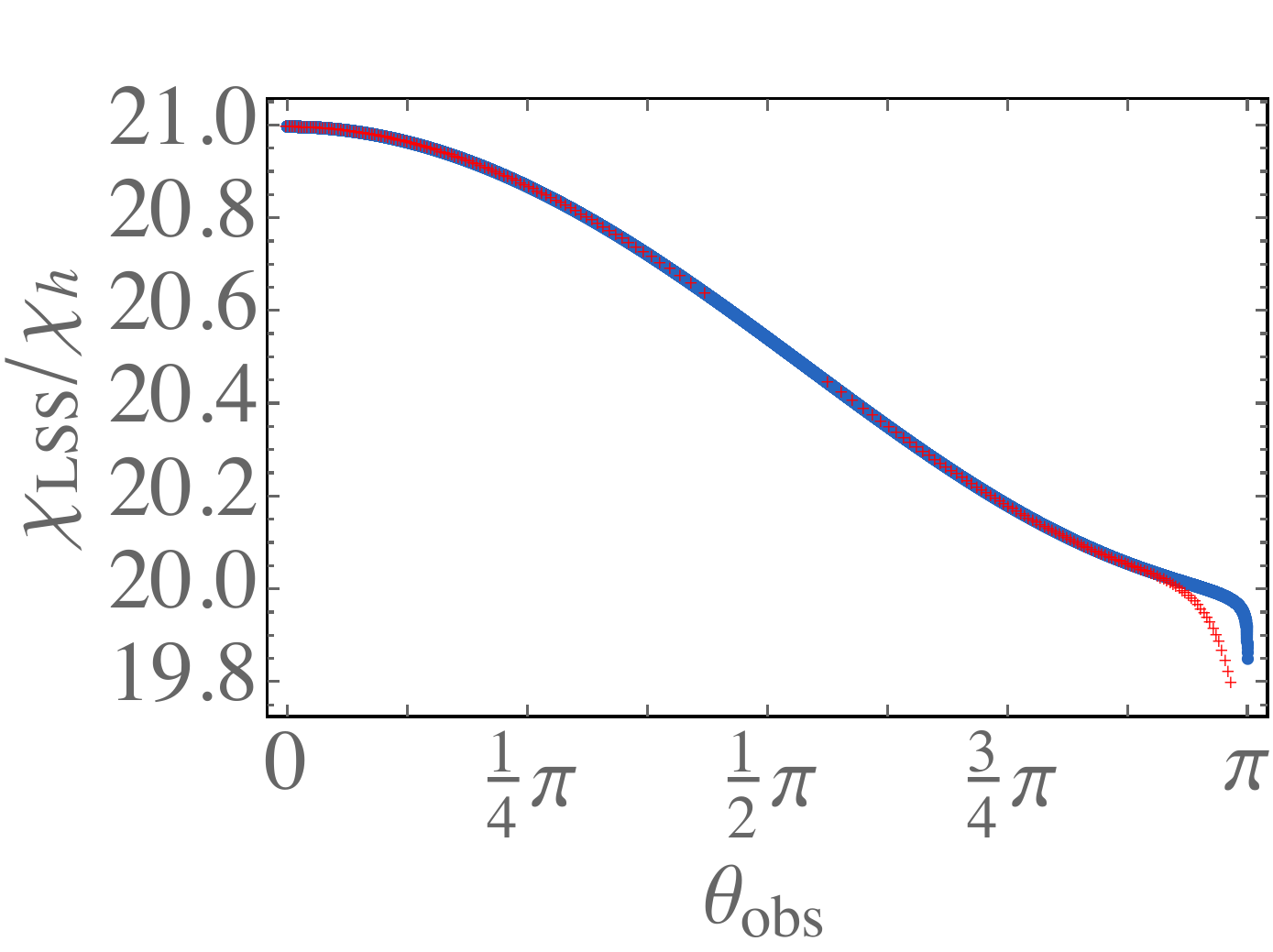}}\quad
      %    \subfigure[\label{dchiLSS1}]
     {\includegraphics[scale=0.34]{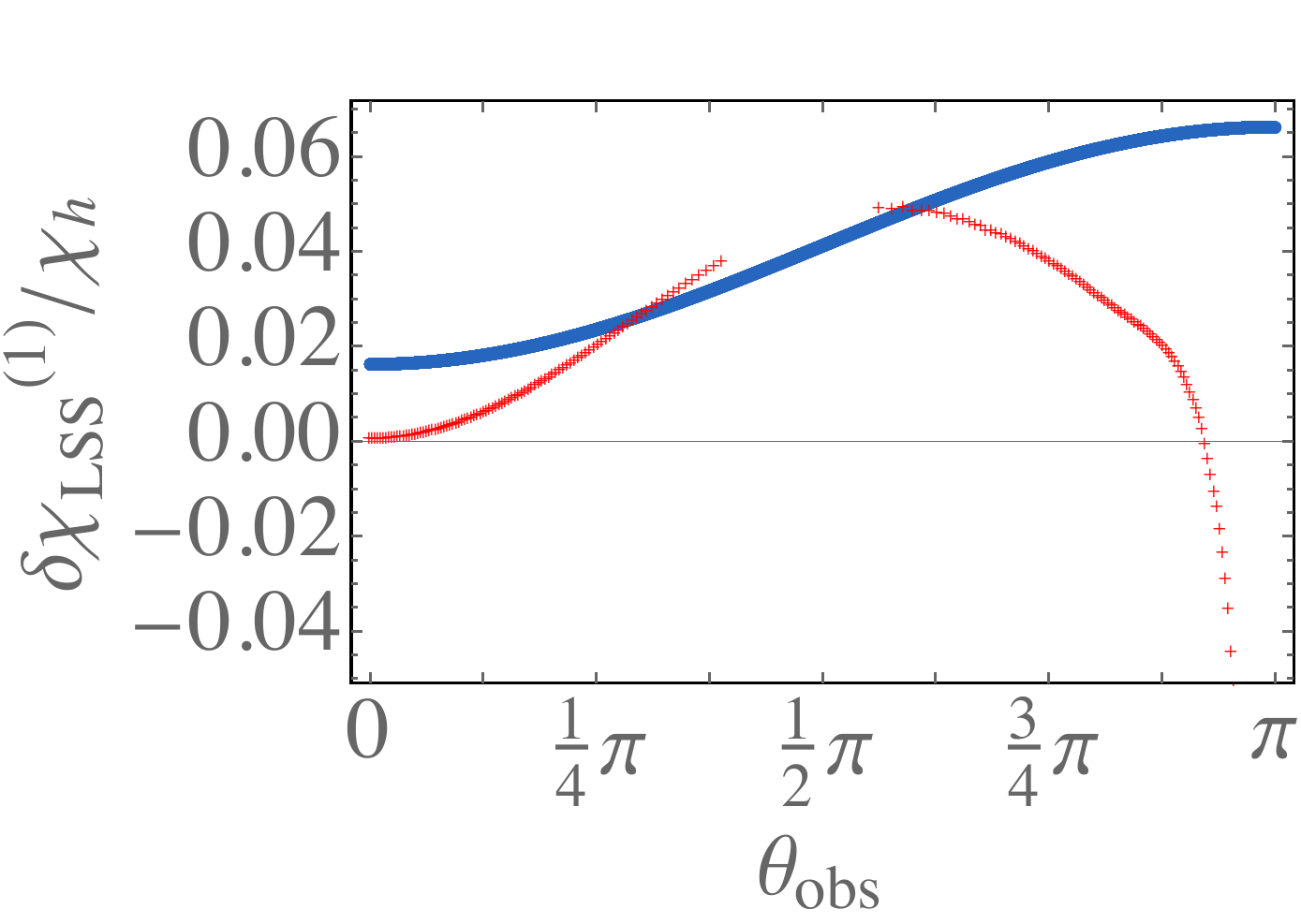}}\quad
         % \subfigure[\label{dchiLSS2}]
     {\includegraphics[scale=0.34]{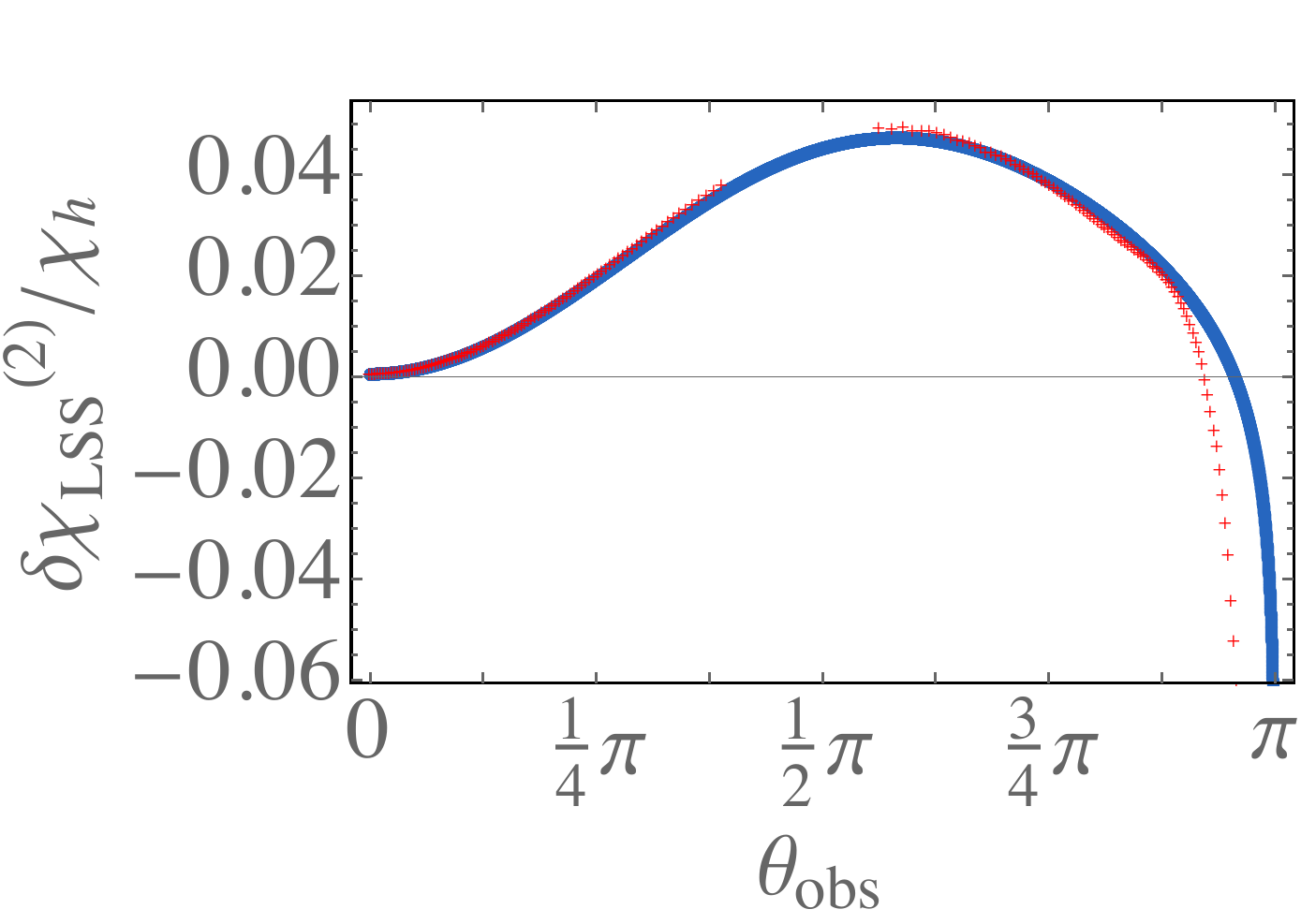}}
     
         % \subfigure[\label{thetaLSS}]
     {\includegraphics[scale=0.34]{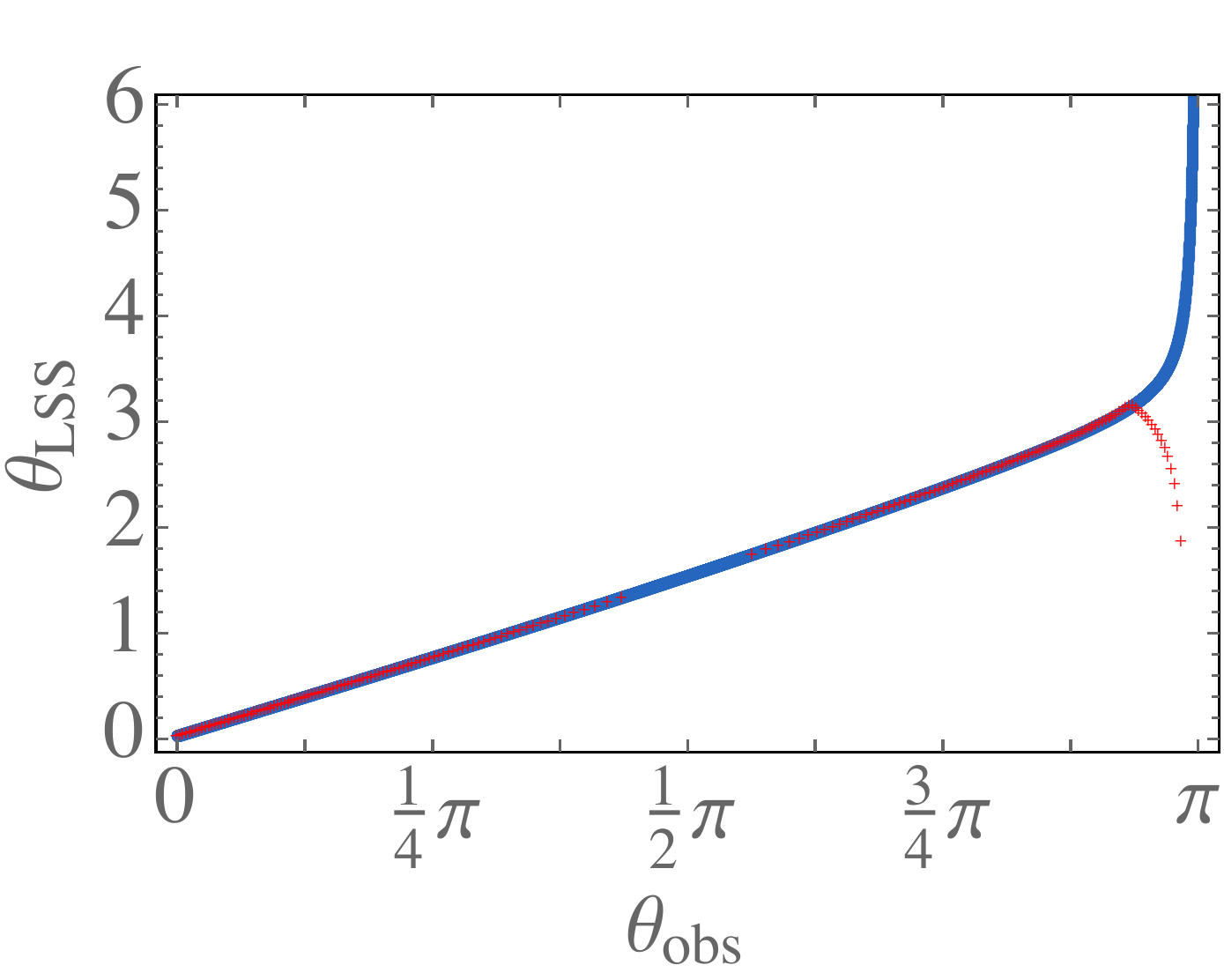}}\quad
         % \subfigure[\label{dthetaLSS1}]
     {\includegraphics[scale=0.34]{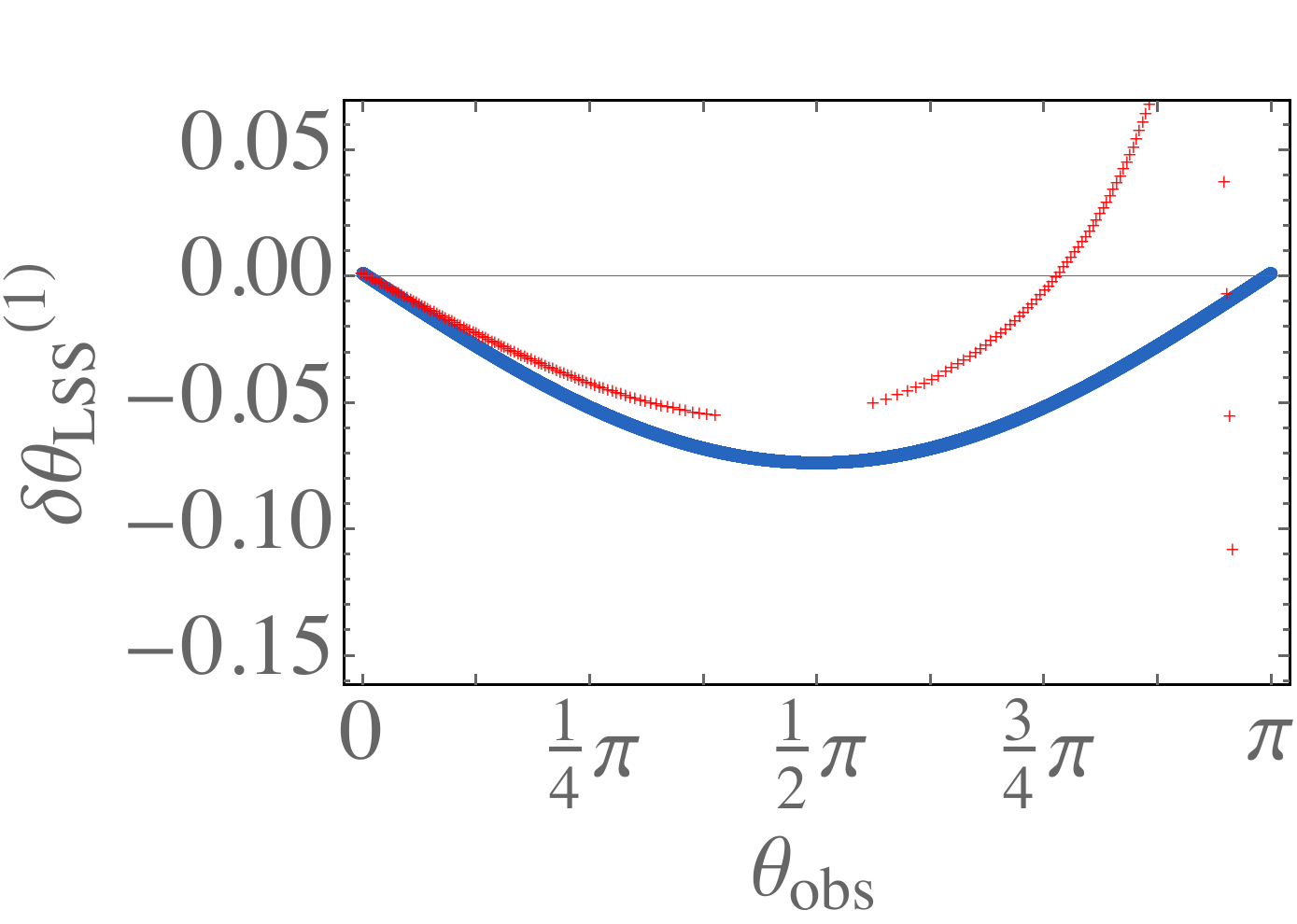}}\quad
         % \subfigure[\label{dthetaLSS2}]
     {\includegraphics[scale=0.34]{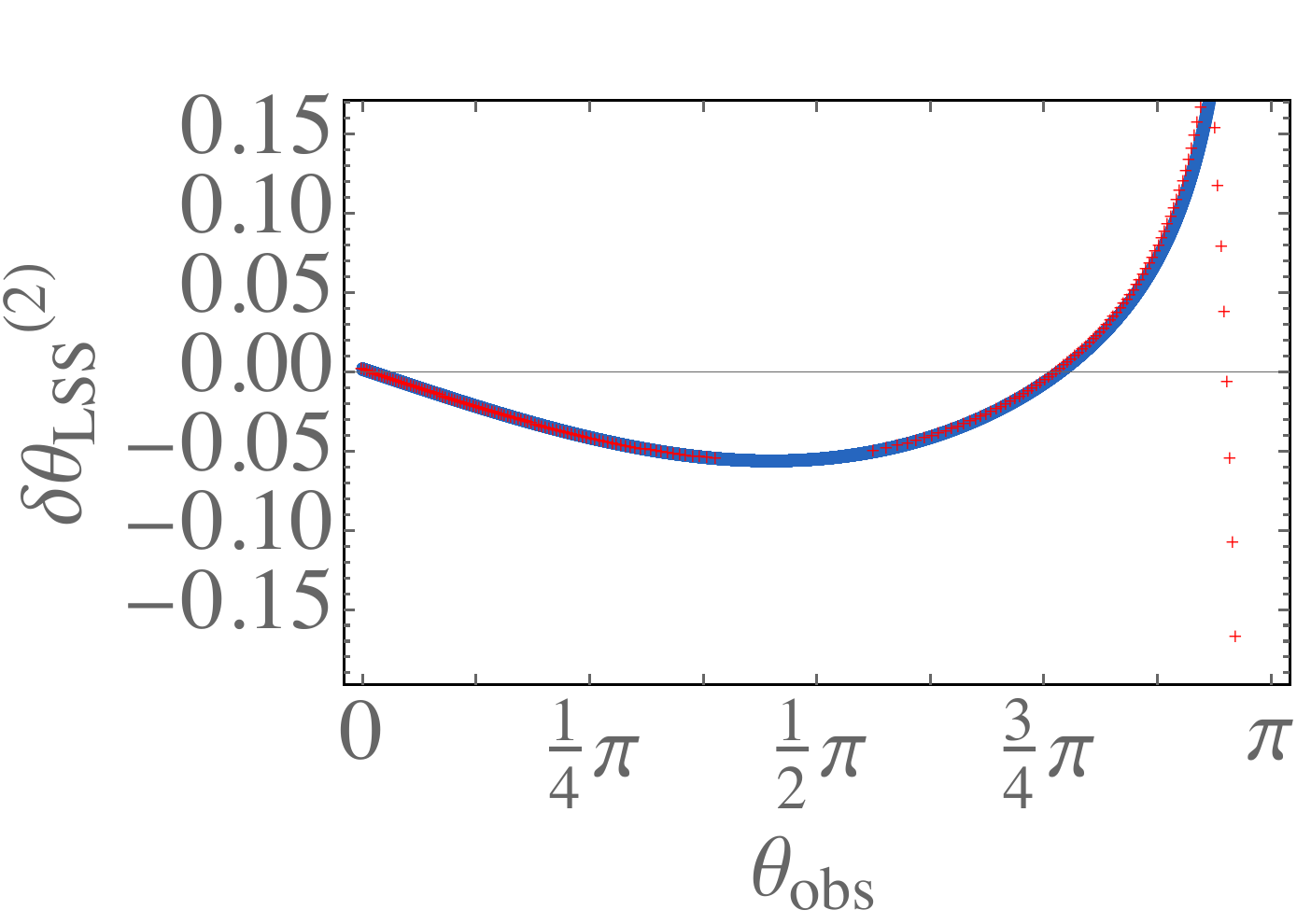}}
 \caption{\label{fig2} \small Coordinates of the event $\mathcal{E}_{\text{LSS}}$ (emitting point on the LSS) as a function of the direction of observation. The blue line is the result of the analytical integration, while the red one represents the solution obtained integrating numerically the system. We present three plots for each quantity of interest: in the first one we present the full analytic solution together with the result of the numerical integration, in the second and third ones we compare the numerical correction to the euclidian result to the analytic one at order $\hat{r}_S^{1/2}$ and up to $\hat{r}_S$, respectively. We have chosen the following values for the parameters describing our model: $r_S=10^{-2} \chi_h$, $D=0.5 \chi_h$.}
  \end{figure}

We integrate numerically the geodesic equation for the photon and compare the results with the analytic solutions found performing a perturbative expansion in $\hat{r}_S^{1/2}$. Fig.~\ref{fig2} compares the coordinates of the event $\mathcal{E}_{\text{LSS}}$ (radial coordinate $\chi_{\text{LSS}}$ and polar angle $\theta_{\text{LSS}}$) as a function of the direction of observation $\theta_{\text{obs}}$. For each quantity, we present three plots: in the first plot we compare the full numerical solution together with the analytical one, in the second and third plots we compare the numerical correction to the Euclidean result (order $(\hat r_S)^0$) to the analytic one at first order in $\hat{r}_S^{1/2}$ and up to order $\hat{r}_S$, respectively. 

We observe that the numerical solution is discontinuous around $\theta_{\text{obs}}\sim \pi/2$. This discontinuity is not physical, and it is simply due to the fact that we have chosen to solve geodesic equations spanning the range of values of the impact parameter, i.e. $b\in [0, D]$.\footnote{The observation direction is connected to the impact parameter by the relation (\ref{btheta}) and because of corrections proportional to $r_S$, for $b=D$, $\theta_{\rm obs}$ is not exactly $\pi/2$.} As expected, both numerically and analytically we find that the crossing angle is diverging when approaching the direction of observation $\theta_{\text{obs}}\sim\pi$ (photon coming across the horizon of the \emph{Sch} region). 
  
 \normalsize 

\subsection{Effects for a boosted observer}\label{boost}

We have considered the energy and direction as measured by a static observer in the \emph{Sch} region at $r=D$. However, a general observer could be boosted with respect to this static observer. In particular, one can consider an observer having a radial velocity outward with respect to the static observer, but being at the same position $r=D$. We investigate how this would affect the measurement of the CMB so as to isolate contributions which are non-degenerate with a boost.

Using the general results for a boost presented in appendix \ref{boost0} with the choice ${\bm{n}}=e_z$, we find that in a frame which is boosted by a factor $\beta$ %in the outward radial direction, 
along the $e_z$ axis, the energy and direction of observation transform as
\be\label{transE}
E' = \gamma (1+\beta \cos \theta) E = \frac{1}{\gamma(1-\beta
  \cos\theta')} E\,,
\ee
\be\label{phii}
\cos \theta' = \frac{\cos \theta + \beta}{1+\beta \cos \theta}\,,\qquad \cos \theta = \frac{\cos \theta' - \beta}{1- \beta \cos \theta'}\,,
\ee
with $\gamma=1/\sqrt{1-\beta^2}$. Expanding in the parameter $\beta$ we get
\begin{align}
E' &\simeq E\left[1+\beta \cos \theta'+\beta^2 \left(\cos^2 \theta'-\frac{1}{2}\right)+\mathcal{O}(\beta^3)\right] \simeq E\left[1+\beta \cos \theta+\frac{1}{2}\beta^2+\mathcal{O}(\beta^3)\right]\,,\nn\\
\theta &\simeq  \theta' +\beta \sin  \theta' +\frac{1}{2}\beta^2\cos \theta' \sin \theta' +\mathcal{O}(\beta^3)\,.
\end{align}
We boost the results found for the radial modulation, lensing and energy, eqs. (\ref{radialmod}),  (\ref{lensing}) and (\ref{vac}). We choose a boost velocity $\beta$ at order $\hat{r}_S^{1/2}$ in such a way to cancel the angular modulation in the energy and lensing-like deflection at order $\hat{r}_S^{1/2}$. We have the freedom to choose the contribution to $\beta$ proportional to $\hat{r}_S$. We choose it in such a way to cancel the dipole in the energy modulation so as to separate completely the effects of geometry from the kinematical effects. Explicitly, our choice is a velocity with modulus
\be\label{bbeta}
\beta=\hat{r}_S^{1/2} \hat{D}+\frac{3}{2}\hat{D}\left(\hat{D}-1\right) \hat{r}_S\,,
\ee
and along the $e_z$ axis. Transforming angles and energies with (\ref{phii}) and (\ref{transE}), respectively and indicating with a prime the boosted quantities, we get up to order $\hat{r}_S\hat{D}^3$
\be\label{cin1}
E_{\text{o}}'=E_{\text{CMB}}\,,\qquad \tilde{\chi}'_{\text{LSS}}=\tilde{\chi}_{\text{LSS}}\,.
\ee
Note that the radial modulation under a boost stays unchanged up to order $\hat{r}_S$ since eq. (\ref{radialmod}) depends on angles only at order $\hat{r}_S$. 
%%Note that not only our choice has eliminated the dipole in the energy modulation, but also all other types of modulation and for this choice of the boost, the observer sees no modulation of energies at all at order $\sqrt{\hat r_S}$. 
For the lensing-like deflection, we find up to order $\hat{r}_S\hat{D}^3$ corrections
\be\label{cin3}
\tilde{\Gamma}'\equiv \tilde{\theta}'_{\text{LSS}}-\theta'_{\text{obs}}=\hat{r}_S\left(-\frac{3}{2}\hat{D} \sin\theta_{\text{obs}}'+\hat{D}^2 \sin\theta_{\text{obs}}'\cos\theta_{\text{obs}}'+\frac{1}{\hat{D}} \tan\frac{\theta_{\text{obs}}'}{2}\right)\,.
\ee

\subsection{Discussion of the physics at order $\sqrt{\hat r_S}$}\label{discussion physics}

We observe that at order $\sqrt{\hat r_S}$ no gravitational effects are present\footnote{The effects of curvature are proportional to $\hat{r}_S$.}, and the corrections to the Euclidean results are entirely due to kinematics  (i.e. to the fact that the boundary of the hole is expanding with Hubble rate). When we boost the results as in  section \ref{boost}, we are choosing a special cosmological observer, who does not see any expansion of the void boundary. This is even more obvious if we consider an observer sitting on the boundary, corresponding to the case $\hat D=1$. In that case the required boost is directly given by the dimensionless Hubble parameter $\hat H = \sqrt{\hat r_S} = H \chi_h$ which is just the recession velocity of the boundary. Inside the hole, the recession velocity is only a fraction $\hat D$, leading to the required boost $\sqrt{\hat r_S} \hat D$. It is therefore clear why this boost is what is needed to remove all effects at order $\sqrt{\hat r_S}$. However, at order $\hat r_S$ we get corrections to the Euclidean results due to true gravitational effects which affect angles, energies and distances. After the boost (\ref{bbeta}), the energy looses any angular dependence, even at order $\hat r_S$. 
%\textcolor{red}{Cyril: why?} 
Therefore, at order $\hat r_S$ only the lensing-like and radial displacement effects of the LSS have to be considered. % \textcolor{red}{Comment on $D(D-1)$}

%%%%%%%%%%%%%%%%%%%%%%%%%%%%%%%%

\section{CMB sky seen by the off-center observer}\label{CMBoff}

We now study how geometrical lensing-like and radial modulations affect the shape of the CMB temperature and polarization angular power spectrum. We consider the boosted observer defined in Sec. \ref{boost}. The CMB temperature and polarization seen by this observer will be lensed and radially modulated, but not modulated by a multiplicative factor~\footnote{The boosted observer defined in Sec. \ref{boost} sees no energy modulation, up to order $\hat{r}_S D^2$. The effect of an energy modulation would enter as an overall multiplicative factor on the right hand side of eq. (\ref{tailm}).}.  In appendix \ref{CMBsky} we collect our definitions for the CMB intensity map. In appendix \ref{CMB void} we detail the calculation of the effects of geometrical lensing-like deflection and radial modulation on the shape of the CMB temperature and polarization angular power spectrum.

The lensed and delayed temperature anisotropy field can be expanded up to first order in lensing-like displacement and radial modulation as
\be\label{tailm}
\tilde{\Theta}({\bm{x_o}}\,, \eta_o\,,{\bm{n}})=\Theta({\bm{x_o}}\,, \eta_o\,,{\bm{n}})+\Theta^{\varphi}({\bm{x_o}}\,, \eta_o\,,{\bm{n}})+\Theta^{d}({\bm{x_o}}\,, \eta_o\,,{\bm{n}})\,,
\ee
where we have explicitly indicated the dependence on the observer position ${\bm{x_o}}$ and reception time $\eta_o$. In eq. (\ref{tailm}), $\Theta({\bm{x_o}}\,, \eta_o\,,{\bm{n}})$ is the zeroth order contribution from the primary anisotropies while $\Theta^{\varphi}$ and $\Theta^d$ are the lensing-like and radial modulation effects, linear in lensing-like deflection (\ref{cin3}) and in the radial modulation (\ref{Shapiro}), respectively.

The right way to proceed to take into account the effects of radial modulation is to write the CMB temperature field on the sky as the projection of sources $S$ which contribute in an optically thin regime, see appendix \ref{CMB temperature} for details. Doing this, after having defined the Fourier transform of the secondary anisotropy contributions as
\be\label{transformtailm}
\Theta^{\varphi\,,d}({\bm{x_o}}\,, \eta_o\,,{\bm{n}})=\int \frac{d^3 k}{(2\pi)^{3/2}}\,\hat{\Theta}^{\varphi\,,d}({\bm{k}}\,,\eta_o\,,{\bm{n}})\,e^{i {\bm{k}}\cdot {\bm{x_o}}}\,,
\ee
we get
\begin{eqnarray}
\hat{\Theta}^{\varphi}({\bm{k}}\,, \eta_o\,,{\bm{n}})&=&\nabla_i\varphi({\bm{n}})\,\sum_{\ell}\sum^{\ell}_{m_i=-\ell} I_{m_i}[j_{\ell}]\,\nabla^i Y_{\ell m_i}({\bm{n}})\,,\label{taillensingm}\\
\hat{\Theta}^{d}({\bm{k}}\,, \eta_o\,,{\bm{n}})&=&d({\bm{n}})\,\sum_{\ell}\sum^{\ell}_{m_i=-\ell} I_{m_i}[j'_{\ell}]\, k(\eta_o-\eta_{\text{LSS}})\,  Y_{\ell m_i}({\bm{n}})\,,\label{taildelaym}
\end{eqnarray}
where the operator $I_{m_i}[j_{\ell}(k\chi)]$ is defined in appendix \ref{CMB temperature}, a prime indicates derivative with respect to the argument of the spherical Bessel function and $\chi(\eta)=\eta_o-\eta$. Analogous results hold for polarization, see appendix \ref{CMB polarization}. 

It is easy to verify from an inspection of  eqs.  (\ref{taillensingm}) and (\ref{taildelaym})  that the contribution of the radial modulation to the temperature anisotropy field is subdominant with respect to the one coming from lensing-like deflection. Indeed, the lensing depends on the angular gradient of the lensing potential and its observable consequences are weighted by a factor of order $\ell$. This has the effect of increasing the magnitude of the effect and shifting it to higher multipoles.\footnote{The fact that the  effect  of radial modulation is negligible with respect to the one of lensing-like deflection can be understood also from geometrical considerations, using the analytic results found, see appendix \ref{SShapiro}.} 

\subsection{Multipoles of lensing-like deflection and radial modulation}\label{yves}

From eq. (\ref{cin1}) we see that after the boost, the energy measured by the comoving observer at the center of the tilde system of coordinates has no angular dependence. In our problem, lensing-like deflection has only gradient modes (see appendix \ref{generic generic}) and we introduce a lensing potential as $\Gamma_a=\nabla_a\varphi$. The radial modulation and lensing-like deflection are decomposed as 
\be\label{above}
 \tilde{d}' /\chi_h=\sum_{\ell} d_{\ell 0}\, Y_{\ell0}\,,\qquad
\tilde{{\bm{\Gamma}}}' =-\sum_{\ell}  \varphi_{\ell 0}\, \sqrt{\ell(\ell+1)} \, _1 Y_{\ell 0} \, {\bm{e}}_{\theta}\,.
\ee
Multipoles in eq. (\ref{above}) can be extracted using the analytic results for (boosted) radial modulation and lensing-like deflection, eqs. (\ref{cin1}) and (\ref{cin3}), respectively, and recalling the definition of radial modulation, eq. (\ref{Shapiro}). 

Until now we have considered a system of coordinates such that the azimuth was aligned with  ${\bm{e}_z}$, where ${\bm{e}_z}$ denotes the direction joining the center of the \emph{Sch} region to the observer.  To generalize our analysis, we now consider a rotated coordinate frame. The rotation is described by a $SO(3)$ matrix $R_1$ characterized by its Euler angles $(\phi_1, \theta_1, 0)$.  In the new coordinate frame the direction observer-hole is described by the unit vector ${\bm{n_1}}= R_1 \gr{e}_z$. A direction described by a unit vector ${\bm{n}}$ in the old reference frame,  is rotated to $R_{1}^{-1} {\bm{n}}$ in the new one.  In this reference frame the lensing potential can be expanded as
\be\label{pot2}
\varphi_{\text{new}}({\bm{n}})=\sum_{\ell m} \varphi_{\ell m} Y_{\ell m} ({\bm{n}})\,,
\ee
with 
\be\label{mess}
\varphi_{\ell m}=\varphi_{\ell 0} \sqrt{\frac{ 4\pi}{2\ell +1}} Y_{\ell m}^*({\bm{n}_1})\,, \quad  \varphi_{\ell 0}=\frac{\hat{r}_S}{\hat{D}}\sqrt{4\pi}\,(-)^{\ell} \frac{\sqrt{2\ell+1}}{\ell(\ell+1)}\,. 
\ee
Similarly, the time delay can be expanded as
\be
d_{\text{new}}({\bm{n}})=\sum_{\ell m} d_{\ell m} Y_{\ell m}\,,
\ee
with 
 \be\label{mess2}
d_{\ell m}=d_{\ell 0} \sqrt{\frac{ 4\pi}{2\ell +1}} Y_{\ell m}^*({\bm{n}_1})\,,\quad d_{\ell 0}=\hat{r}_S^{3/2}\,\sqrt{\pi}\,(-)^{\ell+1} \frac{\sqrt{2\ell+1}}{\ell(\ell+1)}\,.
\ee
Details on these derivations are presented in appendix \ref{generic}. 

\subsection{Correlation functions}\label{corre}

The CMB sky seen by an observer inside the hole is not statistically isotropic. In the absence of statistical isotropy, the correlation function of the lensed temperature anisotropy and polarization (indicated with a tilde) are defined as
\be\label{cic}
\tilde{C}({\bm{n_1}}, {\bm{n_2}})\equiv \langle X ({\bm{n_1}}) Y({\bm{n_2}}) \rangle\,,
\ee
where $X\,,Y=\tilde{\Theta}\,,\tilde{E}\,,\tilde{B}$. Since statistical isotropy is violated, the correlation $\tilde{C}({\bm{n_1}}, {\bm{n_2}})$ is estimated by a single product $X ({\bm{n_1}}) Y({\bm{n_2}})$ and hence it is poorly determined by a single realization.  Anyway, even if the nature of the violation of statistical isotropy is not known, some measuraments of statistical anisotropy of the CMB map can be estimated through suitably weighted angular averages of $ X ({\bm{n_1}}) Y({\bm{n_2}})$, see e.g. Ref. \cite{Hajian:2003qq}.  In the presence of statistical anisotropy, the correlation function (\ref{cic}) can be expanded in Bipolar Spherical Harmonics: the coefficients of this expansion are a complete representation of statistical isotropy violation, see appendix \ref{correlation} for details. 

We define a 2-point correlator as 
\be\label{FFm}
F_{\ell m}^{LM}|_{(X\,Y)}\equiv \langle X_{\ell m}Y_{\ell+L\,m+M}^* \rangle\,.
\ee
In our problem the only source of violation of statistical isotropy has a geometric origin (geometrical lensing-like deflection and radial modulation). To evaluate the 2-point correlators of polarization and temperature anisotropy, we find convenient working separately with lensing and radial modulation and linearly sum the effects at the end. In other words, we decompose the 2-point function (\ref{FFm}) as
\be\label{FFF}
F_{\ell m}^{L\,M}|_{(X\,Y)}=C^{XY}_{\ell m}\delta_{L0}\delta_{M0}+(F_{\ell m}^{L\,M})|_{(X\,Y)}^d+(F_{\ell m}^{L\,M})|_{(X\,Y)}^{\varphi}\,,
\ee
where the first term on the right hand side denotes the contribution coming only from primary anisotropies while $(F_{\ell m}^{L\,M})_d$ and $(F_{\ell m}^{L\,M})_{\varphi}$ denote contributions to the 2-point function, linear in radial modulation and lensing potential, respectively. Explicitly
\begin{align}
(F_{\ell m}^{L\,M})|_{(X\,Y)}^{\varphi}&\equiv \langle X^{\varphi}_{\ell m} Y^*_{\ell+L\,m+M} \rangle + \langle X_{\ell m} Y^{*\varphi}_{\ell+L\,m+M}\rangle\,,\label{FphiP}\\
(F_{\ell m}^{L\,M})|_{(X\,Y)}^{d}&\equiv \langle X^{d}_{\ell m} Y^*_{\ell+L\,m+M} \rangle +\langle X_{\ell m} Y^{*d}_{\ell+L\,m+M} \rangle\label{FdP}\,.
\end{align}
Since the contribution of radial modulation is negligible with respect to lensing-like deflection, from now on we focus only on the latter.

At linear order in lensing, we find the following result for the temperature anisotropy correlation function 
\begin{align}
(F_{\ell m}^{L\,M})|_{(\tilde{\Theta}\tilde{\Theta})}^{\varphi}&=\varphi_{\ell_1 -M} \,\mathcal{C}_{\ell \,\,\ell+L\,\, \ell_1}^{m\,\, m+M\,\, -M}\left(\alpha_+ C_{\ell+L}^{\Theta\Theta}+\alpha_- C_{\ell}^{\Theta \Theta} \right) \,\label{u1m}
\end{align}
where the summation over $\ell_1$ is understood, $\mathcal{C}_{\dots}^{\dots}$ is defined in appendix  \ref{lm} and we defined
\be
\alpha_{\pm}\equiv \frac{1}{2}\left[\ell_1(\ell_1+1)\pm L (L+2\ell+1)\right]\,,
\ee
while the multipoles of the lensing potential are given by eqs. (\ref{mess}).  This is our main result and the detailed derivation is presented in  appendix \ref{CMB temperature}. To obtain (\ref{u1m}) we have used that in the absence of the hole, temperature anisotropy and polarization are stochastic variables characterized by diagonal correlation functions. We assume that primary anisotropies are not generating $B$-modes, i.e. $C_{\ell}^{BB}=C_{\ell}^{EB}=C_{\ell}^{\Theta B}=0$. The results for the polarization and for the  temperature-polarization correlators are collected in appendix \ref{CMB polarization}.

We observe that the result for the correlation matrix (\ref{u1m}) is proportional to the ratio $\hat{r}_S/\hat{D}$.\footnote{The proportionality is through the multipoles of the lensing potential, see eq. (\ref{mess}).} To give en estimate of this effect, we consider that the distance to the CMB is approximately given by eq. (\ref{Tinetain}), which at lowest order in $\hat{r}_S$ reads
\be
\eta_{\text{in}}\simeq 2 \chi_h \frac{1}{\sqrt{\hat{r}_S}}\,.
\ee
Considering that $\eta_{\text{in}}\simeq 14 000$ Mpc and that for a typical cluster $\chi_h\simeq 20$ Mpc, we get
\be        
\hat{r}_S\simeq 4 \left(\frac{\chi_h}{\eta_{\text{in}}}\right)^2\sim 10^{-5}\,.
\ee
In our treatment, the parameter $\hat{D}$ has to be chosen bigger than $\hat{r}_S$ and  $\hat{D}<1$, since it has been used as a perturbative parameter to derive (\ref{u1m}), e.g. typically we have $r_S/D \sim 10^{-3}-10^{-4}$.

\begin{figure}[ht!]
    \centering
    % \subfigure[\label{chiLSS}]
     {\includegraphics[scale=0.43,angle=0]{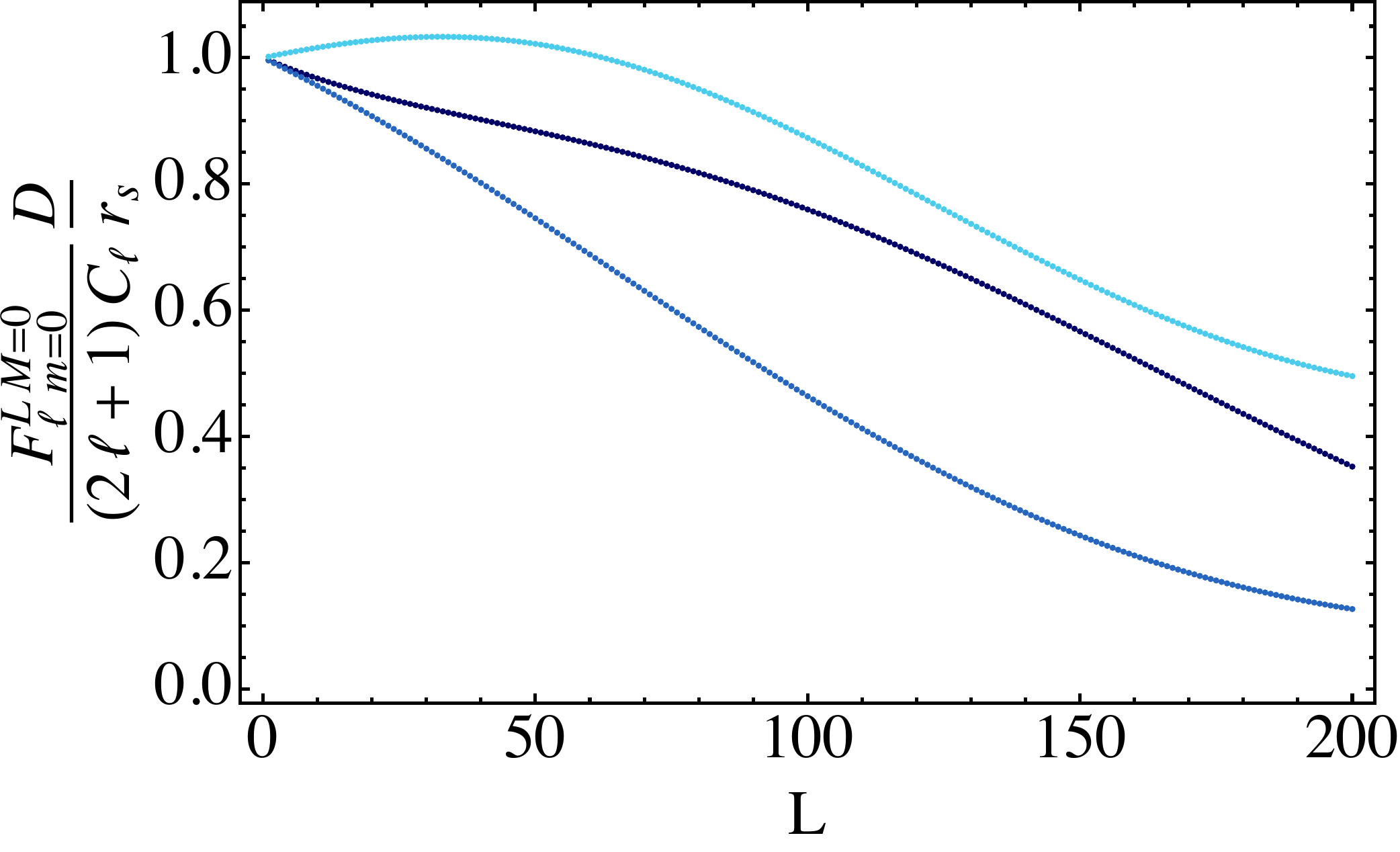}}
     \caption{\label{FlmLM} \small Correlation matrix $( F_{\ell m}^{L\,M})_{(\tilde{\Theta}\tilde{\Theta})}^{\varphi}$ with $M=m=0$ as a function of $L$, for three different values of $\ell$: $\ell=100$ (dark blue), $\ell=200$ (blue), $\ell=500$ (light blue). We have measured the matrix elements in units of $r_S/D$.}
  \end{figure}

The structure of the correlation matrix $( F_{\ell m}^{L\,M})_{(XY)}^{\varphi}$ is quite complex: at linear order in lensing-like deflection the correlation matrix has non-vanishing diagonal elements and  all the off-diagonal terms are excited, i.e. at linear order in lensing we get correlations among all the multipoles $\ell\leftrightarrow \ell\pm L$\,, $\forall L$.  To have an idea of the off-diagonal structure of the correlation matrix, we plot $( F_{\ell m}^{L\,M})_{(\tilde{\Theta}\tilde{\Theta})}^{\varphi}$ with $M=m=0$ as a function of $L$, for different values of $\ell$. The result is shown in Fig. \ref{FlmLM}. We have chosen to rescale  $( F_{\ell 0}^{L\,0})_{(\tilde{\Theta}\tilde{\Theta})}^{\varphi}$ such that the diagonal elements of the effect are unity.\footnote{The normalization factor in Fig. \ref{FlmLM} can be found observing that $( F_{\ell 0}^{00})_{(\tilde{\Theta}\tilde{\Theta})}^{\varphi}=r_S/D\,C_{\ell}^{(\tilde{\Theta}\tilde{\Theta})} (2\ell+1)$.}% (basic properties of the 3j-symbols have been exploited).}    
\,Independently of the $\ell$, the matrix elements more enhanced are those close to the diagonal of the correlation matrix (in the $L-\ell$ space): the correlation is mildly stronger for neighboring multipoles and it slowly decreases going away from the diagonal, as can be seen on Fig. \ref{FlmLM}. We observe that, for a given $\ell$, all the off-diagonal correlators have approximately the same order of magnitude: this is due to the fact that in eq. (\ref{u1m}) the multipoles of the lensing potential are in turn approximately of the same order.\footnote{This fact can be verified equating eq. (\ref{D.10}) defining lensing ${\bm{\Gamma}}$ to the dominant contribution of eq. (\ref{cin3}) (we are choosing a frame with ${\bm{e_z}}$ aliged with the azimuth)  
$${\bm{\Gamma}}{\bm{e_{\theta}}}=-\sum_{\ell}\varphi_{\ell 0} \sqrt{\ell(\ell+1)}_1Y_{\ell0}=\frac{r_S}{D} \tan\frac{\theta_{\text{obs}}'}{2}\,,\nn$$
with $\varphi_{\ell0}$ defined in eq. (\ref{mess}). The series of multipoles converges very slowly: a large range of $\ell$ is need to reconstruct the angular structure $\tan\theta_{\text{obs}}'/2$.}

%%%%%%%%%%%%%%%%%%%%%%%%%%%%%%%%%%%%%%%%

\section{Effect of a peculiar velocity on the CMB sky}\label{Boost}

In this section, we turn to the standard case of a FLRW universe so as to compare the kinematic effect of a boost of the observer in a FLRW universe with the geometrical effects of a local void studied in the previous section. In particular, we want to highlight  the difference between these two cases at the level of the diagonal and off-diagonal correlation functions of temperature and polarization. 

Therefore,  we consider two observers of a pure FLRW universe: the first one comoving with the CMB rest frame and the second one in motion with respect to the first. We relate CMB correlation functions in the CMB rest frame $S$ with the ones in a moving observer frame $\tilde{S}$. We indicate with a tilde quantities in $\tilde{S}$ and without a tilde quantities in the CMB rest frame $S$.  Statistical isotropy in $S$ still leads to violation of statistical isotropy in $\tilde{S}$. \footnote{We define \emph{CMB rest frame} the system of reference in which the temperature dipole vanishes} 

There are two different effects on the CMB sky map due to the motion of the observer:  (a) a modulation of intensity/Stokes parameters and (b) an aberration in the direction ${\bm{n}}$ of incoming photons which leads to a remapping of the intensity map/Stokes parameters on the sky.  The aberration and modulation effects of the CMB  have been recently measured by the  {\em Planck} satellite \cite{Aghanim:2013suk}. 

Under a boost, up to linear order in the boost velocity ${\bm \beta}$, the temperature anisotropy field transforms as 
\be
\tilde{\Theta}({\bm{\tilde{n}}})\simeq \Theta({\bm{\tilde{n}}})\left(1+\zeta({\bm{\tilde{n}}})\right)-\nabla^a \zeta({\bm{\tilde{n}}}) \nabla_a \Theta({\bm\tilde{
  n}})\,,
\ee
where $\zeta({\bm{n}})=\bm{n}\cdot{{\bm \beta}}$. 
An analogous expression holds for polarization, see appendix \ref{lastt} for details. 

At linear order in $\beta$, the correlation function of temperature anisotropy in the boosted frame $\tilde{S}$ has the following expression as a function of the correlator in  $S$ \footnote{We assume that in the CMB rest frame, primary anisotropies are not generating $B$ modes.}
\be
( F_{\ell m}^{L M})|^{\beta}_{(\tilde{\Theta} \tilde{\Theta})}=\alpha_{\ell}\,\beta_{1-M} \,\mathcal{C}_{\ell\,\,\ell+L\,\,1}^{m\,\,m+M\,\,-M}(C_{\ell}^{\Theta\Theta}-C_{\ell+L}^{\Theta\Theta})\,,\label{v1}\qquad \alpha_{\ell}\equiv \frac{L}{2}(L+2\ell+1)\,.
\ee
Results for the  polarization and the temperature-polarization correlation functions are collected in appendix \ref{fame}. We observe that at linear order in $\beta$ all the diagonal terms (i.e. $L=0\,, M=0$) of the correlation matrices $( F_{\ell m}^{L M})|^{\beta}_{(\tilde{X} \tilde{Y})}$ are vanishing. Off-diagonal correlators are non-vanishing only for $L=1$, i.e. we have only correlation among $\ell\leftrightarrow \ell\pm 1$ multipoles.

\section{Conclusions}\label{conclusion}

This article fully characterizes the effect of the local void on the temperature and $E$, $B$- polarization modes measured by an off-centered observer. We have considered  a universe consisting of a spherical void, described by a Kottler spacetime, embedded in a  FLRW universe and a static observer into the void, displaced with respect to the center of symmetry. We have introduced a perturbation scheme which allowed us to analytically calculate the 2-point angular correlation functions and the off-diagonal correlators for both temperature and polarization at leading order and next-to leading order in the perturbation parameter $\hat{r}_S^{1/2}$. We found that the energy shift can be suppressed by a proper choice of the observer velocity, while the lensing-like effect remains. This last effect is a genuinely geometrical effect (non-degenerate with the effect of a boost), which reflects in the structure of the off-diagonal correlators. Indeed the structure of the correlation matrix is quite complex: at linear order in lensing the correlation matrix has non-vanishing diagonal elements and all the off-diagonal terms are excited, i.e. at linear order in lensing we get corrections among all the multipoles $\ell \leftrightarrow \ell \pm L$, $\forall L$. We have explicitly computed the off-diagonal structure of the correlation matrix for $M=m=0$: the correlation is mildly stronger for closer multipoles (i.e. small $L$) and it slowly decreases going away from the diagonal, as can be seen on Fig. \ref{FlmLM}. 

As a second model we have considered a FLRW universe, allowing for a boost of the observer with respect to the CMB rest frame and we have calculated the correlation function of temperature and polarization for such an observer. For this model, the results for the CMB correlators are standard, but  we re-derived them to make a direct comparison with the results of the void model. In this analysis, the small parameter that allows us  to expand the analytic results is the boost velocity $\beta$.   At linear order in $\beta$ all the diagonal terms of the correlation matrices are vanishing. Off-diagonal correlators are non-vanishing only for $L=1$, i.e. we have only correlation among $\ell\leftrightarrow \ell\pm 1$ multipoles. If we repeat the calculation of the correlators up to order $\beta^2$, we get non vanishing diagonal correlators and off-diagonal terms of the correlation  matrix at order $\beta^2$ non-vanishig for $L=2$. In fact this result can be generalized: at a generic order $\beta^n$, we would get off-diagonal terms of the correlation matrix at order $\beta^n$ non vanishing for $L=n$. 

The fact that in the void model, at first order in lensing, we get correlations among all the multipoles is an extremely interesting signature of the void model, which in principle would allow one to distinguish in CMB observations geometrical effects (coming e.g. from the presence of an overdense region) from kinematical effects. 

This work puts for the first time the discussion of the possible geometrical origin of the CMB dipole on a firm ground.  The next step would be to refine our toy-model for the void, considering e.g. an LTB geometry to describe an overdense region. We expect that the results found in this analysis would stay qualitatively the same for the case of a LTB void model.\\

\noindent {\bf Acknowledgements:}  GC is particularly grateful to Prof. Ruth Durrer for having encouraged her to start working in this direction. The work of GC is supported by the Swiss National Science Foundation. The work CP and JPU was supported by French state funds managed by the ANR within the Investissements d'Avenir programme under reference ANR-11-IDEX-0004-02.
\vspace{3cm}

\newpage
\appendix

\section{Photon dynamics inside the hole} \label{AppGeodesicInside}

In this appendix we detail the procedure presented in section \ref{preliminary} to solve the geodesic equation of the photon inside the hole. While the formal solution of the null geodesic equation in a \emph{Sch} geometry is a standard textbook result (see e.g.  \cite{Wald:1984rg}), here there is an additional complication due to the fact that the boundary of the \emph{Sch} void is also expanding with time. 

The equation governing the radial motion of the photon is
\be\label{radialphoton2}
\frac{\dd u}{\dd t}=\pm \frac{u^2}{\epsilon_1}P(u)^{1/2}A(u)\,.
\ee
If the photon is received at a direction $\theta_{\rm obs}\in [0, \pi/2]$ the radial coordinate of the photon is decreasing with time and we need to pick up the plus sign in the differential equation above. The situation is more complicated if the photon is received at an angle $\theta_{\rm obs} \in ]\pi/2, \pi]$: the radial coordinate of the photon is decreasing between the crossing and the radius of minimum approach $r_{\text{min}}$, solution of $\dd r/\dd t=0$ or $P(1/r_{\text{min}})=0$. The radial coordinate is instead increasing between the radius of minimum approach and the observer position (minus sign in the equation above). 

Eq. (\ref{radialphoton2}) and eq. (\ref{dynamicsbounday}) describing the radial motion of the photon and of the boundary respectively, are then integrated as $t_{\text{photon}}(u)$ and $t_{\text{hole}}(u)$. The entrance radius and time can then be found by requiring $t_{\text{photon}}(u_{\text{in}})=t_{\text{hole}}(u_{\text{in}})$. 

Once the entrance radius is known, the angle at entrance can be found integrating the second equation in (\ref{radialphoton})
\be
r_S\,\frac{\dd u}{\dd\theta}=\pm \,\sqrt{P(u)}\,,
\ee
where for an incoming direction $\theta_{\rm obs}\in [0, \pi/2]$ we need to pick up the minus sign
\be
\Delta \theta=\theta_{\text{in}}-\theta_o=-\int _{u_o}^{u_{\text{in}}} {\rm d}u \frac{r_S}{\sqrt{P(u)}}\,.
\ee
For $\theta_{\text{obs}}\in ]\pi/2, \pi]$ we need to pick up the minus sign for $u$ between $u_{\text{in}}$ and $1/r_{\text{min}}$ and the plus sign for $u$ between $u_{\text{min}} = 1/r_{\text{min}}$ and $u_{\text{o}} = 1/D$. Explicitly, 
\be
\Delta \theta=\theta_{\text{in}}-\theta_o=-\int _{u_{\text{min}}}^{u_{\text{in}}} {\rm d}u \frac{r_S}{\sqrt{P(u)}}+\int ^{u_{\text{min}}}_{u_{\text{o}}} {\rm d}u \frac{r_S}{\sqrt{P(u)}}\,.
\ee

\section{Perturbative expansion of the geodesic inside the hole}\label{AppExpand}
In this appendix we detail the analytic method presented in section~\ref{geodesic3} to determine the emission point on the LSS given the reception direction and time. 

We express all quantities related to the geodesic trajectory as a function of the direction of observation $\theta_{\text{obs}}$ and the position $r=D$ of the observer. From eq.~(\ref{EnergyAndDirection}), it follows that the direction of observation is related to the impact parameter $b$ by
\be
\sin \theta_{\text{obs}} = \frac{b}{D}\sqrt{A(D)}\,.
\ee
In order to get tractable results, we perform a perturbative expansion in the parameter $\hat r_S$ and obtain
\be\label{btheta}
\hat{b}=\hat{D}\left(1+\frac{1}{2}\frac{\hat r_S}{\hat D}+\mathcal{O}(\hat{r}_S^2)\right)\sin\theta_{\text{obs}}\,,
\ee
where we have introduced dimensionless quantities $\hat{D}\equiv D/\chi_h$, $\hat{b}\equiv b/\chi_h$ and $\hat{r}_S\equiv r_S/\chi_h$. Using this relation the impact parameter can be traded for the direction of observation and all quantities can be expressed in terms of $\theta_{\text{obs}}$.

\subsection{Crossing time and radius}\label{crossing}

From the condition (\ref{1junction}) we can Taylor expand the radial coordinate of the boundary around the reference geodesic which crosses the boundary at $\bar t _{\text{in}}$ for a radius $\bar{r}_{\text{in}}=\chi_h$ as \small
\begin{align}\label{Taylorr}
r(t) \simeq\, &\rinbar + [H A
r]_{\rinbar,\tinbar}(t-\tinbar)+\frac{1}{2} \left[-\frac{1}{2}H^2A^2 r
  +(\partial_r A)H^2 A r^2 \right]_{\rinbar,\tinbar}(t-\tinbar)^2+\nn\\
  &+\frac{1}{6} \left[\left(\partial_r A\right)^2 r^3 H^3 A-4 \left(\partial_r A \right)A^2 r^2 H^3+r A^3 H^3 \right]_{\rinbar,\tinbar}(t-\tinbar)^3\,,
\end{align}\normalsize
where we have neglected terms $\mathcal{O}(r_S^{2})$.\footnote{We want to have terms which are up to order $r_S^{3/2}$. In this way we would be able to have the crossing time $\eta_{\text{in}}\propto \sqrt{r_\text{in}/r_S}$ up to order $\hat r_S$. } For a given geodesic, we introduce the following quantities $\hat{\delta r}_{\rm in} \equiv \hat r_{\rm in} - \hatrinbar$ and $\hat{\delta t}_{\rm in} \equiv \hat t_{\rm in} - \hattinbar$ which correspond to the dimensionless difference of radial coordinate and time at entrance with respect to the reference geodesic. Using eqs. \eqref{Hofrs} and \eqref{dynamicsbounday}, the expansion (\ref{Taylorr}) leads to
\be\label{Taylordr}
\hat{\delta r}_{\rm in} \simeq \sqrt{\hat r_S} \hat{\delta t}_{\rm
  in}-\hat{r}_S^{3/2} \hat{\delta t}_{\rm in} -\frac{\hat r_S}{4}(\hat{\delta t}_{\rm in})^2+\frac{1}{6} \hat{r}_S^{3/2}(\hat{\delta t}_{\rm in})^3\,.
\ee
We see that we need $\hat{\delta t}_{\rm in}$ up to order $\hat r_S^{n-1/2}$ in order to get $\hat{\delta r}_{\rm in}$ up to order $\hat r_S^n$.  

The expression (\ref{Taylorr}) has been obtained from  a Taylor expansion of the equation describing the evolution of the hole boundary, around $\bar{r}_{\text{in}}=\chi_h$ (or $\hat{\bar r}_{\text{in}} = 1$ in dimensionless form). Since we are interested in radius and crossing times, this expansion has to be compared with the evolution of the radial coordinate of the photon. This latter is described by
\begin{align}\label{nonrad}
\frac{\dd t}{\dd r}=\pm \frac{1}{A(r)}\left[1-A(r)\left(\frac{b}{r}\right)^2\right]^{-1/2}\,.
\end{align}
For an observation angle $0 \leq \theta_{\text{obs}} \leq \pi/2$, we need to pick up the minus sign in the equation above. On the other hand for $\theta_{\text{obs}} > \pi/2$ the radial coordinate of the photon is decreasing for the photon going from the boundary to the point of minimum approach, characterized by a radial coordinate $r_{\text{min}}$, solution of $\dd r/\dd t=0$. The radius increases then from the point of minimum approach to the observer's position.\footnote{The validity of the perturbative expansion breaks down in a region characterized by $r<r_S$, i.e. our approach is not suitable to describe  geodesics intersecting a sphere centered in the origin of the coordinate system and of radius $r_S$ as it corresponds to the inner part of the black hole described by the \emph{Sch} metric.}  Let us consider the case  $0 \leq \theta_{\text{obs}} \leq \pi/2$ first. Expanding the right hand side of eq. (\ref{nonrad}) up to first order in $r_S/r$, the equation can be solved exactly. We get \small
\be\label{tin2}
t_{\text{in}}=\sqrt{D^2-b^2}-\sqrt{r_{\text{in}}^2-b^2}+\frac{r_S}{2}\left(\frac{D}{\sqrt{D^2-b^2}}-\frac{r_{\text{in}}}{\sqrt{r_{\text{in}}^2-b^2}}\right)+r_S\log\left(\frac{D+\sqrt{D^2-b^2}}{r_{\text{in}}+\sqrt{r_{\text{in}}^2-b^2}}\right)\,,
\ee \normalsize
which happens to be valid also for $\theta_{\text{obs}} > \pi/2$ as shown in appendix~\ref{proof}. 
The smallest crossing time is the one of the photon propagating radially outward, that is the one of the reference geodesic, and  can be found solving eq. (\ref{nonrad}) with $b=0$ and initial condition $r(t_o=0)=D$. At lowest order in $\hat{r}_S^{1/2}$ it is given by 
\be
\hattinbar = \hat D -1 + \hat r_S \log \hat D+\mathcal{O}(\hat{r}_S^2)\,.
\ee

The idea is now to evaluate eq. (\ref{tin2}) at lowest order in $\hat{r}_S$ (which is $\hat{r}_S^0$). Inserting the result in eq. (\ref{Taylordr}), we obtain then the crossing radius at order $\hat{r}_S^{1/2}$. This quantity in turn has to be substituted in eq. (\ref{tin2}) to obtain $t_{\text{in}}$ (and hence $\hat{\delta t}_{\rm in}$) up to order $\hat{r}_S^{1/2}$. Substituting this last expression for the crossing time in  (\ref{Taylordr}) we get the crossing radius at order $\hat{r}_S$. This procedure can be iterated until we obtain the crossing radius and crossing time at the desired order in powers of $\hat{r}_S^{1/2}$. 

Applying this procedure, from eq. (\ref{tin2}) we get at lowest order in $\hat r_S$ (that is $\hat r_S^0$) 
\be\label{deltat0}
\hat{\delta t}_{\rm in}=\hattin-\hattinbar\simeq  \hat D
(\cos\theta_{\rm obs}-1)- \left[\sqrt{1-\hat D^2 \sin^2 \theta_{\rm
      obs}}-1\right]\equiv \left(\hat{\delta t}_{\rm in}\right)^0\,.
\ee
Using this in \eqref{Taylordr} we can get $\hat{\delta r}_{\rm in}$ up
to $\sqrt{\hat r_S}$
\be\label{rin0}
\delta \hat{r}_{\text{in}}\simeq \sqrt{\hat{r}_S}(\delta t_{\text{in}})^{(0)}+\mathcal{O}(\hat{r}_S)\,.
\ee
We  use  this expression in eq. (\ref{tin2}) to get $\hat{\delta t}_{\text{in}}$ up to order $\hat{r}_S$. For this we use eq. (\ref{btheta}) and
\be
\sqrt{D^2-b^2} \simeq D \cos \theta_{\rm
  obs}\left(1-\frac{r_S}{2D}\tan^2 \theta_{\rm obs}\right)+\mathcal{O}(\hat{r}_S^2)\,.
\ee
We need also to expand $\sqrt{\rin^2 -b^2}
$ in eq. (\ref{tin2}). Using eq. (\ref{rin0}) we get
\be
\frac{\sqrt{\rin^2 -b^2}}{\chi_h} \simeq\sqrt{1-\hat D^2 \sin^2 \theta_{\rm
      obs}} - \sqrt{\hat r_S} \left[1-\frac{1+\hat D (1+\cos \theta_{\rm obs})}{\sqrt{1-\hat D^2 \sin^2 \theta_{\rm obs}} }\right]+\mathcal{O}(\hat{r}_S)\,.
\ee
Finally we obtain 
\be\label{t1}
\hat{\delta t}_{\rm in} \simeq \left(\hat{\delta t}_{\rm in}\right)^0 + \sqrt{\hat r_S} \left[1-\frac{1+\hat D (\cos \theta_{\rm obs}-1)}{\sqrt{1-\hat D^2 \sin^2 \theta_{\rm obs}} }\right] +\mathcal{O}(\hat{r}_S)\,.
\ee
No expansion in $\hat{D}$ has been performed until this point and thus the factors of order $\sqrt{\hat r_S}$ in (\ref{t1}) and (\ref{rin0}) are exact. Now using eq. (\ref{t1}) in eq. (\ref{Taylordr}) we can correct $\hat{\delta r}_{\rm in}$ up to order $\hat r_S$ \small
\be\label{rr}
\hat{\delta r}_{\rm in} = \sqrt{\hat{r}_S}\left(\hat{\delta t}_{\rm
    in}\right)^0 + \hat r_S \left[-\frac{1}{4}\hat D^2 (\cos \theta_{\rm obs}-1)^2 -\hat D (\cos \theta_{\rm obs}-1)
   -\frac{\hat D^2}{2}\sin^2 \theta_{\rm obs} \right]+\mathcal{O}(\hat{r}_S\,\hat{D}^3)\,.
\ee \normalsize
In this expansion we have expanded in powers of $\hat D$ the terms which are of order $\hat r_S$ and kept only the lowest powers. Our goal is indeed to get exact results at order $\sqrt{\hat r_S}$ but approximate results at order $\hat r_S$. We can iterate the procedure and using eq. (\ref{rr}) in eq. (\ref{tin2}) we get\small
\begin{align}\label{hatdt}
\hat{\delta t}_{\rm in} &\simeq \left(\hat{\delta t}_{\rm in}\right)^0 - \sqrt{\hat r_S} \left(\hat{\delta t}_{\rm in}\right)^0 \frac{1}{\sqrt{1-\hat D^2 \sin^2 \theta_{\rm obs}}} +\nn\\
&+\hat r_S\left[
   \frac{ (\cos \theta_{\rm obs} -1)}{2}+\hat D\left(\frac{\sin^2 \theta_{\rm obs}}{2}+\cos \theta_{\rm obs}-1\right)+\right.\nn\\
   &\hspace{3 em}\left.+2 \log \cos \left(\frac{\theta_{\rm obs}}{2}\right)-\frac{1}{2} \hat{D}^2 \left(\cos\theta_{\text{obs}}-1-\frac{\sin^2\theta_{\text{obs}}}{2}\right)\right] +\mathcal{O}(\hat{r}_S \hat{D^3})\,.
\end{align}\normalsize
We underline that the first correction in $\hat{r}^{1/2}_S$ both to the crossing time and to the crossing radius is an exact expression of $\hat{D}$. The next to leading order corrections of order $\hat r_S$ includes only terms up to order $\hat{r}_S \hat{D}^2$ as we have expanded in $\hat D$. Finally, substituting this result (\ref{hatdt}) in eq. (\ref{Taylordr}), we can consistently obtain $\hat{\delta r}_{\text{in}}$ up to order $\hat{r}_S^{3/2}$. We do not report here its explicit expression, which will be used in the calculation of $\eta_{\text{in}}$ in section \ref{eta} (see equation (\ref{eta final})). 

\subsection{Crossing angle}

The angular dynamics of the photon is described by
\begin{align}
\frac{\dd\theta}{\dd u}=\pm \sqrt{\frac{b^2}{1-b^2 u^2+r_S b^2 u^3}}\,.
\end{align}
The angle is always decreasing with time. The double sign above, as explained in appendix  \ref{AppGeodesicInside} is due to the fact that the radial coordinate of the photon decreases with time for $\theta_{\text{obs}} \leq \pi/2$. For $\theta_{\text{obs}} > \pi/2$, the radial coordinate of the photon decreases with time between the hole crossing and the point of minimum approach and it increases between the point of minimum approach to the position of the observer. Let us focus on the case $\theta_{\rm obs}<\pi/2$. Expanding the right hand side in $r_S/r = r_S u$ up to first order, the
equation reads simply
\be
\frac{\dd\theta}{\dd u}=-\frac{b}{\sqrt{1-b^2u^2}}+\frac{b^3 u^3 r_S}{2 (1-b^2u^2)^{(3/2)}}\,,
\ee 
and its integral is just
\be
\theta = {\rm const} - \arcsin(bu)+\frac{r_S(2-b^2u^2)}{2b \sqrt{1-b^2u^2} }\,.
\ee
Imposing initial conditions $\theta(u_o)=0$ with $u_o=1/D$, we get
\be
\theta(u) =
\theta(b/D)-\arcsin(bu)+\frac{r_S}{2b}\left(\frac{2-b^2u^2}{\sqrt{1-b^2u^2}}-\frac{2-b^2
  /D^2}{\sqrt{1-b^2/D^2}}\right)\,.
\ee
Performing an expansion in $\hat{D}$, the first and last term can be simplified and we get 
\be\label{thetaincompact}
\theta_{\rm in} \simeq \theta_{\rm obs} + \frac{\hat r_S}{\hat D} \tan (\theta_{\rm
  obs}/2)-\arcsin (b/r_{\rm in}) + \frac{1}{8}\hat r_S (\hat D \sin
\theta)^3 + {\cal O}(\hat r_S \hat D^5)\,.
\ee
We now use the following expansion in power of $\hat{r}_S^{1/2}$ that is deduced from (\ref{btheta})
\be
\frac{b}{r_{\rm in}} \simeq \hat D \sin \theta_{\rm obs} \left[1-
  \hat{\delta r}_{\rm in}+(\hat{\delta r}_{\rm in})^2+\frac{\hat
r_S}{2 \hat D}+\mathcal{O}(\hat{r}_S^2)\right]\,.
\ee
To simplify this expression at order $\hat{r}_S$, we keep only terms which are quadratic in $\hat D$ (but being
general at order $\sqrt{\hat r_S}$). We find 
\be\label{bonrin}
\frac{b}{r_{\rm in}} \simeq \hat D \sin \theta_{\rm obs} \left[1-\sqrt{r_S}\left(\hat{\delta t}_{\rm
    in}\right)^0 +\hat r_S \hat D (\cos \theta_{\rm obs}-1)+\frac{\hat
r_S}{2 \hat D}\right]+\mathcal{O}(\hat{r}_S \hat{D}^3) \,.
\ee
Plugging eq. (\ref{bonrin}) in eq. (\ref{thetaincompact}) and expanding in $\hat{r}_S^{1/2}$, we get\small
\begin{align}\label{thetainnonpert}
\theta_{\text{in}}&=\theta_{\text{obs}}-\arcsin(\hat{D}\sin\theta_{\text{obs}})-\hat{r}_S^{1/2}\hat{D} \sin\theta_{\text{obs}}\left[1+\frac{\hat{D}-1-\hat{D}\cos \theta_{\text{obs}}}{\sqrt{1-\hat{D}^2 \sin\theta_{\text{obs}}^2}}\right]+\nn\\
&+\hat{r}_S\left[\frac{1}{\hat{D}}\tan\left(\frac{\theta_{\text{obs}}}{2}\right)-\frac{1}{2}\sin\theta_{\text{obs}}+\hat{D}^2 \sin\theta_{\text{obs}}\left(1-\cos\theta_{\text{obs}}-\frac{\sin^2\theta_{\text{obs}}}{4}\right)\right]+\mathcal{O}(\hat{r}_S \hat{D}^3)\,.
\end{align} \normalsize
We verified that we get exactly the same result for the case $\theta_{\rm  obs}>\pi/2$.

\subsection{Propagation in the FLRW region}\label{eta}

In order to compute the geodesic in the FLRW region we need the conformal time at entrance and the angle $\alpha$ defined in eq. (\ref{aalpha}). Eq. (\ref{Tinetain}) is rewritten in dimensionless form as
\be\label{etaindef}
\hat \eta_{\text{in}}=2\sqrt{\frac{\hat r_{\text{in}}}{\hat r_S}}\,.
\ee
Having calculated the expression for $\hat{r}_{\text{in}}$ up to order $\hat{r}_S^{3/2}$, we can consistently find the expression for $\hat \eta_{\text{in}}$ up to order $\hat{r}_S$, just taking the definition (\ref{etaindef}) and performing a Taylor expansion in $\hat{r}_S^{1/2}$ up to second order.  Up to $\mathcal{O}(\hat{r}_S \hat{D}^3)$ we get \small
\begin{align}\label{eta final}
\hat{ \eta}_{\rm in}&=\frac{2}{\sqrt{\hat{r}_S}}+1+\hat{D}(\cos\theta_{\text{obs}}-1)-\sqrt{1-D^2\sin\theta_{\text{obs}}^2}+\hat{r}_S^{1/2} -\nn\\
&-\hat{r}_S^{1/2}\,\,\frac{1+\hat{D}(\cos\theta_{\text{obs}}-1)}{\sqrt{1-\hat{D}^2\sin\theta_{\text{obs}}^2}} 
-\frac{\hat{r}_S^{1/2}}{2}\left[1+\hat{D}(\cos\theta_{\text{obs}}-1)-\sqrt{1-D^2\sin\theta_{\text{obs}}^2}\right]^2+\\
&+\hat{r}_S\left[\frac{1}{2}\left(\cos\theta_{\text{obs}}-1\right)+2\log \left(\cos\frac{\theta_{\text{obs}}}{2}\right)+\frac{5}{4} \hat{D}^2\left(\cos\theta_{\text{obs}}-1\right)^2+\frac{\hat{D}}{2}\sin\theta_{\text{obs}}^2\right]+\mathcal{O}(\hat{r}_S \hat{D}^3)\,.\nn
\end{align}
\normalsize
This result is exact in $\hat{D}$  at order $\hat{r}_S^{1/2}$, but it is truncated at quadratic order in $\hat{D}$ at order $\hat{r}_S$. 
Alternatively, this result could have been obtained by a Taylor expansion of $\eta(t)$ around $\bar t_{\text{in}}$, using $\dd \eta/\dd t = A(r_h(t))/a(T(t))$.

Furthermore, in order to expand $\alpha$, we use the following relations, valid up to $\hat{r}_S$
\be
a_{\text{in}} k_{\text{in}}^\eta  =  k_{\text{in}}^t-\left(\sqrt{\frac{r_S}{r
    A}}\right)_{\rm in}k_{\text{in}}^r \simeq
E\left(1+\frac{r_S}{r}+\sqrt{\frac{r_S}{r}} \sqrt{1-\frac{b^2A}{r^2}}\right)_{\rm in}\,,
\ee
from which we find up to order $\mathcal{O}(\hat{r}_S \hat{D}^3)$
\be
\tilde{E}^{-1} a_{\text{in}} k_{\text{in}}^\eta \simeq 1+\sqrt{\hat r_S}\sqrt{1-\hat D^2 \sin^2
  \theta_{\rm obs}}+\hat r_S\left[1-\frac{\hat D}{2} (\cos \theta_{\rm obs}-1)-\frac{1}{4} \sin^2 \theta_{\text{obs}} \hat{D}^2\right] \,,
\ee
where at order $\hat r_S$ only terms up to quadratic order in $\hat D$ have been retained. Therefore we get
\begin{align}
\sin \alpha = \frac{b}{r_{\rm in}}& \left[1-\sqrt{\hat r_S}\sqrt{1-\hat D^2 \sin^2 \theta_{\rm obs}}+\right. \nn\\
& \left.+ \hat r_S \frac{\hat D}{2} (\cos \theta_{\rm obs}-1)-\hat{r}_S \frac{3}{4} \hat{D}^2 \sin^2 \theta_{\text{obs}}+\mathcal{O}(\hat{r}_S \hat{D}^3)\right]\,.
\end{align}
Substituting $b/r_{\rm in}$ given by eq. (\ref{bonrin}) and retaining only leading terms in $\hat{r}_S$, we get
\begin{align}
\alpha=\arcsin (\hat{D}&\sin\theta_{\text{obs}})-\hat{D}\sqrt{\hat{r}_S}\sin\theta_{\text{obs}}\,\frac{1+\hat{D}(\cos\theta_{\text{obs}}-1)}{\sqrt{1-\hat{D}^2\sin\theta_{\text{obs}}^2}}+\nn\\
&+\frac{\hat{r}_S}{2} \sin\theta_{\text{obs}}\left[1+5 \hat{D}^2 \left(\cos\theta_{\text{obs}}-1\right)+\frac{1}{2} \hat{D}^2 \sin^2\theta_{\text{obs}}\right]+\mathcal{O}(\hat{r}_S \hat{D}^3)\,.
\end{align}
\normalsize

\subsection{Position of the last scattering surface}

With the results of the previous section, we calculate the radial distance to the CMB and the angle of the emission point, defined in eqs. (\ref{phiLSS})\small
\begin{align}\label{chiLSSS}
\hat{\chi}_{\text{LSS}}&=1-\hat \eta_{\text{LSS}}+\frac{2}{\sqrt{\hat r_S}}+\hat{D}\left[1-\sqrt{\hat
  r_S}(1-\hat D)\right]\left(\cos\theta_{\text{obs}}-1\right)+\frac{5}{4} \sqrt{\hat{r}_S} \hat{D}^2\sin\theta_{\text{obs}}^2+\mathcal{O}( \hat{r}_S^{1/2} \hat{D}^3)+\nn\\
  &+\frac{\hat{r}_S}{2}\left[\left(\cos\theta_{\text{obs}}-1\right)(1-5 \hat{D}^2)+\frac{\hat{D}^2}{4}\hat\eta_{\text{LSS}} \sin^2\theta_{\text{obs}}-\frac{19}{4} \hat{D}^2 \sin^2\theta_{\text{obs}}+4 \log\left(\cos\frac{\theta_{\text{obs}}}{2}\right)\right]+\nn\\
  &+\mathcal{O}( \hat{r}_S \hat{D}^3)\,,
\end{align}
\normalsize and
\small
\begin{align}\label{phiLSSS}
\theta_{\text{LSS}}&=\theta_{\text{obs}} - \frac{3}{2}\hat D \sqrt{\hat {r}_S} \sin \theta_{\text{obs}}+\mathcal{O}( \hat{r}_S^{1/2} \hat{D}^3)+\nn\\
&+\hat{r}_S \tan \frac{\theta_{\text{obs}}}{2}\left[\frac{1}{\hat{D}}+\frac{3}{4} \hat{D}\left(1+\cos\theta_{\text{obs}}\right)-\frac{9}{4}\hat{D}^2 \sin^2\theta_{\text{obs}}-\frac{1}{4} \hat{D}\, \hat\eta_{\text{LSS}} \left(1+\cos\theta_{\text{obs}}\right)\right]+\nn\\
&+\mathcal{O}( \hat{r}_S \hat{D}^3)\,.
\end{align}

\subsection{Energy shift}\label{Eobsobs}

Each trajectory is univocally characterized by the constant of motion $\tilde{E}$ and by the impact parameter $b$ or $\theta_{\text{obs}}$. For static observers in the hole having velocity $1/\sqrt{A(r)}\left(\partial_t\right)^{\mu}$, the energy measured is given by (\ref{EnergyAndDirection}). For the observer at $r=D$, and keeping only the lowest orders in $\hat r_S$, this energy reads simply
\be\label{eo}
E_o=\tilde{E}\left(1+\frac{1}{2}\frac{\hat{r}_S}{\hat{D}}+\mathcal{O}(\hat{r}_S^2)\right)\,.
\ee
For static observers of the FLRW region having velocity $1/a(\eta)\left(\partial_{\eta}\right)^{\mu}$, the energy measured is $a k^\eta$. In particular, for a static observer of the FLRW region sitting exactly on the boundary of the hole at the point where the geodesic intersects the boundary, from eqs. (\ref{kcompin}) and (\ref{kofk}), we find that the energy measured is
\be\label{EnergyBoostatin}
E_{\rm in} = a_{\rm in} k_{\rm in}^\eta\simeq \tilde E \left\{1+
  \sqrt{\hat r_S}\sqrt{1-\hat D^2 \sin^2 \theta_{\rm obs}}+ \hat{r}_S\left[1-\frac{\hat{D}}{2}\left(\cos\theta_{\text{obs}}-1\right)-\frac{\hat{D}^2}{4} \sin^2\theta_{\text{obs}}\right]\right\}\,.
\ee
Finally, a comoving observer on the LSS where the photon was emitted would have measured an energy
\begin{align}\label{ELSS}
E_{\text{LSS}}&=\left(\frac{a_{\text{in}}}{a_{\text{LSS}}}\right)E_{\text{in}} = \left(\frac{1+\hat{\delta r}_{\text{in}}}{a_{\text{LSS}}}\right)E_{\text{in}}\,.
\end{align}
Gathering eqs. (\ref{eo}, \ref{EnergyBoostatin}, \ref{ELSS}) and using (\ref{rr}) we can relate the observed energy to the energy of emission. We get
\be
E_o=E_{\text{LSS}}\,a _{\text{LSS}}\left(1+\delta E+\mathcal{O}(\hat{r}_S \hat{D}^3)\right)\,,
\ee
\begin{align}
\delta E = &-\hat{r}_S^{1/2}\left[\hat{D}\left(\cos\theta_{\text{obs}}-1\right)+1\right]+\nn\\
&+\hat{r}_S\left[\frac{1}{2\hat{D}}-\frac{5 \hat{D}}{2} \left(1-\cos\theta_{\text{obs}}\right)+\frac{5 \hat{D}^2}{2}\left(1-\cos\theta_{\text{obs}}\right)-\hat{D}^2 \sin^2\theta_{\text{obs}}\right]\,.
\end{align}
Normalizing the result above with respect to the average on angles, we get up to $\mathcal{O}(\hat{r}_S \hat{D}^3)$\small
\begin{align}\label{Eobs}
\PotE(\theta_{\text{obs}}) &\equiv \frac{E_o}{\langle E_o\rangle}=1-\hat{D}\, \hat{r}_S^{1/2} \cos\theta_{\text{obs}}+\frac{\hat{r}_S}{2}\left[3 \hat{D} (1-\hat D)\cos\theta_{\text{obs}}+2 \hat{D}^2 (2\cos^2\theta_{\text{obs}}-1)\right]\,.
\end{align}
\normalsize
This quantity gives the extra angular modulation introduced by the local void, which multiplies the standard CMB signal. Indeed we can identify the angular averaged signal $\langle E_o\rangle$ with the signal $E_{\text{CMB}}(\theta_{\text{obs}})$ that would have been measured without the local void,  with the usual angular dependence characterizing photon energies in standard cosmology. Hence, the effect of the hole is given by
\be
E_{\text{o}}(\theta_{\text{obs}}) = \PotE (\theta_{\text{obs}}) E_{\text{CMB}}(\theta_{\text{obs}})\,.
\ee 
In particular we note that at order $\sqrt{\hat r_S}$ only a dipolar modulation is present.

\section{Proof of generality of the analytic solution ($\forall\, \theta{_\text{obs}}$)}\label{proof}

In this appendix we show that the equation (\ref{tin2}) given in Sec. \ref{crossing}   is valid for every direction of observation and not just for $0\leq \theta_{\text{obs}}\leq \pi/2$ for which it was derived.

The radial evolution of the photon is described by
\be\label{first}
\frac{\dd t}{\dd r}=\frac{1}{A(r)}\left[1-A(r) \left(\frac{b}{r}\right)^2\right]^{-1/2}\,,
\ee
for $\cos\theta_{\text{obs}}<0$ and for the radial coordinate evolving from $D$ to $r_{\text{min}}$, with $r_{\text{min}}$ being the radius of minimum approach, solution of $\dd r/\dd t=0$. The radial evolution of the photon is described by
\be\label{second}
\frac{\dd t}{\dd r}=-\frac{1}{A(r)}\left[1-A(r) \left(\frac{b}{r}\right)^2\right]^{-1/2}\,,
\ee
for $\cos\theta_{\text{obs}}\geq0$ and for $\cos\theta_{\text{obs}}<0$ and $r$ evolving from $r_{\text{min}}$ to the boundary, $r_{\text{in}}$. We focus now on equation (\ref{first}). If we expand the right hand side up to first order in $r_S$, we find the following primitive
\be\label{pr1}
{\rm I}=\sqrt{r^2-b^2}+\frac{r_S}{2} \frac{r}{\sqrt{r^2-b^2}}+r_S \log\left(\hat{r}+\sqrt{\hat{r}^2-\hat{b}^2}\right)+cnst\,.
\ee
Substituting
\be\label{btheta2}
\frac{b}{r}=\frac{\sin\theta}{\sqrt{A(r)}}\,,
\ee
the primitive becomes
\be\label{pr2}
{\rm I}=|\cos\theta|\left(r+\frac{r_S}{2}\right)+r_S \log\left[r\left(1+|\cos\theta|\right)\right]+cnst\,.
\ee
Obviously, the primitive of (\ref{second}) is just $-{\rm I}$. We fix the constant of integration in (\ref{pr2}) in such  way that ${\rm I}(r_{\text{min}})=0$. Using the equation (\ref{btheta2}), it is easy to verify that $\cos\theta_{\text{min}}=0$ and $r_{\text{min}}=b-r_S/2+\mathcal{O}(r_S)$. Once the constant of integration is fixed, we get up to $\mathcal{O}(r_S^2)$
\be\label{pr3}
{\rm I}=|\cos\theta|\left(r+\frac{r_S}{2}\right)+r_S \log\left[r\left(1+|\cos\theta|\right)\right]-r_S \log (\hat{D}\sin\theta_{\text{obs}})\,.
\ee
For $\cos\theta_{\text{obs}}\geq0$, we simply have
\be
t_{\text{in}}\equiv \Delta t ={\rm I(}D)-{\rm I}(r_{\text{in}})\,,
\ee
hence
\begin{align}  \label{end}
t_{\text{in}}=&\cos\theta_{\text{obs}}\left(D+\frac{r_S}{2}\right)+r_S \log\left(\frac{1+\cos\theta_{\text{obs}}}{\sin\theta_{\text{obs}}}\right)-\nn\\
&-|\cos\theta_{\text{in}}|\left(D+\frac{r_S}{2}\right)-r_S \log\left(\frac{r_{\text{in}}(1+|\cos\theta_{\text{in}}|)}{D \sin\theta_{\text{obs}}}\right)\,.
\end{align} \normalsize
Using eq. (\ref{btheta2}) expanded as in (\ref{btheta}), it is possible to show that this is exactly the solution~(\ref{tin2}).

For $\cos\theta_{\text{obs}}<0$ the situation is more complex. In this case the radial dynamics of the photon is described by eq. (\ref{second}) for the coordinate $r$ evolving between $D$ and $r_{\text{min}}$ and by  eq. (\ref{first}) for $r$ evolving from $r_{\text{min}}$ to the boundary. We need therefore to add the corresponding time intervals
\be
t_{\text{in}}\equiv \Delta t=\Delta t_1+\Delta t_2\,,
\ee
with
\be
\Delta t_1={\rm I}(r_{\text{in}})-{\rm I}(r_{\text{min}})={\rm I}(r_{\text{in}})\,,
\ee
\be
\Delta t_2={\rm I}(r_{\text{min}})-{\rm I}(D)=- {\rm I}(D)\,.
\ee
Performing an explicit calculation and using that in this case $\cos\theta_{\text{obs}}<0$,  we recover exactly eq. (\ref{end}) which is equivalent to eq. (\ref{tin2}). Hence we conclude that eq. (\ref{tin2}) is valid for all values of $\theta_{\text{obs}}$.

\section{Multipoles of lensing-like deflection, energy shift and radial modulation}\label{generic generic}

In this appendix we detail the calculations presented in section \ref{yves}. 

\subsection{Tools for multipoles}\label{tools}

The symmetry of the void model implies that when we decompose in spherical harmonics a given function on the sphere of photon directions, a natural choice is to take the azimuthal direction to be the one defined by the hole center
and the observer.  Then in the decomposition, contributions are non-vanishing only for $m=0$ due to the axial symmetry. The energy shift and the radial modulation are scalar quantities on the sphere, which can be decomposed as  
\be\label{alpha}
\varrho = \sum_{\ell m} \varrho_{\ell m} Y_{\ell m}\,,
\ee
where $\varrho$ stands for either for energy or for radial modulation. 
Only $\varrho_{\ell0}$ is non-vanishing so writing $\cos \theta = \mu$
\be\label{result1}
\varrho_{\ell0} = \sqrt{\pi(2 \ell+1)}\int_{-1}^1 \varrho(\theta) P_\ell(\mu) \dd \mu\,,
\ee
and we used
\be
Y_{\ell 0} = \sqrt{\frac{2\ell+1}{4\pi}}P_\ell(\cos \theta)\,.
\ee

In full generality, the lensing-like deflection $\Gamma$ has to be decomposed in its gradient-mode and curl-mode as
\be\label{lensing2}
\Gamma_a({\bm{n}})=\nabla_a \varphi ({\bm{n}}) +\epsilon_a^b\, \nabla_b \omega ({\bm{n}})\,,
\ee
where
\be
\nabla_a \varphi=\sum_{\ell m} \varphi_{\ell m} \nabla_a Y_{\ell m}\,,\hspace{2 em} \epsilon_a^b\,\nabla_b \omega=\sum_{\ell m} \omega_{\ell m} \epsilon_a^b \nabla_b Y_{\ell m}\,.
\ee
After standard manipulations, it is possible to verify the following identities
\be
\nabla_a Y_{\ell m}=\sqrt{\frac{\ell(\ell+1)}{2}}\left[-  _1 Y_{\ell m} ({\bm{e}}_+)_a+_ {-1}Y_{\ell m} ({\bm{e}}_-)_a\right]\,,
\ee
\be
\epsilon_a^{\,b}\,\nabla_b Y_{\ell m}=i \sqrt{\frac{\ell(\ell+1)}{2}}\left[ _1 Y_{\ell m} ({\bm{e}}_+)_a+_ {-1}Y_{\ell m} ({\bm{e}}_-)_a\right]\,,
\ee
where we have introduced the helicity basis as
\be
{\bm{e}}_+ \equiv \frac{{\bm{e}}_{\theta}-i {\bm{e}}_{\phi}}{\sqrt{2}}\,,\hspace{2 em} {\bm{e}}_- \equiv \frac{{\bm{e}}_{\theta}+i {\bm{e}}_{\phi}}{\sqrt{2}}\,.
\ee
For $m$=0, using $_{-1} Y_{\ell 0}=-(\, _1 Y_{\ell 0})$, the expressions above simply reduces to
\be
\nabla_a Y_{\ell 0}= -  \sqrt{\ell(\ell+1)} \,_1 Y_{\ell 0}\left({\bm{e}}_{\theta} \right)_a\,,\hspace{2 em}\epsilon_a^{\,b }\,\nabla_b Y_{\ell 0}= \sqrt{\ell(\ell+1)} \,_1 Y_{\ell 0}\left({\bm{e}}_{\phi} \right)_a\,.
\ee
We conclude that being the lensing in our problem totally along ${\bm{e}}_{\theta}$, we do not have curl-mode excited in the lensing angular power spectrum, i.e. $\omega_{\ell m}=0$. Eq. (\ref{lensing2}) simplifies to 
\be\label{D.10}
\Gamma {\bm{e}}_{\theta}=-\sum_{\ell}  \varphi_{\ell 0}\, \sqrt{\ell(\ell+1)} \, _1 Y_{\ell 0} \, {\bm{e}}_{\theta}\,.
\ee
We recall the orthonormality condition
\be
\int {\rm d}\Omega\, \, {}_sY^*_{\ell m}({\bm{n}})\, {}_sY_{l' m'} ({\bm{n}})=\delta_{\ell \ell'} \delta_{m m'}\,,
\ee
and the following relations between spin-0 and spin-1 spherical harmonics 
\be
\spart Y_{\ell m} = \sqrt{\ell(\ell+1)} \,{}_1 Y_{\ell m}\,,\qquad \spartb Y_{\ell m} = -\sqrt{\ell(\ell+1)} \,{}_{-1} Y_{\ell m}\,,
\ee
where all the conventions are those of Ref. \cite{Ruth_book}. Making use of the relations above, we immediately get 
\be\label{result2}
\varphi_{\ell 0 } =
\frac{2 \pi}{\ell(\ell+1)}\sqrt{\frac{2\ell+1}{4\pi}}\int_{-1}^{1}\partial_\theta
P_\ell(\cos \theta) \Gamma(\theta) \dd\mu\,.
\ee

\subsection{Multipoles of lensing and radial modulation in the boosted frame}\label{generic}

We write now the first harmonic coefficients of radial modulation and lensing after the boost, eqs. (\ref{cin1}) and (\ref{cin3}), respectively. From eq. (\ref{cin1}) we see that after the boost, the energy measured by the comoving observer at the center of the tilde system of coordinates has no angular dependence. For the radial modulation, writing
\be
 \tilde{d}' /\chi_h=\sum_{\ell} d_{\ell 0}\, Y_{\ell0}\,,
\ee
and using the result of the previous section, eq. (\ref{result1}), we get apart from the trivial monopole \small
\be
 d_{1 0}=0\,,\quad d_{2 0}=\frac{\sqrt{5 \pi}}{30}\left(-5 +3 \hat{D}^2 \right)\hat{r}_S^{3/2}\,,\quad d_{3 0}= \frac{\sqrt{7 \pi}}{12}\hat{r}^{3/2}_S\,,\quad d_{4 0}=-\frac{3}{20}\,\sqrt{\pi}\,\hat{r}_S^{3/2}\,.
 \ee \normalsize
From this we can easily infer the general formula, valid at the leading order $\hat{r}_S/\hat{D}$ and for $\ell\geq 2$
 \be
d_{\ell 0}=\hat{r}_S^{3/2}\,\sqrt{\pi}\,(-)^{\ell+1} \frac{\sqrt{2\ell+1}}{\ell(\ell+1)}\,.
 \ee

For lensing, writing
\be
\tilde{\Gamma}' \, {\bm{e}}_{\theta} =-\sum_{\ell}  \varphi_{\ell 0}\, \sqrt{\ell(\ell+1)} \, _1 Y_{\ell 0} \, {\bm{e}}_{\theta}\,,
\ee
and using eq. (\ref{result2}) we get for the first multipoles \small
\be\small
\hspace{-1 em}\varphi_{1 0}=-\sqrt{3 \pi}\,\hat{r}_S \left(\frac{1}{\hat{D}}-\hat{D}\right)\,,\,\,\, \varphi_{2 0}=\sqrt{5 \pi}\,\hat{r}_S \left(\frac{1}{3\hat{D}}-\frac{2}{15} \hat{D}^2\right)\,,\,\,\, \varphi_{3 0}=-\frac{\sqrt{7\pi}}{6} \frac{\hat{r}_S}{\hat{D}}\,,\,\,\,  \varphi_{4 0}=\frac{3 \sqrt{\pi}}{10}\frac{\hat{r}_S}{\hat{D}}\,.%,\,\,\,\varphi_{5 0}=-\frac{\sqrt{11 \pi}}{15} \frac{\hat{r}_S}{\hat{D}}\,.
 \ee \normalsize
 From this we can easily infer the general formula, valid at the leading order $\hat{r}_S/\hat{D}$ 
 \be
 \varphi_{\ell 0}=\frac{\hat{r}_S}{\hat{D}}\sqrt{4\pi}\,(-)^{\ell} \frac{\sqrt{2\ell+1}}{\ell(\ell+1)}\,. 
 \ee

\subsection{Generalization to a generic azimuthal direction}

Until now we have considered a system of coordinates such that the azimuth was aligned with  ${\bm{e}_z}$, where ${\bm{e}_z}$ denotes the direction observer-hole.  With this choice, when we decompose in spherical harmonics, the only non-vanishing multipoles of the lensing potential and radial modulation are the $m=0$ ones. For the lensing potential we got
\be
\varphi({\bm{n}})=\sum_{\ell} \varphi_{\ell 0} Y_{\ell 0}({\bm{n}})\,.
\ee
We now want to generalize the results found to the  case of a general orientation of the observer-hole axis . We consider a rotated coordinate frame. The rotation is described by a $SO(3)$ matrix $R_1$ and it is characterized by Euler angles $(\phi_1, \theta_1, 0)$.  In the new coordinate frame the direction observer-hole is described by the unit vector ${\bm{n_1}}=R_1 \gr{e}_z$ and a direction described by a unit vector ${\bm{n}}$ in the old reference frame, is given by $R_{1}^{-1} {\bm{n}}$ in the new one. Then, we use
\be
Y_{\ell0}(R_1^{-1}{\bm{n}})=\sum_m D_{m0}^{(\ell)}(R_1) Y_{\ell m}({\bm{n}})=\sqrt{\frac{4\pi}{2\ell+1}} \sum_m Y_{\ell m}^* ({\bm{n_1}}) Y_{\ell m} ({\bm{n}})\,,
\ee
and we get for the lensing potential 
\be\label{pot1}
\varphi_{\text{new}}({\bm{n}})=\varphi(R_1^{-1} {\bm{n}})=\sum_{\ell} \varphi_{\ell 0} \sqrt{\frac{4\pi}{2\ell+1}} \sum_m Y_{\ell m}^* ({\bm{n_1}}) Y_{\ell m} ({\bm{n}})\,.
\ee
Now we need to extract multipoles. We write
\be\label{pot2}
\varphi_{\text{new}}({\bm{n}})=\sum_{m \ell} \varphi_{\ell m} Y_{\ell m} ({\bm{n}})\,,
\ee
and from a comparison of  (\ref{pot2}) and (\ref{pot1}) we can immediately extract the multipoles $\varphi_{\ell m}$ 
\be
\varphi_{\ell m}=\varphi_{\ell 0} \sqrt{\frac{ 4\pi}{2\ell +1}} Y_{\ell m}^*({\bm{n}_1})\,.
\ee

Analogously, for the time delay defined in eq. (\ref{Shapirodef}), if we consider a system of coordinates in which the azimuth is aligned with  ${\bm{e}_z}$, we have that  the only non vanishing multipoles are the  $m=0$ ones
\be
d({\bm{n}})=\sum_{\ell} d_{\ell 0} Y_{\ell 0}\,.
\ee
If we choose a system of coordinates in which the axis-observer hole is described by the unit vector ${\bm{n_1}}$, writing
\be
d_{\text{new}}({\bm{n}})=d\left(R_{1}^{-1} {\bm{n}}\right)=\sum_{\ell m} d_{\ell m} Y_{\ell m}\,,
\ee
 and repeating the same steps presented above for the lensing potential, we get
 \be
d_{\ell m}=d_{\ell 0} \sqrt{\frac{ 4\pi}{2\ell +1}} Y_{\ell m}^*({\bm{n}_1})\,.
\ee

\section{CMB sky: general definitions}\label{CMBsky}

In this appendix we collect our definitions for the CMB intensity map, used in sections \ref{CMBoff} and \ref{Boost}. 

We consider an electromagnetic wave propagating in direction ${\bm{n}}$. We define the polarization direction ${\bm{\epsilon^{(1)}}}$ and ${\bm{\epsilon^{(2)}}}$ is such  way that $({\bm{\epsilon^{(1)}}}\,,{\bm{\epsilon^{(2)}}}\,, {\bm{n}})$ form a right-handed orthonormal system. The electric field of the wave is of the form
\be
{\bm{e}}=E_1 {\bm{\epsilon^{(1)}}}+ E_2 {\bm{\epsilon^{(2)}}}\,.
\ee
The polarization tensor of an electromagnetic wave is defined as
\be
P_{ij}=\tilde{\mathcal{P}}_{a b}\, \epsilon_i^{(a)} \epsilon_j^{(b)}\,, \qquad \text{with}\quad \tilde{\mathcal{P}}_{a b}=E_a^* E_b\,.
\ee
$\tilde{\mathcal{P}}_{a b}$ is a Hermitian $2 \times 2$ matrix and therefore it can be written as
\begin{align}
\tilde{\mathcal{P}}_{a b}=&\frac{1}{2}\left[ I\,\sigma_{ab}^{(0)}+U\,\sigma_{ab}^{(1)}+V\,\sigma_{ab}^{(2)}+Q\,\sigma_{ab}^{(3)}\right]= \frac{1}{2} \,I \,\sigma_{ab}^{(0)}+\mathcal{P}_{ab}\,,
\end{align}
where $\sigma^{(\alpha)}$ with $\alpha=1\,, 2\,, 3$ denote the Pauli matrices and $\sigma^{(0)}=1_2$.  The objects $I\,, U\,, Q\,, V$ are  four real functions of the photon direction ${\bm{n}}\equiv (\theta_{\text{obs}}\,, \phi_{\text{obs}})$ and are called Stokes parameters. In terms of the electric field, the Stokes parameters are given by
\begin{align}
I=|E_1|^2+|E_2|^2\,,\quad Q=|E_1|^2-|E_2|^2\,,\quad U=2 \text{Re}(E_1^*E_2)\,,\quad  V=2 \text{Im} (E_1^* E_2)\,.
\end{align}

Since Thomson scattering does not introduce circular polarization, we expect the $V$ Stokes parameter of the CMB radiation to vanish. In the following we therefore set $V=0$.  The intensity $I$ is proportional to the energy density of the CMB, $8\pi \rho= I$ and therefore it is related to the temperature anisotropy field as
\be\label{deftemperature}
\Theta({\bm{n}})\equiv\frac{T({\bm{n}})-\langle T\rangle}{\langle T\rangle}=\frac{1}{4} \frac{\rho({\bm{n}})-\langle \rho\rangle}{\langle \rho\rangle}=\frac{1}{4}  \frac{I({\bm{n}})-\langle I\rangle}{\langle I\rangle}\,.
\ee

We define the following quantities (complex Stokes parameters)
\be\label{complexstokes1}
P_+\equiv \mathcal{P}_{++}=2 \mathcal{P}^{a b} \epsilon_a^{(+)}\epsilon_b^{(+)}=Q + i U\,,
\ee
\be\label{complexstokes2}
P_-\equiv \mathcal{P}_{++}=2 \mathcal{P}^{a b} \bar{\epsilon}_a^{(+)}\bar{\epsilon}_b^{(+)}=2 \mathcal{P}^{a b} \epsilon_a^{(-)}\epsilon_b^{(-)}=Q - i U\,,
\ee
where we have introduced the helicity basis
\be
{\bm{\epsilon}}^{(\pm)}=\frac{1}{\sqrt{2}}\left({\bm{\epsilon}}^{(1)}\pm i\,{\bm{\epsilon}}^{(2)}\right)\,.
\ee
Up to a factor $2$, the complex Stokes parameters are the components of the polarization tensor in the helicity basis ($\mathcal{P}_{+\,-}=\mathcal{P}_{-\,+}=0$ with $V=0$). Under a rotation $O\in SO(3)$ the complex Stokes parameters transform as
\be
\left(P\right)_{\pm}'({\bm{n}})=e^{\pm i \alpha_O({\bm{n}})}\,P\left(O^{-1}{\bm{n}}\right)\,.
\ee
where $\alpha_O({\bm{n}})$ are rotation angles associated to $O$.  We see that $P_{\pm}$ transform like spin-2 variables with magnetic quantum number $\pm 2$ under rotation around the ${\bm{n}}$ axis. 

%It is convenient to define a frame-independent representation to expand the complex Stokes parameters, making use of spin-weighted spherical harmonics (with spin $2$). The spin-weighted spherical harmonics are the components of an asymmetric rank $|s|$ tensor field defined on the tangent space of the sphere in the canonical basis ${\bm{e}}_{\theta}=\partial_{\theta}$ and ${\bm{e}}_{\phi}=\partial_{\phi}/\sin\theta$. 

With respect to the helicity basis
\be
{\bm{e}}_{\pm}=\frac{{\bm{e}}_{\theta}\pm i {\bm{e}}_{\phi}}{\sqrt{2}}\,,
\ee
the complex Stokes parameter can be expanded as 
\be\label{pizza2}
P_{\pm}({\bm{n}})=\sum_{\ell m} A_{\ell m}^{(\pm 2)}\, _{\pm 2}Y_{\ell m}({\bm{n}})=\sum_{\ell m} \left(E_{\ell m}\pm B_{\ell m}\right)\, _{\pm 2}Y_{\ell m}({\bm{n}})\,,
\ee
where we have defined
\be
E_{\ell m}\equiv \frac{1}{2}\left(A_{\ell m}^{(2)}+A_{\ell m}^{(-2)}\right)\,,\qquad
B_{\ell m}\equiv -\frac{i}{2}\left(A_{\ell m}^{(2)}-A_{\ell m}^{(-2)}\right)\,.
\ee
To conclude this section, we define the following scalar quantities
\be\label{EE}
E({\bm{n}})=\sum_{\ell m}E_{\ell m} Y_{\ell m}({\bm{n}})\,,\qquad
B({\bm{n}})=\sum_{\ell m}B_{\ell m} Y_{\ell m}({\bm{n}})\,.
\ee
$E$ and $B$  are invariant under rotation and under parity they transfer as a scalar and a pseudo scalar, respectively. $E$ measures gradient contributions, while $B$ curl contributions to the electric field considered as a function on the sphere of photon directions.

\section{CMB sky from the void: technical aspects}\label{CMB void}

In this appendix, we detail the calculation of the effects of geometrical lensing and time-delay on the shape of the CMB temperature and polarization angular power spectrum.  The main results are summarized in the body of the paper, section \ref{CMBoff}.

The temperature anisotropy field  defined in eq. (\ref{deftemperature}), is a function on the sphere of photon directions and can be decomposed in spherical harmonics as 
\be\label{decomp1}
\Theta({\bm{x_o}}\,, \eta_o\,, {\bm{n}})=\sum_{\ell m} \Theta_{\ell m}({\bm{x_o}}, \eta_o) Y_{\ell m}({\bm{x_o}})\,,
\ee
where we have explicitly indicated the dependence on the observer position ${\bm{x_o}}$ and reception time $\eta_o$.  It is convenient to work in Fourier space, defining the following Fourier decomposition 
\be
\Theta({\bm{x_o}}\,, \eta_o\,, {\bm{n}})=\int \frac{d^3k}{(2\pi)^{3/2}}\,\hat{\Theta}({\bm{k}}, \eta_o\,, {\bm{n}})\,e^{i{\bm{k}}\cdot {\bm{x_o}}}\,.
\ee
The complex Stokes parameters are functions on the sphere and can be decomposed as
\begin{align}
P_{\pm}({\bm{x_o}}\,, \eta_o\,, {\bm{n}})&=\sum_{\ell m} (P_{\pm} ({\bm{x_o}}\,, \eta_o))_{\ell m} \, _{\pm 2} Y_{\ell m} ({\bm{n}})\,,\\
&=\sum_{\ell m} (E_{\ell m}({\bm{x_o}}\,, \eta_o)\pm i B_{\ell m}({\bm{x_o}}\,, \eta_o))\, _{\pm 2} Y_{\ell m}({\bm{n}})\,.
\end{align}
We decompose the complex Stokes parameters in Fourier modes
\be
P_{\pm}({\bm{x_o}}\,, \eta_o\,, {\bm{n}})=\int \frac{d^3 k}{(2 \pi)^{3/2}}\, \hat{P}_{\pm}({\bm{k}}\,, \eta_o\,, {\bm{n}})\,e^{i {\bm{k}}\cdot {\bm{x_o}}}\,.
\ee

We assume statistical isotropy of the CMB without the hole. It follows that $\Theta({\bm{x_o}}\,,\eta_o\,, {\bm{n}})$ is a stochastic variable, which can be characterized by its correlation function 
\be
C(\vartheta)=\langle \Theta ({\bm{x_o}}\,,\eta_o\,, {\bm{n_1}}) \Theta({\bm{x_o}}\,,\eta_o\,, {\bm{n_2}}) \rangle\,,
\ee
where $\cos\vartheta={\bm{n_1}}\cdot {\bm{n_2}}$.  The statistical local isotropy implies that this correlation function only depends on the relative angle between the two directions of observation ${\bm{n}_1}$ and  ${\bm{n}_2}$. It is convenient to expand this correlation function in a basis of Legendre polynomials as
\be
C(\vartheta)=\sum_{\ell} \frac{2\ell+1}{4\pi} C^{\Theta\Theta}_{\ell} \,P_{\ell}({\bm{n}_1}\cdot{\bm{n}_2})\,,
\ee
which defined the angular power spectrum $C^{\Theta\Theta}_{\ell}$. 
%A multipole $\ell$ corresponds approximately to an angular scale $\pi/\vartheta$ so that $C^{\Theta\Theta}_{\ell}$ is a measure of the variance of the temperature fluctuation at this scale. 
If the temperature fluctuation has a Gaussian statistics, this function entirely characterizes the temperature distribution. Using eq. (\ref{decomp1}) it is easy to check that
\be\label{cln}
\langle \Theta_{\ell m} ({\bm{x_o}}\,,\eta_o)\,\Theta^*_{\ell' m'}({\bm{x_o}}\,,\eta_o)\rangle=C_{\ell}^{\Theta\Theta} \delta_{\ell \ell'} \delta_{mm'}\,.
\ee
Since  $\Theta({\bm{x_o}}\,,\eta_o\,, {\bm{n}})$ is a stochastic variable, in Fourier space we can write
\be
\hat{\Theta}({\bm{k}}, \eta_o\,, {\bm{n}})\equiv \hat{ \Theta}(k, \eta_o\,, {\bm{n}}) a({\bm{k}})\,,
\ee
where $a({\bm{k}})$ is a unit Gaussian random variable satisfying
\be
\langle  a({\bm{k}}) a({\bm{k'}})^*\rangle=\delta^3({\bm{k}}-{\bm{k'}})\,.
\ee
Hence
\be
\int \frac{dk}{2\pi^2}\,k^2 \langle \hat{\Theta}_{\ell' m'}(k\,, \eta_o)\hat{\Theta}^*_{\ell m}(k\,, \eta_o)\rangle=\delta_{\ell \ell'}\delta_{m m'} C_{\ell}^{\Theta\Theta}\,.
\ee
Completely analogous results hold for polarization. 

In the absence of the hole, temperature anisotropy and polarization fields $P_{\pm}$ are stochastic variables, fully described by the following correlators
\be\label{correlators1}
\hspace{-1 em} \langle \Theta_{\ell m}\Theta_{\ell'm'}^*\rangle =C_{\ell}^{\Theta \Theta}\delta_{\ell \ell'} \delta_{m m'}\,,\,\,  \langle E_{\ell m} E^*_{\ell' m'}\rangle=C_{\ell}^{EE} \delta_{\ell \ell'} \delta_{m m'}\,, \,\, \langle B_{\ell m} B^*_{\ell' m'}\rangle=C_{\ell}^{BB} \delta_{\ell \ell'} \delta_{m m'}\,,
\ee 
together with the correlators with the temperature anisotropy field 
\be\label{correlators}
\langle E_{\ell m} \Theta^*_{\ell' m'}\rangle=C_{\ell}^{E\Theta} \delta_{\ell \ell'} \delta_{m m'}\,, \qquad \langle B_{\ell m} \Theta^*_{\ell' m'}\rangle=C_{\ell}^{B\Theta} \delta_{\ell \ell'} \delta_{m m'}\,.
\ee
Parity is conserved and we consider that primary anisotropy do not generarate $B$-mode polarization. From now on we indicate with a tilde the temperature anisotropy field and Stokes parameters seen by an observer inside the hole.

\subsection{Correlation functions in the absence of statistical isotropy}\label{correlation}

 The CMB sky seen by an observer into the hole is not statistically isotropic. In the absence of statistical isotropy, the correlation function of the lensed temperature anisotropy and polarization are defined as
\be
\tilde{C}({\bm{n_1}}, {\bm{n_2}})\equiv \langle X ({\bm{n_1}}) Y({\bm{n_2}}) \rangle\,,
\ee
where $X\,,Y=\tilde{\Theta}\,,\tilde{E}\,,\tilde{B}$. Since statistical isotropy is violated, the correlation $\tilde{C}({\bm{n_1}}, {\bm{n_2}})$ is estimated by a single product $X ({\bm{n_1}}) Y({\bm{n_2}})$ and hence it is poorly determined by a single realization.  Anyway, even if the nature of the violation of statistical isotropy is not known, some measures of statistical anisotropy of the CMB map can be estimated trough suitably weighted angular averages of $ X ({\bm{n_1}}) Y({\bm{n_2}})$, see e.g. Ref.\cite{Hajian:2003qq}.  It is useful to expand the 2-point correlator in terms of the orthonormal set of Bipolar Spherical Harmonics (BipoSH) as\footnote{The BipoSH transform just as standard spherical harmonics under rotation.}
\be
\tilde{C}({\bm{n_1}}, {\bm{n_2}})=\sum_{\ell_1 \ell_2 L M} A_{\ell_1 \ell_2}^{L M}|_{(X Y)}\,\left\{ Y_{\ell_1}({\bm{n_1}})\otimes Y_{\ell_2}({\bm{n_2}})\right\}_{L M}\,,
\ee
where $A_{\ell_1 \ell_2}^{L M}|_{(X Y)}$ are the coefficients of the expansion and they are called Bipolar Spherical Harmonics coefficients. These coefficients can be directly expressed as a function of the product $X_{\ell_1 m_1}Y^*_{\ell_2 m_2}$ as
\be\label{AAm}
A_{\ell_1 \ell_2}^{L M}|_{(X Y)}=\sum_{m_1 m_2}\langle X_{\ell_1 m_1}Y^{*}_{\ell_2 m_2}\rangle (-)^{m_2}  \,\mathfrak{C}_{\ell_1\,m_1\,\ell_2\,-m_2}^{LM}\,,
\ee
where $\mathfrak{C}_{\ell_1\,m_1\,\ell_2\,-m_2}^{LM}$ are Clebsch-Gordan coefficients. Under the hypothesis of statistical isotropy, the covariance matrix in the spherical harmonic space is diagonal, which implies only $A_{\ell\ell}^{00}|_{XY}\neq 0$. It is convenient to recast eq. (\ref{AAm}) in the following form 
\be\label{Am}
A_{\ell \,\ell+L}^{\ell'\, m'}|_{(X Y)}=\sum_{m M} F_{\ell \,m}^{L \,M} |_{(X Y)}(-)^{m+M} \,\mathfrak{C}_{\ell\,m\,\ell+L\,\,-(m+M)}^{\ell'\,m'}\,,
\ee
where $F_{\ell m}^{L\,M}|_{(X\,Y)}$ are defined in eq. (\ref{FFm}) 
and the Clebsch-Gordan coefficients $\mathfrak{C}_{\ell\,m\,\ell+L\,-(m+M)}^{\ell'\,m'}$ are explicitly given by 
\be
\mathfrak{C}_{\ell\,m\,\ell+L\,-(m+M)}^{\ell'\,m'}=(-)^{-L+m'}\sqrt{2\ell'+1}
\left(
\begin{array}{ccc}
\ell&\ell+L&\ell'\\
m&-m-M&-m'\\
\end{array}
\right)\,.
\ee

In the presence of statistical non-isotropy, the BipoSH coefficients are a complete representation of statistical isotropy violation. In the remaining of this work, we focus on the calculation of the 2-point correlation function $F_{\ell m}^{L\,M}|_{(XY)}$. Once this last is known, the BipoSH coefficients are straightforwardly determined by eq. (\ref{Am}).

\subsection{CMB temperature}\label{CMB temperature}

The lensed and delayed temperature anisotropy field can be expanded up to first order in lensing and time delay as
\be\label{tail}
\tilde{\Theta}({\bm{x_o}}\,, \eta_o\,,{\bm{n}})=\Theta({\bm{x_o}}\,, \eta_o\,,{\bm{n}})+\Theta^{\varphi}({\bm{x_o}}\,, \eta_o\,,{\bm{n}})+\Theta^{d}({\bm{x_o}}\,, \eta_o\,,{\bm{n}})\,,
\ee
where $\Theta({\bm{x_o}}\,, \eta_o\,,{\bm{n}})$ is the zeroth order contribution from the primary anisotropies while $\Theta^{\varphi}$ and $\Theta^d$ are the lensing and delay effects, linear in the lensing potential and in the radial modulation, respectively. We define  the Fourier transform of the various contributions in (\ref{tail}) as 
\be\label{transformtail}
\Theta^{\bullet}({\bm{x_o}}\,, \eta_o\,,{\bm{n}})=\int \frac{d^3 k}{(2\pi)^{3/2}}\,\hat{\Theta}^{\bullet}({\bm{k}}\,,\eta_o\,,{\bm{n}})\,e^{i {\bm{k}}\cdot {\bm{x_o}}}\,,
\ee
where $\bullet=\left(\text{nothing}\,, \varphi\,, d\right)$  for the primary anisotropy contribution, the lensed one and the time-delayed one, respectively.

The CMB temperature field on the sky may be written implicitly as the projection of sources $S$ which contribute in the optically thin regime and are so weighted by $e^{-\tau}$ where $\tau$ is the optical depth. In general, these sources have an intrinsic angular structure on their own  and are characterized by the spherical harmonic moments of their Fourier amplitude $S_{\ell_i}^{m_i}(k)$. Explicit forms for the sources are given in Ref. \cite{Hu:1997hp}. It is convenient to chose a specific frame where $\hat{{\bm{z}}} \parallel \hat{{\bm{k}}}$. We focus on the zeroth order contribution from the primary anisotropies in eq. (\ref{tail}). The contribution from a given wave number $k$ to the temperature anisotropy field in the sky today, can be formally expressed as
\be\label{F.23}
\hat{\Theta}({\bm{k}}, \eta_o\,, {\bm{n}})=\int _{\eta_{\text{LSS}}}^{\eta_o} d\eta \,e^{-\tau}\,\sum_{\ell_i m_i} S_{\ell_i}^{m_i}(\eta, k)\,G_{\ell_i}^{m_i}(\chi{\bm{n}}\,, {\bm{k}})\,,
\ee
where $\chi(\eta)=\eta_o-\eta$ and 
\be
G_{\ell}^{m}({\bm{x}}, {\bm{k}})\equiv (-i)^{\ell} \sqrt{\frac{4\pi}{2\ell+1}}\,Y_{\ell m}({\bm{n}})\,e^{i{\bm{k}}\cdot{\bm{x}}}\,, 
\ee
where 
%The angular structure of these relations can be simplified by considering a specific frame where $\hat{{\bm{z}}} \parallel \hat{{\bm{k}}}$ and
\be
e^{i{\bm{k}}\cdot{\bm{x}}}=\sum_{\ell } (-i)^{\ell} \sqrt{4\pi(2\ell+1)}\,j_{\ell}(k\chi) Y_{\ell0}({\bm{n}})\,.
\ee
The separation of the mode function $G_{\ell}^{m}$ into an intrinsic angular dependence and plane-wave spatial dependence is essentially a division into spin $_sY_{\ell m}$ and orbital $Y_{\ell 0}$ angular momentum. Since only the total angular dependence is observable, it is instructive to employ Clebsch-Gordan relations to add the angular momenta \cite{Ruth_book}. In this specific frame with  for $\hat{{\bm{z}}} \parallel \hat{{\bm{k}}}$,  the temperature field can be written as 
\begin{align} \label{senzaI}
\hat{\Theta}^{\parallel}({\bm{k}}\,, \eta_o\,,{\bm{n}})&=\sum_{\ell m}\hat{\Theta}^{\parallel}_{\ell m} ({\bm{k}}\,,\eta_o) Y_{\ell m}({\bm{n}})\\
&=\sum_{\ell m_i} I^{\parallel}_{m_i}[j_{\ell}]\,Y_{\ell m_i}({\bm{n}})\,,
\end{align}
where we have introduced the operator
\be
I^{\parallel}_{m_i}[j_{\ell}]\equiv \int_{\eta_{\text{LSS}}}^{\eta_o} d\eta\,e^{-\tau}\,\sqrt{4\pi(2\ell+1)}\,\sum_{\ell_i} S_{\ell_i}^{m_i}(\eta, k)\,j_{\ell}^{\ell_im_i}(k\chi)\,,\label{F.28}
\ee
where $j_{\ell}^{\ell_i m_i}$ are linear combinations of $j_{\ell}$ weighted by the Clebsch-Gordan coefficients of the couplings \cite{Hu:1997hp}.  This result can be generalized to a generic reference frame by a simple rotation of the result  (\ref{senzaIfinal}). We get 
\begin{align} \label{senzaIfinal}
\hat{\Theta}({\bm{k}}\,, \eta_o\,,{\bm{n}})&=\sum_{\ell m}\hat{\Theta}_{\ell m} ({\bm{k}}\,,\eta_o) Y_{\ell m}({\bm{n}})\\
&=\sum_{\ell m_i} I_{m_i}[j_{\ell}]\,Y_{\ell m_i}({\bm{n}})\,,\label{a above}
\end{align}
with
\begin{align}
&\hat{\Theta}_{\ell m} ({\bm{k}}\,,\eta_o)\equiv \sum_{m'}D_{mm'}^{\ell}(\bm{\hat{k}})\hat{\Theta}^{\parallel}_{\ell m'} ({\bm{k}}\,,\eta_o)\,,\\
& I_{m_i}[j_{\ell}]\equiv \sum_{m'}D_{m_im'}^{\ell}(\bm{\hat{k}})I^{\parallel}_{m_i}[j_{\ell}]\,,
\end{align}
where $D_{m m'}^{\ell}$ are the matrix elements of the spin representations of the rotation group (see e.g. \cite{Ruth_book}).

The lensed contribution in eq. (\ref{tail}) is defined as 
\be\label{thetaphi}
\Theta^{\varphi}({\bm{x_o}}\,, \eta_o\,,{\bm{n}})=\nabla_i \Theta({\bm{x_o}}\,, \eta_o\,,{\bm{n}})\nabla^i \varphi({\bm{x_o}}\,, \eta_o\,,{\bm{n}})\,.
\ee
Using  eq. (\ref{a above}) in  eq. (\ref{thetaphi}), we get 
\be
\hat{\Theta}^{\varphi}({\bm{k}}\,, \eta_o\,,{\bm{n}})=\nabla_i\varphi({\bm{n}})\,\sum_{\ell m_i} I_{m_i}[j_{\ell}]\,\nabla^i Y_{\ell m_i}({\bm{n}})\,.\label{taillensing}\\
%\hat{\Theta}^{d}({\bm{k}}\,, \eta_o\,,{\bm{n}})&=&d({\bm{n}}) \sum_{\ell m_i} I_{m_i}[j'_{\ell}]\, k(\eta_o-\eta_{\text{LSS}})\,  Y_{\ell m_i}({\bm{n}})\,,\label{taildelay}
\ee

To calculate the time-delay contribution in eq. (\ref{tail}), we take into account that in the fixed time-interval since last scattering, the distance travelled by the photon is perturbed as $\chi\rightarrow \chi(1+d({\bm{n}}))$.\footnote{We underline that the time-delay contribution can not be deduced from the unlensed and undelayed temperature anisotropy field.} We can therefore repeat the same passages (\ref{F.23})-(\ref{F.28}) for $\chi\rightarrow \chi(1+d({\bm{n}}))$ and keep only contributions linear in $d({\bm{n}})$. We get 
\be\label{taildelay}
\hat{\Theta}^{d}({\bm{k}}\,, \eta_o\,,{\bm{n}})=\sum_{\ell m_i} J_{m_i}[j_{\ell}]\,Y_{\ell m_i}({\bm{n}})\,,
\ee
with
\begin{align}
& J_{m_i}[j_{\ell}]\equiv \sum_{m'}D_{m_im'}^{\ell}(\bm{\hat{k}})J^{\parallel}_{m_i}[j_{\ell}]\,,\\
& J^{\parallel}_{m_i}[j_{\ell}]\equiv  d({\bm{n}}) \int_{\eta_{\text{LSS}}}^{\eta_o} d\eta\,e^{-\tau}\,\sqrt{4\pi(2\ell+1)}\,\sum_{\ell_i} S_{\ell_i}^{m_i}(\eta, k)\,\frac{d\,j_{\ell}^{\ell_im_i\,}(k\chi)}{d(k\chi)} \,k(\eta_o-\eta)\,,
\end{align}
The result in eq. (\ref{taillensing}) and (\ref{taildelay}) reduces to the one of~\cite{Hu:2001yq} for a reference frame $\hat{{\bm{z}}} \parallel \hat{{\bm{k}}}$. 
It is useful to introduce the following parametrization
\be\label{par}
\Theta^{d}({\bm{x_o}}\,, \eta_o\,,{\bm{n}})=\bar{\Theta}({\bm{x_o}}\,, \eta_o\,,{\bm{n}})d(\bm{n})\,,
\ee
where the function $\bar{\Theta}({\bm{x_o}}\,, \eta_o\,,{\bm{n}})$ is independent of $\Theta({\bm{x_o}}\,, \eta_o\,,{\bm{n}})$ and its explicit expression can be derived  simply dividing eq. (\ref{taildelay}) by $d({\bm{n}})$. 
%\be
%\bar{\Theta}({\bm{x_o}}\,, \eta_o\,,{\bm{n}})\equiv \int \frac{d^3 k}{(2\pi)^{3/2}}\,e^{i {\bm{k}}\cdot {\bm{x_o}}}\, \sum_{\ell m_i} J_{m_i}[j_{\ell}]\,  Y_{\ell m_i}({\bm{n}})\,.
%\ee

We decompose in multipoles the temperature anisotropy field lensed-delayed (\ref{tail}) and  the single contributions of lensing and time delay, as in (\ref{decomp1}). We note the corresponding multipoles as $\tilde{\Theta}({\bm{x_o}}\,, \eta_o)_{\ell m}$\,, $\Theta^{\varphi}({\bm{x_o}}\,, \eta_o)_{\ell m}$ and  $\Theta^{d}({\bm{x_o}}\,, \eta_o)_{\ell m}$\,, respectively.  From now on, to make the notation compact, we omit the dependence of the temperature anisotropy field on the observer position and reception time. We will reintroduce it explicitly when ambiguities may arise. 

In our problem the only source of violation of statistical isotropy has a geometric origin (geometrical lensing and time-delay). We can therefore consider an ensemble average of the 2-point correlator  of the lensed and delayed temperature anisotropy field (\ref{tail}) considering the fact that we are dealing with non-stochastic effects (the lensing potential and the radial modulation go out from ensemble averages). To evaluate the 2-point correlator of temperature anisotropy, we find convenient working separately with lensing and time-delay and linearly sum the effects at the end, as explained in section \ref{corre}. 

\subsubsection{Contribution from lensing-like displacement}

We switch-off the radial modulation in eq. (\ref{tail}) and we study the effects of lensing. We use the decomposition
\be
\Gamma_a({\bm{n}})=\nabla_a \varphi ({\bm{n}})=\sum_{\ell m} \varphi_{\ell m} \nabla_a Y_{\ell m} ({\bm{n}})\,.
\ee
From eqs. (\ref{transformtail})-(\ref{taillensing}) we get the following decomposition
\be
\Theta^{\varphi}({\bm{n}})=\sum_{\ell m \ell_1 m_1} \varphi_{\ell m} \Theta_{\ell_1 m_1} \nabla_a Y_{\ell m} ({\bm{n}}) \nabla^a Y_{\ell_1 m_1} ({\bm{n}})\,.
\ee
Decomposing the temperature anisotropy (lensed and unlensed) in multipoles, after standard manipulations we can extract the multipoles $\Theta^{\varphi}_{\ell m}$ as 
\be\label{mtem}
\Theta^{\varphi}_{\ell m}=\sum_{\ell_1 m_1 \ell_2 m_2} \varphi_{\ell_2 m_2} \Theta_{\ell_1 m_1}\, \mathcal{I}_{\ell \,\ell_1 \ell_2}^{m\, m_1 m_2}\,,
\ee
where $\mathcal{I}_{\dots}^{\dots}$ is defined in appendix \ref{lm}. 
We therefore find for the 2-point  correlator of (\ref{FphiP}) 
\begin{align}
( F_{\ell m}^{L\, M})|_{(\tilde{\Theta} \tilde{\Theta})}^{\varphi}=&
\sum_{\ell_1 m_1 \ell_2 m_2} \varphi_{\ell_2 m_2} \,\,\mathcal{I}_{\ell \,\,\ell_1\,\, \ell_2}^{m\,\, m_1\,\, m_2} 
\langle \Theta_{\ell_1 m_1}  \Theta^*_{\ell+L\,m+M}\rangle+\nn\\
&+
\sum_{\ell_1 m_1 \ell_2 m_2} \varphi^*_{\ell_2 m_2} \,\mathcal{I}_{\ell+L \,\,\ell_1\,\, \ell_2}^{m+M\,\, m_1 m_2}
\langle \Theta_{\ell m}  \Theta^*_{\ell_1 m_1}\rangle\,,
\end{align}
where we have neglected terms quadratic in lensing to be consistent with the expansion (\ref{tail}), which stopped at first oder in the lensing potential. Using eq. (\ref{cln}) it simplifies to
\be\label{very mess}
( F_{\ell m}^{L\, M})|_{(\tilde{\Theta} \tilde{\Theta})}^{\varphi}=C_{\ell+L}^{\Theta\Theta}
\, \varphi_{\ell_1 -M} \,\,\mathcal{I}_{\ell \,\,\ell+L\,\, \ell_1}^{m\,\, m+M\,\, -M} + C_{\ell}^{\Theta\Theta}
\,\varphi^*_{\ell_1 M}\,\mathcal{I}_{\ell+L \,\,\ell\,\, \ell_1}^{m+M\,\, m\,\, M} \,,
\ee
where the summation over $\ell_1$ is understood and the multipoles of the lensing potential are listed in eq. (\ref{mess}). 
The corresponding BipoSH coefficients defined in eq. (\ref{Am}) can be written in the following compact form 
\be
A_{\ell\,\ell+L}^{\ell'\,m'}|_{(\tilde{\Theta} \tilde{\Theta})}=\frac{(-)^{\ell'}}{\sqrt{2\ell'+1}}\,\varphi_{\ell'm'} \left(\mathcal{F}_{\ell\,\ell+L\,\ell'}\right)^*\left[\alpha_+ C^{\Theta\Theta}_{\ell+L}+\alpha_- C^{\Theta\Theta}_{\ell}\right]\,,
\ee
with $\mathcal{F}_{\dots}$  defined in Appendix \ref{lm}.

\subsubsection{Contribution from radial modulation}\label{SShapiro}

We now calculate the contribution to the correlator of  (\ref{FFF}) linear in the radial modulation. The procedure is straightforward: we use eq. (\ref{taildelay}) in (\ref{FdP}) and we introduce the diagonal correlator\footnote{Using the expressions for $\Theta_{\ell m}$ and $\bar{\Theta}_{\ell m}$ derived at the beginning of Sec. \ref{CMB temperature}, it is possible to verify that only diagonal elements of the correlation matrix $\langle \Theta_{\ell m} \bar{\Theta}_{\ell'm'}^*\rangle$ are excited.}
\be
\langle \Theta_{\ell m} \bar{\Theta}_{\ell'm'}^*\rangle=\delta_{\ell\ell'}\delta_{mm'} C_{\ell}^{\Theta\bar{\Theta}}\,.
\ee
We find
\be\label{very mess 2}
( F_{\ell m}^{L\, M})|_{(\tilde{\Theta} \tilde{\Theta})}^{d}=C_{\ell+L}^{\Theta\bar{\Theta}}
\, d_{\ell_1 -M} \,\,\mathcal{C}_{\ell \,\,\ell+L\,\, \ell_1}^{m\,\, m+M\,\, -M} + C_{\ell}^{\Theta\bar{\Theta}}
\,d^*_{\ell_1 M}\,\mathcal{C}_{\ell+L \,\,\ell\,\, \ell_1}^{m+M\,\, m\,\, M} \,,
\ee
where the summation over $\ell_1$ is understood and the multipoles of the time-delay potentials are listed in eq. (\ref{mess2})

 It is easy to verify from an inspection of (\ref{taillensing}) and (\ref{taildelay}) that the contribution of the time-delay to the temperature anisotropy field is subdominant with respect to the one coming from lensing. Indeed, the lensing depends on the angular gradient of the lensing potential and its observable consequences are weighted by $\ell(\ell+1)$. This has the effect of increasing the magnitude of the effect and shifting it to higher multipoles. The fact that the  effect  of the time-delay is negligible with respect to the one of lensing can be understood also from geometrical considerations, using the analytic results found for time delay potential and lensing. In the boosted frame defined in section \ref{boost}, we found
\be
\tilde{d}'\simeq \hat{r}_S^{3/2}\,,\qquad  \tilde{\Gamma}' \simeq \nabla \varphi \simeq \hat{r}_S\,,\qquad \langle \tilde{\chi}'_{\text{LSS}} \rangle\simeq \frac{2}{\hat{r}_S^{1/2}}\, \chi_h,.
\ee
It follows that the geometrical displacement on the LSS generated by lensing is of order $\tilde{\Gamma}'  \langle\tilde{\chi}'_{\text{LSS}} \rangle \simeq \hat{r}_S^{1/2}\, \chi_h$ while the one generated by time delay is $\tilde{d}' \langle \tilde{\chi}'_{\text{LSS}} \rangle\simeq \chi_h \hat{r}_S$. Therefore the geometrical effect of time-delay  is suppressed by a factor $\hat{r}_S^{1/2}$. For these reasons, from now on we neglect the effect of time delay on the CMB sky. 

\subsection{CMB polarization}\label{CMB polarization}

A single Fourier mode of the complex Stokes parameters can be decomposed in spherical harmonics as in eq. (\ref{pizza2}). 
%\be
% \hat{P}_{\pm}({\bm{k}}\,, \eta_o\,, {\bm{n}})=\sum_{\ell m}\left[\hat{E}_{\ell m}({\bm{k}}\,, \eta_o)\pm i \hat{B}_{\ell m}({\bm{k}}\,, \eta_o)\right]\, _{\pm 2} Y_{\ell m}({\bm{n}})\,,
%\ee
Taking into account the angular structure of the sources in the LSS surface, following a similar reasoning to the one presented for temperature anisotropy field, we get
\be\label{above}
 \hat{P}_{\pm}({\bm{k}}\,, \eta_o\,, {\bm{n}})=\sum_{\ell m_i}P_{m_i}[_{\pm} \alpha_{\ell}]\, _{\pm 2} Y_{\ell m}({\bm{n}})\,,
\ee
where $P_{m_i}[_{\pm} \alpha_{\ell}]$ is the generalization of the operator defined in Ref. \cite{Hu:1997hp} for the case $\bm{k}\parallel \bm{z}.$\footnote{The operator (\ref{above}) for a generic reference frame is  obtained  rotating the corresponding result presented in  \cite{Hu:1997hp} for the special case ${\bm{k}}\parallel {\bm{z}}$, analogously to what we did in Sec. \ref{CMB temperature}}. It contains information on the angular structure of the sources and it is a function of $_{\pm} \alpha_{\ell}=\epsilon_{\ell}^{m_i}\pm i \beta_{\ell}^{m_i}$. The latter are linear combinations of spherical Bessel functions $j_{\ell}$ defined in Ref. \cite{Hu:1997hp} which defines the projection of the source onto the $E$ and $B$ polarization modes. In our derivation we do not need the explicit expression of this operator and the interested reader can go to Ref. \cite{Hu:1997hp} for details on its derivation.

We indicate with $\tilde{P}_{\pm}$ the lensed and delayed Stokes parameters and we expand them up to first order in lensing potential and time delay as
\be\label{ccic}
 \tilde{P}_{\pm}({\bm{x_o}}\,, \eta_o\,, {\bm{n}})= P_{\pm}({\bm{x_o}}\,, \eta_o\,,{\bm{n}})+P^{\varphi}_{\pm}({\bm{x_o}}\,, \eta_o\,,{\bm{n}})+P^d_{\pm}({\bm{x_o}}\,, \eta_o\,,{\bm{n}})\,,
 \ee
where $P_{\pm}$ is the zeroth order contribution from the primary anisotropies while $P_{\pm}^{\varphi}$ and $P_{\pm}^d$ are the lensing and delay effects, linear in the lensing potential and in the radial modulation, respectively. Defining the Fourier transform of the secondary anisotropy contributions
\be\label{transformtailP}
P_{\pm}^{\varphi\,,d}({\bm{x_o}}\,, \eta_o\,,{\bm{n}})=\int \frac{d^3 k}{(2\pi)^{3/2}}\,\hat{P}_{\pm}^{\varphi\,,d}({\bm{k}}\,,\eta_o\,,{\bm{n}})\,e^{i {\bm{k}}\cdot {\bm{x_o}}}\,,
\ee
we get
\begin{align}\label{taillensingP}
&\hat{P}_{\pm}^{\varphi}({\bm{k}}\,, \eta_o\,,{\bm{n}})=\sum_{\ell m_i} P_{m_i}[_{\pm} \alpha_{\ell}]\,\nabla_i\varphi({\bm{n}})\,\nabla^i \, _{\pm 2}Y_{\ell m_i}({\bm{n}})\,,\\
&\hat{P}^{d}_{\pm}({\bm{k}}\,, \eta_o\,,{\bm{n}})=\sum_{\ell m_i} P_{m_i}[_{\pm} \alpha'_{\ell}]\, k(\eta_o-\eta_{\text{LSS}})\, d({\bm{n}}) \, _{\pm 2}Y_{\ell m_i}({\bm{n}})\,,\label{taildelayP}
\end{align}
where a prime indicates derivative with respect to the argument of the spherical Bessel function.  The same reasoning as in Sec. \ref{CMB temperature} can be applied here to show that in (\ref{ccic}) the contribution of time-delay is subdominant with respect to the one of lensing. In the following, we will focus only on this latter.  To simplify the notations, from now on we omit the dependence on the observer position and on the reception time, re-introducing it only when needed.

We introduce the following decomposition
\begin{align}
P_{\pm}^{\varphi}({\bm{n}})&=\sum_{\ell m}\, _{\pm 2}Y_{\ell m} ({\bm{n}}) \,(P_{\pm})_{\ell m}^{\varphi}\,,\\
&=\sum_{\ell_1 m_1 \ell_2 m_2}\, (P_{\pm})_{\ell_1 m_1} \varphi_{\ell_2 m_2}\,\nabla_i Y_{\ell_2 m_2} \nabla^i (_{\pm 2}Y_{\ell_1 m_1})\,,
\end{align}
from which we get
\be\label{mpol}
(P_{\pm})_{\ell m}^{\varphi}=\sum_{\ell_1 m_1 \ell_2 m_2} \varphi_{\ell_2 m_2} (P_{\pm})_{\ell_1 m_1} (\mathcal{I}_{\ell\,\,\ell_1\,\,\ell_2}^{m\,\,m_1\,\,m_2})_{\pm2}\,.
\ee
with $(\mathcal{I}_{\ell\,\,\ell_1\,\,\ell_2}^{m\,\,m_1\,\,m_2})_{\pm2}$ defined in appendix \ref{lm}. 
We calculate the following correlators \small
\begin{align}\label{ci}
\langle (P_{\pm})_{\ell m} (P_{\pm})^*_{\ell+L\,m+M}\rangle^{\varphi}&=C_{\ell+L}^{EE} \varphi_{\ell_1 -M} (\mathcal{I}_{\ell\,\,\ell+L\,\,\ell_1}^{m\,\,m+M\,\,-M})_{(\pm 2)}+C_{\ell}^{EE} \varphi^*_{\ell_1 M} (\mathcal{I}_{\ell+L\,\,\ell\,\,\ell_1}^{m+M\,\,m\,\,M})_{(\pm 2)}\,.
\end{align}
\begin{align}\label{cip}
\langle (P_{\pm})_{\ell m} (P_{\mp})^*_{\ell+L\,m+M}\rangle^{\varphi}&=C_{\ell+L}^{EE} \varphi_{\ell_1 -M} (\mathcal{I}_{\ell\,\,\ell+L\,\,\ell_1}^{m\,\,m+M\,\,-M})_{(\pm 2)}+C_{\ell}^{EE}  \varphi^*_{\ell_1 M} (\mathcal{I}_{\ell+L\,\,\ell\,\,\ell_1}^{m+M\,\,m\,\,M})_{(\mp 2)}\,,
\end{align} \normalsize
where a summation over $\ell_1$ is understood. 
Using that 
\be
(P^{\varphi}_{\pm})_{\ell m}=\tilde{E}_{\ell m}\pm i \tilde{B}_{\ell m}\,,
\ee
taking combinations of (\ref{ci}) and (\ref{cip}), we get
\begin{align}
( F_{\ell m}^{L\,M})|_{(\tilde{E}\,\tilde{E})}^{\varphi}&=C^{EE}_{\ell+L}\sum_{\ell_1}  \varphi_{\ell_1 -M} \,\mathcal{R}_{\ell\,\,\ell+L\,\,\ell_1}^{m\,\,m+M\,\,-M}+C^{EE}_{\ell}\sum_{\ell_1} \varphi_{\ell_1 M}^* \,\mathcal{R}_{\ell+L\,\,\ell\,\,\ell_1}^{m+M\,\,m\,\,M}\,,\\
(F_{\ell m}^{L\,M})|_{(\tilde{B}\,\tilde{B})}^{\varphi}&=0\,,\\
(F_{\ell m}^{L\,M})|_{(\tilde{E}\,\tilde{B})}^{\varphi}&=-i\,C^{EE}_{\ell+L}\sum_{\ell_1}  \varphi_{\ell_1 } \,\mathcal{Q}_{\ell\,\,\ell+L\,\,\ell_1}^{m\,\,m+M\,\,-M}-i\,C^{EE}_{\ell}\sum_{\ell_1} \varphi_{\ell_1 M}^* \,\mathcal{Q}_{\ell+L\,\,\ell\,\,\ell_1}^{m+M\,\,m\,\,M}\,,
\end{align}
where we have defined
\begin{align}\label{R}
&(\mathcal{R}_{\ell\,\,\ell+L\,\,\ell_1}^{m\,\,m+M\,\,m_1})\equiv \frac{1}{2}\left(\mathcal{I}_{\ell\,\,\ell+L\,\,\ell_1}^{m\,\,m+M\,\,m_1})_{+2}+(\mathcal{I}_{\ell\,\,\ell+L\,\,\ell_1}^{m\,\,m+M\,\,m_1})_{-2}\right)=\delta^{(+)}_{L+\ell_1} \left(\mathcal{I}_{\ell\,\,\ell+L\,\,\ell_1}^{m\,\,m+M\,\,m_1}\right)_{+2}\,,\\
&(\mathcal{Q}_{\ell\,\,\ell+L\,\,\ell_1}^{m\,\,m+M\,\,m_1})\equiv \frac{1}{2}\left(\mathcal{I}_{\ell\,\,\ell+L\,\,\ell_1}^{m\,\,m+M\,\,m_1})_{+2}-(\mathcal{I}_{\ell\,\,\ell+L\,\,\ell_1}^{m\,\,m+M\,\,m_1})_{-2}\right)=\delta^{(-)}_{L+\ell_1} \left(\mathcal{I}_{\ell\,\,\ell+L\,\,\ell_1}^{m\,\,m+M\,\,m_1}\right)_{+2}\label{Q}\,,
\end{align}
and 
\be
\delta^{(\pm)}_{L+\ell_1}\equiv\frac{1}{2}\left[1\pm (-)^{L+\ell_1}\right]\,.
\ee

The last step is to calculate the correlation between temperature and polarization. Neglecting radial modulation and using eqs. (\ref{mtem}) and (\ref{mpol}) and the correlators (\ref{correlators}) we get
\begin{align}
\langle (\tilde{P}_{\pm})_{\ell m} \tilde{\Theta}^*_{\ell+L m+M}\rangle =&C_{\ell}^{\Theta E} \delta_{\ell \ell+L}\delta_{m +M}+\langle (P_{\pm})_{\ell m} (\Theta^{\varphi})^*_{\ell+L m+M}\rangle+\langle (P_{\pm}^{\varphi})_{\ell m} (\Theta^{\varphi})^*_{\ell+L m+M}\rangle\,\nn\\
=&C_{\ell}^{\Theta E} \delta_{\ell \ell+L}\delta_{m +M}+C_{\ell}^{\Theta E} \sum_{\ell_1} \varphi^*_{\ell_1 M} \,\mathcal{I}_{\ell+L\,\,\ell\,\,\ell_1}^{m+M\,\,m\,\,M} +\nn\\
&+C^{\Theta E}_{\ell+L}\sum_{\ell_1} \varphi_{\ell_1 M} \left(\mathcal{I}_{\ell+L\,\,\ell\,\,\ell_1}^{m+M\,\,m\,\,M}\right)_{\pm 2} \,,
\end{align}
from which we immediately find
\begin{align}
(F_{\ell m}^{L\,M})|_{(\tilde{E}\tilde{\Theta})}^{\varphi}&=C_{\ell}^{\Theta E}\sum_{\ell_1} \varphi_{\ell_1 M}\,\mathcal{I}_{\ell+L\,\, \ell\,\,\ell_1}^{m+M\,\, m\,\,M}+ C_{\ell+L}^{\Theta E}\sum_{\ell_1} \varphi_{\ell_1-M } \,\mathcal{R}_{\ell\,\, \ell+L\,\,\ell_1}^{m\,\, m+M\,\, -M}\,,\nn\\
(F_{\ell m}^{L\,M})|_{(\tilde{B}\tilde{\Theta})}^{\varphi}&=-i \, C_{\ell+L}^{\Theta E}\sum_{\ell_1} \varphi_{\ell_1 -M} \,\mathcal{Q}_{\ell\,\, \ell+L\,\,\ell_1}^{m\,\, m+M\,\, -M}\,.
\end{align}

We can further simplify the expressions recalling the symmetry properties collected in appendix \ref{lm} and using the result (\ref{mess}) for the lensing potential from which we find
\be
\varphi_{\ell\, m}=(-)^m\varphi_{\ell -m}\,.
\ee
We get the final and most important result of this article for the off-diagonal correlators \small
\begin{align}
(F_{\ell m}^{L\,M})|_{(\tilde{\Theta}\tilde{\Theta})}^{\varphi}&=\varphi_{\ell_1-M}\, \mathcal{C}_{\ell \,\,\ell+L\,\, \ell_1}^{m\,\, m+M\,\, -M}\left(\alpha_+ C_{\ell+L}^{\Theta\Theta}+\alpha_- C_{\ell}^{\Theta \Theta} \right) \,\\
(F_{\ell m}^{L\,M})|_{(\tilde{E}\tilde{E})}^{\varphi}&= \delta^{(+)}_{L+\ell_1}\varphi_{\ell_1-M}\,( \mathcal{C}_{\ell \,\,\ell+L\,\, \ell_1}^{m\,\, m+M\,\,-M})_{+2} \left(\alpha_+ C_{\ell+L}^{EE}+\alpha_- C_{\ell}^{EE} \right)\,,\\
(F_{\ell m}^{L\,M})|_{(\tilde{B}\tilde{B})}^{\varphi}&=0\,,\\
(F_{\ell m}^{L\,M})|_{(\tilde{E}\tilde{B})}^{\varphi}&=-i\,\delta^{(-)}_{L+\ell_1}\varphi_{\ell_1-M}\,( \mathcal{C}_{\ell \,\,\ell+L\,\, \ell_1}^{m\,\, m+M\,\, -M})_{+2} \left(\alpha_+ C_{\ell+L}^{EE}+\alpha_- C_{\ell}^{EE} \right)\,,\\
(F_{\ell m}^{L\,M})|_{(\tilde{E}\tilde{\Theta})}^{\varphi}&=C^{E\Theta}_{\ell+L}  \,\varphi_{\ell_1 -M} \,\delta^{(+)}_{L+\ell_1}\alpha_+\,(\mathcal{C}_{\ell\,\,\ell+L\,\,\ell_1}^{m\,\,m+M\,\,-M})_{+2}+C^{E\Theta}_{\ell} \,\varphi_{\ell_1-M}\,\alpha_- \,\mathcal{C}_{\ell\,\,\ell+L\,\,\ell_1}^{m\,\,m+M\,\,-M}\,,\\
(F_{\ell m}^{L\,M})|_{(\tilde{B}\tilde{\Theta})}^{\varphi}&=-i\,C^{E\Theta}_{\ell+L}  \,\varphi_{\ell_1-M} \,\delta^{(-)}_{L+\ell_1}\alpha_+\,(\mathcal{C}_{\ell\,\,\ell+L\,\,\ell_1}^{m\,\,m+M\,\,-M})_{+2}\,,
\end{align} \normalsize
where a summation over $\ell_1$ is understood and the result for the angular power spectra of the lensing potential are listed in eq. (\ref{mess}).  To make the notation compact we have defined
\be
\alpha_{\pm}\equiv \frac{1}{2}\left[\ell_1(\ell_1+1)\pm L (L+2\ell+1)\right]\,.
\ee

\section{Effect of a boost on the CMB}\label{lastt}

In this appendix we detail the derivation presented in section \ref{Boost} of the correlation function of temperature and polarization in a reference frame in motion with respect to the CMB rest frame. 
\subsection{General formalism for boosts}\label{boost0}

We consider a frame $e_a^\mu=(u^\mu,{e_i}^\mu)$ and a boosted frame
$\tilde e_a^\mu =(\tilde
u^\mu,{\tilde e_i}^\mu)$ (two tetrads) related by a boost such that 
\be
\tilde e^a_\mu = {\Lambda^a}_b e^b_\mu\,, \qquad \tilde e_a^\mu = e_b^\mu
{({\Lambda^{-1}})^b}_a = {\Lambda_a}^b e_b^\mu\,,
\ee
\be
{\Lambda^0}_0 = \gamma \,,\,\quad {\Lambda^0}_i = {\Lambda^i}_0 = -
\gamma v_i\,,\qquad {\Lambda^i}_j = \delta^i_j +
\frac{\gamma^2}{1+\gamma}v^i v_j\,,
\ee
with $\gamma^{-2}=1-v_i v^i$ and $\beta^2 \equiv v_i v^i$.
For a given photon momentum $k^\mu$, the energies measured by the observer ${\bm{u}}$ and ${\bm{\tilde{u}}}$ are related by 
\be\label{TRuleE}
\tilde E = -\tilde {\bm u} \cdot {\bm k} = \gamma (1+ {\bm n}\cdot
{\bm v}) E \equiv \lambda E\,,
\ee
where the Doppler shift factor is
\be\label{alphadef}
\lambda \equiv  \gamma (1+ {\bm n}\cdot
{\bm v})=\frac{1}{\gamma(1-{\bm {\tilde{n}}}\cdot {\bm v})}\,.
\ee
Eq. (\ref{TRuleE}) is more often written in the form (\ref{transE}).
The aberration is the projection in the tilde basis of the tilde direction 
\be
\widetilde{n^i}= \frac{1}{\lambda}\left(n^i
  +\frac{\gamma^2}{1+\gamma} {\bm n} \cdot {\bm v} v^i + \gamma v^i\right)\,.
\ee
Here $n_i \equiv {\bm n} \cdot {\bm e}_i$ and $\widetilde{n_i} \equiv
\tilde{\bm n} \cdot \tilde{\bm e}_i$. The expression for the aberration is more often rewritten in
terms of components along and orthogonally to $v^i$ as (noting $\hat
v^i$ the unit vector in direction of $v^i$)
\beqa
\widetilde{n^i} &=& \frac{\hat v^i}{1+{\bm n}\cdot{\bm v}}\left(n_j \hat
  v^j + \beta\right)+\frac{1}{\gamma(1+{\bm n}\cdot{\bm v})}(n^i-{\bm n}\cdot \textcolor{black}{{\bm{\hat{v}}}} \textcolor{black}{\hat{v}}^i) \,,
\eeqa
which leads to the usual aberration formula
\be
\widetilde{\bm n}\cdot {\bm{\hat v}} =\frac{{\bm n}\cdot {\bm{ \hat v}}+\beta}{1+{\bm n}\cdot {\bm v}} \,,
\ee
which is more often written in the form (\ref{phii}). Finally the number density and the temperature transform respectively as 
\be\label{TT}
\widetilde{f}(\tilde E, \widetilde{n^i}) = f(E,n^i)\,,\qquad\widetilde{T}(\widetilde{n^i}) = \lambda T(n^i)\,.
\ee
Temperature is thus affected by aberration but also by the Doppler shift.

\subsection{CMB correlators in a boosted frame}\label{fame}

To make the results derived in this section compact, let us define a  potential $\zeta$ as
\be
\zeta \equiv  v_i n^i\,.
\ee
Using eq. (\ref{TT}) at linear order in $\beta$,  it follows that the temperature anisotropy field is aberrated and modulated by 
\beqa\label{thetaboost}
\widetilde{\Theta}({\bm{\tilde{n}}}) \simeq \Theta({\bm{\tilde{n}}})(1+\zeta({\bm{\tilde{n}}}))-\nabla^a \zeta({\bm{\tilde{n}}}) \nabla_a \Theta ({\bm{\tilde{
  n}}})+\dots
\eeqa
The Stokes parameters $Q$ and $U$ transform in a completely analogous way \cite{Mukherjee:2013zbi}. It follows that for the complex Stokes parameters up to linear order in $\zeta$ the following expansion holds
\be\label{polboost}
(\widetilde{P_{\pm}})({\bm{\tilde{n}}})\simeq P_{\pm}({\bm{\tilde{n}}})(1+\zeta({\bm{\tilde{n}}}))-\nabla^a \zeta({\bm{\tilde{n}}}) \nabla_a P_{\pm}({\bm{\tilde{n}}})+\dots
\ee
From now on, to make the notation more compact, we omit the tilde over the normal vector $\bm{n}$ indicating the direction of the incoming photon. We will reintroduce it if ambiguities may arise. 
  
We introduce the following spherical harmonic decomposition. For quantities in the $S$ reference frame
\begin{align}
\Theta({\bm{n}})&=\sum_{\ell m} \Theta_{\ell m} Y_{\ell m}({\bm{n}})\,,\\
(P_{\pm})({\bm{n}})&=\sum_{\ell m} (P_{\pm})_{\ell m}\,_ {\pm 2}Y_{\ell m}({\bm{n}})=\sum_{\ell m} (E_{\ell m}\pm i B_{\ell m})\,_ {\pm 2}Y_{\ell m}({\bm{n}})\,.
\end{align} 
For the CMB observables in the $\tilde{S}$ frame
\begin{align}
\tilde{\Theta}({\bm{n}})&=\sum_{\ell m} \tilde{\Theta}_{\ell m} Y_{\ell m}({\bm{n}})\,,\\
(\tilde{P}_{\pm})({\bm{n}})&=\sum_{\ell m} (\tilde{P}_{\pm})_{\ell m}\,_ {\pm 2}Y_{\ell m}({\bm{n}})=\sum_{\ell m} (\tilde{E}_{\ell m}\pm i \tilde{B}_{\ell m})\,_ {\pm 2}Y_{\ell m}({\bm{n}})\,,
\end{align}
and finally
\be
\zeta ({\bm n})= v_i n^i =\sum_{M} \beta_{1M} Y_{1M}({\bm n})\,.
\ee
From eqs. (\ref{thetaboost}) and (\ref{polboost}) we get
\begin{align}
\widetilde{\Theta}_{\ell m} &= \Theta_{\ell m}+\sum_{\ell_1 m_1 m_2} \Theta_{\ell_1 m_1}\,\beta_{1 m_2} \left[\mathcal{C}_{\ell \, \ell_1\,1}^{m\,m_1\, m_2}-\mathcal{I}_{\ell \, \ell_1\,1}^{m\,m_1\, m_2} \right]\,,\\
(\widetilde{P}_{\pm})_{\ell m} &= (P_{\pm})_{\ell m}+\sum_{\ell_1 m_1 m_2}(P_{\pm})_{\ell_1 m_1}\,\beta_{1 m_2} \left[(\mathcal{C}_{\ell \, \ell_1\,1}^{m\,m_1\, m_2})_{\pm2}-(\mathcal{I}_{\ell \, \ell_1\,1}^{m\,m_1\, m_2} )_{\pm 2}\right]\,,
\end{align}
where the objects $\mathcal{C}_{\dots}^{\dots}$ and $\mathcal{I}_{\dots}^{\dots}$ are defined in appendix \ref{lm}. 

Statistical isotropy in the CMB rest frame $S$ still leads to  observable statistical non-isotropy in the observer frame $\tilde{S}$. In $S$ we consider parity to be conserved and primary anisotropy not to generarate B-mode polarization, i.e. we consider correlators (\ref{correlators1}) and (\ref{correlators}). 
Since in the $\tilde{S}$ frame statistical isotropy is broken, we introduce BipoSH coefficients to characterize the CMB sky defined in appendix \ref{correlation}.  We decompose the boosted 2-point function $\widetilde{F}_{\ell m}^{L\,M}|_{(X\,Y)}$ in the following way
\be
\tilde{F}_{\ell m}^{L M}|_{(X\,Y)}=F_{\ell m}^{L M}|_{(X\,Y)}+(F_{\ell m}^{L M})^{\beta}_{(X\, Y)}\,,
\ee
where $F_{\ell m}^{L M}|_{(X\, Y)}$ is the 2-point function in the CMB rest frame,  $(F_{\ell m}^{L M})^{\beta}_{(X\, Y)}$ indicates the contribution linear in $\beta$ and $X\,, Y=\tilde{\Theta}\,,\tilde{E}\,,\tilde{B}$. 

For the temperature anisotropy field we simply get
\be
(F_{\ell m}^{L M})|^{\beta}_{(\tilde{\Theta}\tilde{\Theta})}=\frac{L}{2}(L+2\ell+1)\,\beta_{1 -M}\, \mathcal{C}_{\ell\,\ell+L\,1}^{m\,m+M\,-M}\left(C_{\ell}^{\Theta\Theta}-C_{\ell+L}^{\Theta\Theta}\right)\,.
\ee
For the polarization, we start calculating the correlators of the complex Stokes parameters. Using the properties listed in appendix \ref{lm}, we get \small
\begin{align}\label{pm}
\hspace{-0 em}\langle (P_{\pm})_{\ell m} (P_{\pm})_{\ell+L m+M}^*\rangle^{\beta}&=\frac{L}{2}(L+2\ell+1) \beta_{1 -M} \left(\mathcal{C}_{\ell\,\ell+L\,1}^{m\,m+M\,-M}\right)_{\pm 2}\left(C_{\ell}^{EE}-C_{\ell+L}^{EE}\right)\,,\\
\hspace{-0 em}\langle (P_{\pm})_{\ell m} (P_{\mp})_{\ell+L m+M}^*\rangle^{\beta}&=\frac{L}{2}(L+2\ell+1)\beta_{1 -M}\left[C_{\ell}^{EE} \left(\mathcal{C}_{\ell\,\ell+L\,1}^{m\,m+M\,-M}\right)_{\mp 2}-C_{\ell+L}^{EE}\left(\mathcal{C}_{\ell\,\ell+L\,1}^{m\,m+M\,-M}\right)_{\pm 2}\right]\,.\label{mp}
\end{align} \normalsize
Taking linear combinations of (\ref{pm}) and (\ref{mp}), we get
\begin{align}
(F_{\ell m}^{L M})|^{\beta}_{(\tilde{E} \tilde{E})}&=\frac{L}{2}(L+2\ell+1)\, \beta_{1 -M} \mathcal{Y}_{\ell\,\ell+L\,1}^{m\,m+M\,-M}\left(C_{\ell}^{EE}-C_{\ell+L}^{EE}\right)\,,\\
(F_{\ell m}^{L M})|^{\beta}_{(\tilde{B} \tilde{B})}&=0\,,\\
(F_{\ell m}^{L M})|^{\beta}_{(\tilde{E} \tilde{B})}&=i \,\frac{L}{2}(L+2\ell+1)\, \beta_{1 -M} \mathcal{Z}_{\ell\,\ell+L\,1}^{m\,m+M\,-M}\,C_{\ell}^{EE}\,,
\end{align}
where we have defined 
\begin{align}
\mathcal{Y}_{\ell\,\ell+L\,1}^{m\,m+M\,m_2}&\equiv \frac{1}{2}\left[\left(\mathcal{C}_{\ell\,\ell+L\,1}^{m\,m+M\,m_2}\right)_{+2}+\left(\mathcal{C}_{\ell\,\ell+L\,1}^{m\,m+M\,m_2}\right)_{-2}\right]=\delta_{L+1}^{(+)}\left(\mathcal{C}_{\ell\,\ell+L\,1}^{m\,m+M\,m_2}\right)_{+2}\,,\\
\mathcal{Z}_{\ell\,\ell+L\,1}^{m\,m+M\,m_2}&\equiv \frac{1}{2}\left[\left(\mathcal{C}_{\ell\,\ell+L\,1}^{m\,m+M\,m_2}\right)_{+2}-\left(\mathcal{C}_{\ell\,\ell+L\,1}^{m\,m+M\,m_2}\right)_{-2}\right]=\delta_{L+1}^{(-)}\left(\mathcal{C}_{\ell\,\ell+L\,1}^{m\,m+M\,m_2}\right)_{+2}\,.
\end{align}
For the polarization-temperature cross-correlation, we get 
\be
\hspace{-0.6 em} \langle (P_{\pm})_{\ell m} \Theta_{\ell+L m+M}^*\rangle=\frac{L}{2}(L+2\ell+1)\,\beta_{1 .M}\left[\mathcal{C}_{\ell\,\ell+L\,1}^{m\,m+M\,-M} C_{\ell}^{E\Theta}-(\mathcal{C}_{\ell\,\ell+L\,1}^{m\,m+M\,-M})_{\pm 2}\,C_{\ell+L}^{E\Theta}\right]\,,
\ee
from which we immediately obtain
\begin{align}
( F_{\ell m}^{L M})|^{\beta}_{(\tilde{B} \tilde{\Theta})}&=i\,\frac{L}{2}(L+2\ell+1)\, \beta_{1 -M} \mathcal{Z}_{\ell\,\ell+L\,1}^{m\,m+M\,-M}\,C_{\ell+L}^{E\Theta}\,,\\
(F_{\ell m}^{L M})|^{\beta}_{(\tilde{E} \tilde{\Theta})}&=\frac{L}{2}(L+2\ell+1)\, \beta_{1 -M} \left[\mathcal{C}_{\ell\,\ell+L\,1}^{m\,m+M\,-M} C_{\ell}^{E\Theta}-\mathcal{Y}_{\ell\,\ell+L\,1}^{m\,m+M\,-M}\,C_{\ell+L}^{E\Theta}\right]\,.
\end{align}
The breaking of statistical isotropy becomes most notable at higher multipoles and therefore it can be used to determine our velocity with respect to the CMB rest frame using high angular resolution data from \emph{Planck} without relying on the amplitude and direction of the CMB dipole, see Refs.  \cite{Amendola:2010ty, Chluba:2011zh}. This allows to constraint cosmological models in which the cosmic dipole arises partly from large-scale isocurvature perturbations instead of being fully motion-induced.

\section{Tools for products of spherical harmonics}\label{lm}

The $3-j$ symbol satisfies the following properties
\begin{align}\label{11}
\left(
\begin{array}{ccc}
\ell_1&\ell_2&\ell_3\\
m_1&m_2&m_3\\
\end{array}
\right)&=
\left(
\begin{array}{ccc}
\ell_2&\ell_3&\ell_1\\
m_2&m_3&m_1\\
\end{array}
\right)=
\left(
\begin{array}{ccc}
\ell_3&\ell_1&\ell_2\\
m_3&m_1&m_2\\
\end{array}
\right)\\
&=(-)^{\ell_1+\ell_2+\ell_3}
\left(
\begin{array}{ccc}
\ell_1&\ell_3&\ell_2\\
m_1&m_3&m_2\\
\end{array}
\right)\\
&=(-)^{\ell_1+\ell_2+\ell_3}
\left(
\begin{array}{ccc}
\ell_1&\ell_2&\ell_3\\
-m_1&-m_2&-m_3\\
\end{array}\label{111}
\right)\,.
\end{align}
Moreover, they are identically zero whenever any of the following conditions are violated
\be
m_1+m_2+m_3=0\,,\qquad |\ell_i-\ell_j|\leq \ell_k\leq \ell_i+\ell_j\,,\qquad \{i\,,j\}=\{1,2,3\}\,.
\ee
We recall that the integral of three spin-weighted spherical harmonics can be written as
\begin{align}\label{1111}
\int d\Omega \,_{s_1}&Y_{\ell_1 m_1}\,_{s_2}Y_{\ell_2 m_2}\,_{s_3}Y_{\ell_3 m_3}=\nn\\
&=\sqrt{\frac{(2\ell_1+1)(2\ell_2+1)(2\ell_3+1)}{4\pi}}\left(
\begin{array}{ccc}
\ell_1&\ell_2&\ell_3\\
-s_1&-s_2&-s_3
\end{array}
\right)\left(
\begin{array}{ccc}
\ell_1&\ell_2&\ell_3\\
m_1&m_2&m_3
\end{array}
\right)\,.
\end{align}
We introduce the following objects and their properties, extensively used in this work 
\beqa
C^{m_1 m_2 m_3}_{\ell_1 \ell_2 \ell_3} &\equiv& \int \dd \Omega\,
Y^{\star}_{\ell_1 m_1} Y_{\ell_2 m_2} Y_{\ell_3 m_3}\,,\\
\mathcal{I}^{m_1 m_2 m_3}_{\ell_1 \ell_2 \ell_3} &\equiv& \int \dd \Omega\,
Y^{\star}_{\ell_1 m_1} \nabla^a Y_{\ell_2 m_2} \nabla_a Y_{\ell_3 m_3}\,,
\eeqa
which are related by
\begin{align}
 \mathcal{I}_{\ell_1 \ell_2 \ell_3}^{m_1 m_2 m_3} &= \frac{1}{2}\left[\ell_3
  (\ell_3+1)+\ell_2 (\ell_2+1)-\ell_1 (\ell_1+1)\right]C_{\ell_1 \ell_2 \ell_3}^{m_1 m_2 m_3}\,,\label{EqItoC}\\
C_{\ell_1 \ell_2 \ell_3}^{m_1 m_2 m_3} &=
(-1)^{m_1}\troisj{\ell_1}{\ell_2}{\ell_3}{-m_1}{m_2}{m_3}\,{\cal F}_{\ell_1\ell_2\ell_3}\,,\label{CC}\\
{\cal F}_{\ell \ell_1 \ell_2} &= \sqrt{\frac{(2\ell+1)(2
    \ell_1+1)(2 \ell_2+2)}{4\pi}}\troisj{\ell}{\ell_1}{\ell_2}{0}{0}{0}\,,
\end{align}
while
\begin{align}
 \mathcal{I}_{\ell_1 \ell_2 \ell_3}^{m_1 m_2 m_3} &=
(-1)^{m_1}\troisj{\ell_1}{\ell_2}{\ell_3}{-m_1}{m_2}{m_3}\,{F}_{\ell_1\ell_2\ell_3}\,,\\
F_{\ell \ell_1 \ell_2} &=
\frac{1}{2}[\ell_1(\ell_1+1)+\ell_2(\ell_2+1)-\ell(\ell+1)]  {\cal F}_{\ell \ell_1 \ell_2} \,.
\end{align}
Moreover, it is easy to verify that
\be
 \mathcal{I}_{\ell_1 \ell_2 \ell_3}^{-m_1\, -m_2\, -m_3}=(-)^{\ell_1+\ell_2+\ell_3} \mathcal{I}_{\ell_1 \ell_2 \ell_3}^{m_1 m_2 m_3}\,.
 \ee
 We introduce the following spin-dependent quantities
\beqa
(C^{m_1 m_2 m_3}_{\ell_1 \ell_2 \ell_3})_{\pm s} &\equiv& \int \dd \Omega\,
_{\pm s}Y^{\star}_{\ell_1 m_1}\, _{\pm s}Y_{\ell_2 m_2}\, Y_{\ell_3 m_3}\,,\\
(\mathcal{I}^{m_1 m_2 m_3}_{\ell_1 \ell_2 \ell_3})_{\pm s} &\equiv& \int \dd \Omega\,
_{\pm s} Y^{\star}_{\ell_1 m_1} D^a _{\pm s}Y_{\ell_2 m_2} D_a Y_{\ell_3 m_3}\,,
\eeqa
related by
\be\label{EqItoCspin}
 (\mathcal{I}_{\ell_1 \ell_2 \ell_3}^{m_1 m_2 m_3})_{\pm s} = \frac{1}{2}\left[\ell_3
  (\ell_3+1)+\ell_2 (\ell_2+1)-\ell_1 (\ell_1+1)\right](C_{\ell_1 \ell_2 \ell_3}^{m_1 m_2 m_3})_{\pm s}\,.
\ee
Using eqs. (\ref{11})-(\ref{111}), (\ref{1111}) and recalling 
\be
\left(Y_{\ell m}\right)^*=(-)^m\,Y_{\ell-m}\,,\qquad (_sY_{\ell m})^*=(-)^m \,_{-s} Y_{\ell -m}\,,
\ee
it is easy to verify that
\begin{align}
&\left(C_{\ell_1\,\ell_2\,\ell_3}^{m_1\,m_2\,m_3}\right)_{(\pm s)}=(-)^{m_1+m_2}\left(C_{\ell_2\,\ell_1\,\ell_3}^{-m_2\,-m_1\,m_3}\right)_{(\mp s)}=(-)^{m_1+m_2}\left(C_{\ell_2\,\ell_1\,\ell_3}^{m_2\,m_1\,-m_3}\right)_{(\pm s)}\,,\\
& (\mathcal{I}_{\ell_1 \ell_2 \ell_3}^{-m_1\, -m_2\, -m_3})_{\mp s}= (\mathcal{I}_{\ell_1 \ell_2 \ell_3}^{m_1 m_2 m_3})_{\pm s}\,,\\
& ( \mathcal{I}_{\ell_1 \ell_2 \ell_3}^{m_1 m_2 m_3})_{- s}=(-)^{\ell_1+\ell_2+\ell_3}  ( \mathcal{I}_{\ell_1 \ell_2 \ell_3}^{m_1 m_2 m_3})_{s}\,.
 \end{align}

\newpage
\bibliographystyle{utphys}
\bibliography{myrefs_massive}

\end{document}